\documentclass[usenatbib]{mnras}
\usepackage[T1]{fontenc}

\DeclareRobustCommand{\VAN}[3]{#2}
\let\VANthebibliography\thebibliography
\def\thebibliography{\DeclareRobustCommand{\VAN}[3]{##3}\VANthebibliography}

\usepackage{graphicx}	% Including figure files
\usepackage{amsmath}	% Advanced maths commands
\usepackage{amssymb}	% Extra maths symbols
\usepackage{xspace}

\RequirePackage{lineno}
\usepackage[utf8]{inputenc}
\usepackage{ae,aecompl}
\usepackage{comment}
\usepackage{natbib}
\usepackage{float}
\usepackage{graphicx}
\usepackage{subcaption}
\usepackage{amssymb}
\usepackage{rotating,times,pictex,graphicx,latexsym}
\usepackage{color}
\usepackage{amsmath}
\usepackage{threeparttable}
\usepackage{lipsum}% to automatically generate some text
\usepackage{amsmath}
\usepackage[all]{hypcap}
\usepackage[title,titletoc]{appendix}
\usepackage[]{hyperref}
\PassOptionsToPackage{pdfpagelabels=false}{hyperref}
\usepackage[T1]{fontenc}
\usepackage{textcomp}
\usepackage{gensymb}
\usepackage{newtxtext,newtxmath}
\newcommand{\Ms}{M\ensuremath{_{\odot}}}

\newcommand{\beq}{\begin{equation}}
\newcommand{\eeq}{\end{equation}}

\newcommand{\hii}{\ion{H}{ii}\xspace}
\newcommand{\mum}{\,\ensuremath{\mu}m\xspace}

\newcommand{\arcs}{\hbox{$^{\prime\prime}$}}

\newcommand{\lsun}{\mbox{\rm L$_{\odot}$}}

\newcommand{\kms}{\hbox{km~s$^{-1}$}}
\newcommand{\cms}{\hbox{cm$^{-2}$~}}

\newcommand{\cmq}{\hbox{cm$^{-3}$}}

\newcommand{\nht}{\hbox{N$(\mathrm H_2)$}}

\newcommand{\thco}{\hbox{$^{13}$CO}~}
\newcommand{\thcos}{\hbox{$^{13}$CO}}
\newcommand{\twco}{\hbox{$^{12}$CO}~}
\newcommand{\twcos}{\hbox{$^{12}$CO}}
\newcommand{\eico}{\hbox{C$^{18}$O}~}
\newcommand{\eicos}{\hbox{C$^{18}$O}}

\newcommand{\cloud}{{\rm G148.24+00.41}}

\title[Filamentary Flows in GMC G148.24+00.41]{
\center
 The Giant Molecular Cloud G148.24+00.41: Gas Properties, Kinematics, and Cluster Formation at the Nexus of Filamentary Flows}

\author[Rawat et al.]{
Vineet Rawat,$^{1,2}$\thanks{E-mail: vineet@prl.res.in}
M. R. Samal,$^{1}$
D. L. Walker,$^{3}$
D.K. Ojha,$^{4}$
A. Tej,$^{5}$
A. Zavagno,$^{6,7}$
C.P. Zhang,$^{8,9}$ \newauthor
Davide Elia,$^{10}$
S. Dutta,$^{11}$
J. Jose,$^{12}$
C. Eswaraiah,$^{12}$
E. Sharma,$^{1}$
\\
$^{1}$Physical Research Laboratory, Navrangpura, Ahmedabad, Gujarat 380009, India\\
$^{2}$Indian Institute of Technology Gandhinagar Palaj, Gandhinagar 382355, India\\
$^{3}$Jodrell Bank Centre for Astrophysics, Department of Physics and Astronomy, University of Manchester, Oxford Road, Manchester M13 9PL, UK\\
$^{4}$Department  of  Astronomy  and  Astrophysics,  Tata  Institute  of  Fundamental  Research,  Mumbai  400005, India\\
$^{5}$Indian Institute of Space Science and Technology (IIST), Thiruvananthapuram 695 547, Kerala, India\\
$^{6}$Aix-Marseille Universite, CNRS, CNES, LAM, 38 rue F. Joliot Curie, 13388 Marseille Cedex 13, France\\
$^{7}$Institut Universitaire de France, Paris, 1 rue Descartes, 75231 Paris Cedex 05, France\\
$^{8}$National Astronomical Observatories, Chinese Academy of Sciences, Beijing 100101, People’s Republic of China\\
$^{9}$Guizhou Radio Astronomical Observatory, Guizhou University, Guiyang 550000, People’s Republic of China\\
$^{10}$Istituto di Astrofisica e Planetologia Spaziali, INAF, Via Fosso del Cavaliere 100, I-00133 Roma, Italy\\
$^{11}$ Institute of Astronomy and Astrophysics, Academia Sinica, Roosevelt Rd, Taipei 10617, Taiwan, R.O.C.\\
$^{12}$Indian Institute of Science Education and Research (IISER) Tirupati, Rami Reddy Nagar, Karakambadi Road, Tirupati 517 507, India\\
}
\begin{document}
\label{firstpage}
\pagerange{\pageref{firstpage}--\pageref{lastpage}}
\maketitle
\begin{abstract}

Filamentary flows toward the centre of molecular clouds have been recognized as a crucial process in the formation and evolution of stellar clusters. %These filaments act as channels for the inflow of gas and dust, driving the mass accretion toward the central hub, and they exhibit a hierarchical structure, with larger-scale filaments giving rise to smaller substructures, ultimately leading to the formation of dense cores and protostars. 
%Understanding the role of filamentary flows towards the molecular cloud center is crucial for deciphering the mechanisms of star formation and the formation of stellar clusters. 
In this paper, we present a comprehensive observational study that investigates the gas properties and kinematics of the Giant Molecular Cloud G148.24+00.41 using the observations of CO (1-0) isotopologues.
%and shows the role of longitudinal filamentary flows in gathering the matter at the hub location of the cloud. 
We find that the cloud is massive (10$^5$ \Ms) and is one of the most massive clouds of the outer Galaxy. We identified six likely velocity coherent filaments in the cloud having length, width, and mass in the range of 14$-$38 pc, 2.5$-$4.2 pc, and (1.3$-$6.9) $\times$ 10$^3$ \Ms,  respectively. We find that the filaments
are converging towards the central area of the cloud, and the longitudinal accretion flows along the filaments are in the range of $\sim$ 26$-$264 \Ms~Myr$^{-1}$. The cloud has fragmented into 7 clumps having mass in the range of $\sim$ 260$-$2100 \Ms~and average size around $\sim$ 1.4 pc, out of which the most massive clump is located at the hub of the filamentary structures, 
near the geometric centre of the cloud. Three filaments are found to be directly connected to the massive clump and transferring matter at a  rate of $\sim$ 675 \Ms~Myr$^{-1}$. The clump hosts a near-infrared cluster. Our results show that large-scale filamentary
accretion flows towards the central region of the collapsing cloud is an important mechanism for supplying the matter necessary to form the central high-mass clump and subsequent stellar cluster.

%The fact that filaments of cloud exhibit a hierarchical structure, showing accretion flows towards the hub, and being fragmented into several sub-structures like clumps and protostars within them, points towards the cluster formation via dynamical hierarchical collapse and assembly of both gas and stars.  

\end{abstract}


%%%%%%%%%%%%%%%%%%%%%%%%%%%%%%%%%%%%%%%%%%%%%%%%%%

%%%%%%%%%%%%%%%%% BODY OF PAPER %%%%%%%%%%%%%%%%%%

\section{Introduction}
\label{int}
It is largely established that a large fraction of stars form in stellar clusters \citep{lada_lada_2003}. However, how exactly stellar clusters form, in particular intermediate to high-mass clusters,  remains largely unknown
and has been the subject of several recent reviews \citep{Longmore_2014, Krumholz_2019, Krause_2020, Adamo_2020}.  Massive to intermediate-mass clusters play an important role in the evolution and chemical enrichment
of the Galaxy through radiation and winds. As they contain a large number of 
stars from the same parental cloud, they also serve
as an important astrophysical laboratory for studying the stellar initial 
mass function, stellar evolution, and stellar dynamics.  

%Despite their importance, we still have limited knowledge of their formation mechanisms. 
%Two main cluster formation mechanisms have been proposed – a monolithic “in situ" mode and a hierarchical “conveyor belt" type collapse mode \citep[e.g.][]{Longmore_2014}. In the "in-situ" mode,   a sufficient amount of gas is hypothesized to be condensed into the cluster volume before star formation commences and thus predicts the formation of a massive clump before the onset of star formation. In this scenario, the star-formation is expected to occur at a high efficiency over a short span of time \citep{Ban_Kru_2015}. In the conveyor belt model, molecular clouds are in a state of global hierarchical collapse, and 
%, they form structures such as filaments and dense clumps/cores during their evolution and each level in the hierarchy of density structures is accreting from its parent structure due to large-scale gravity. 
The different mechanisms which have been proposed for cluster formation are \emph{monolithic} collapse mode and flow-driven models like \emph{global hierarchical collapse} \citep[GHC;][]{sema2019} and \emph{inertial inflow} \citep[I2;][]{Padoan_2020}. In the \emph{monolithic} mode, a sufficient amount of gas is hypothesized to be condensed into the cluster volume before star formation commences and thus predicts the formation of a massive clump before the onset of star formation \citep{Ban_Kru_2015}. %In this scenario, the star-formation is expected to occur at a high efficiency over a short span of time \citep{Ban_Kru_2015}. 
In flow-driven models, the matter flows hierarchically in the cloud from large-scale regions down to its collapse centre in a ‘conveyor belt’ fashion, eventually forming a stellar cluster at the bottom of the potential of the cloud \citep[][]{Longmore_2014,walk_2016, Barnes_2019, sema2019, krum-mck_2020}. In this multi-scale dynamical mass transfer, each level in the hierarchy of density structures is accreting from its parent structure.
%The large-scale collapse involves the generation of filamentary flows that funnel material from the large-scale region down to its collapse centre. And the small-scale collapsing regions and stars formed within them  ‘ride’ along these large-scale filamentary flows in a ‘conveyor belt’ fashion, forming a stellar cluster at the bottom of the potential of the cloud \citep[][]{walk_2016, Barnes_2019, sema2019, krum-mck_2020}.
%the large-scale gas accretes to the bottom of the potential
%well of the cloud and star formation in the small-scale structures such as in dense clumps/cores of the cloud occur simultaneously \citep[see also][]{walk2016,barnes_2019,krum-mck2020}. 
%In this scenario, both the extended gas cloud and the protostellar population formed within the  structures undergo global gravitational collapse simultaneously, eventually \textcolor{red}{forming a stellar cluster at the bottom of the potential of the cloud} \citep[][]{walk_2016, Barnes_2019,krum-mck_2020}.
%\textcolor{red}{The stars or group of stars formed in the early share the infall motion of the global larger potential \citep[see also][]%{walk2016,barnes_2019,krum-mck2020}.
%eventually leading to the formation of a massive cluster in the cloud.} 

Understanding the dominant mode(s) of massive cluster formation requires studying massive molecular clouds that are at the early stages of their evolution, because i) a massive bound cloud with a significant dense gas reservoir is required to form a high-mass cluster, and ii) once star formation is underway, the massive members of the cluster can erase/alter the initial conditions and structure of the parental gas on a very short timescale via feedbacks such as radiation, jets, and stellar winds.

Over the last decade, various dust continuum and molecular line observations suggest that the interstellar medium is filamentary, consisting of filamentary structures of different shapes and sizes at all scales \citep[][]{Andre2010,mol2010,Sch_2014,shima2019,Liu_2021,Li_2022,zav23}. Depending on densities and scales, they are often called filaments, fibres, and streamers \citep[for details, see review articles by ][]{hacar22,pine22}. %In particular, studies based on $\it{Herschel}$ far-infrared dust continuum maps have shown the ubiquitous presence of filaments in the Galactic molecular clouds \citep[][]{Andre2010,mol2010,Sch_2014,zav23}. 
The filaments are the preferred sites of active star formation \citep{ kon2015,Andre2017}, with high-mass stars and stellar clusters preferentially forming in the high-density regions of the clouds such as hubs and ridges \citep{myers2009, Motte_2018, kumar2020, kumar2022, Beltran_2022, Yang_2023, zhang_2023}, where converging flows found to be funnelling the cold matter to the hub through the filamentary networks \cite[e.g.][]{sch2010, Morales_2019}. Thus, evaluating the physical conditions of the gas in molecular clouds and characterizing structures,  such as filaments, ridges, and hubs, and investigating their kinematics using molecular line data, are crucial steps for understanding the evolution of molecular clouds and associated cluster formation.

\begin{figure}
    \centering
    \includegraphics[width=8.5cm]{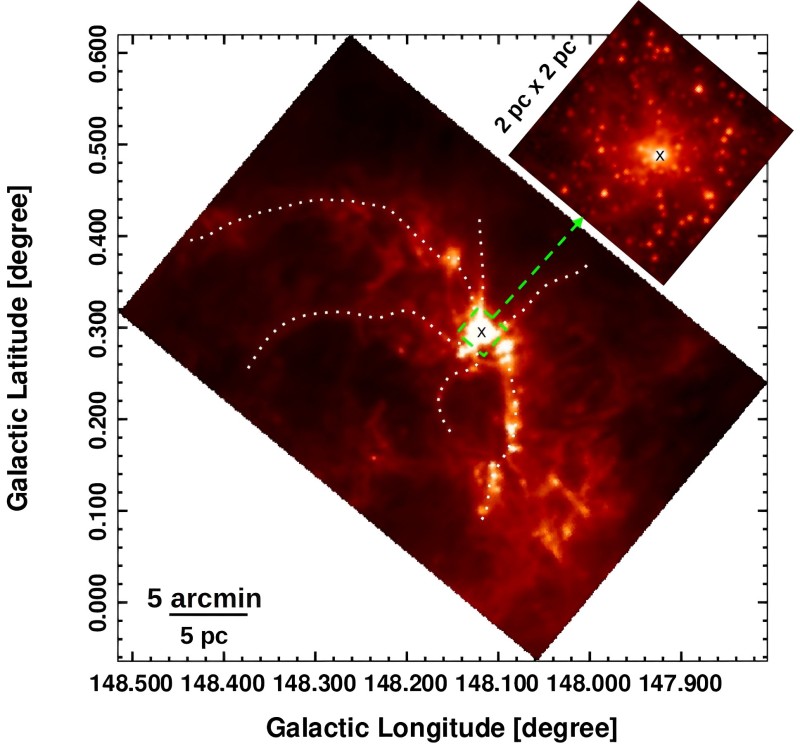}
    \caption{Central area of the \cloud~cloud as seen in $\it{Herschel}$ 250 $\mu$m band, showing the hub-filamentary morphology. The inset image shows the presence of an embedded cluster within the hub region (shown by a green box) at 3.6 $\mu$m. The filamentary structures are from \protect\cite{Rawat_2023}. We note that for a better presentation of the molecular data, in this work, this figure, as well as the subsequent figures, are presented in the galactic coordinates, whereas figures of \protect\cite{Rawat_2023} are in the FK5 system.} 
\label{fig_hub}
\end{figure}

In a recent work, \cite{Rawat_2023} characterized and investigated one such aforementioned type of cloud, "\cloud", in order to find out its cluster formation potential and mechanism(s) by which an eventual cluster may emerge. \cite{Rawat_2023} found that
\cloud~is a bound,  massive (mass $\sim$ 10$^5$ \Ms), and cold (dust temperature $\sim$ 14.5 K)  giant molecular cloud (GMC) located at a kinematic distance of $\sim$ 3.4 $\pm$ 0.3 kpc. The cloud is still in the early stages of its evolution, such that stellar feedback is not yet significant. Comparing with nearby molecular clouds as well as the massive clouds of the Galactic centre, they conclude
that the total gas mass content and dense gas fraction ($\sim$ 18\%) of \cloud~is similar to the Orion-A cloud \citep{lada_2010}. %Based on  $\it{ Herschel}$ observations, \cite{Rawat_2023} found out that the cloud hosts a massive clump at its centre of potential, which lies at the nexus of several large-scale (5$-$10 pc) filament-like structures (shown in Fig. \ref{fig_hub}). 
Based on  $\it{ Herschel}$ observations, \cite{Rawat_2023} visually identified several large-scale (5$-$10 pc) filament-like structures (shown in Fig. \ref{fig_hub}) in the cloud, which appear to merge near the geometric centre of the cloud. This configuration is found to resemble the hub-filamentary systems of molecular clouds \cite[e.g.][]{myers2009}, where star-cluster formation takes place. Using  $\it{Spitzer}$ mid-infrared images, the authors observed the presence of an embedded cluster at the hub location (shown in Fig. \ref{fig_hub}). The cluster is not visible in optical and barely visible in near-infrared 2MASS images, suggesting that the young cluster is still forming. The cluster location corresponds to an Infrared source, RAFGL 5107, identified 
by IRAS based on far-infrared observations \citep{Wouter_1989}. 
Using various observational metrics of the cloud  (such as enclosed mass over radius, density profile, fractalness, spatial and temporal distribution of protostars, degree and scales of mass-segregation, and distribution and structure of the cold gas) and comparing them with the prediction of aforementioned models of cluster formation, \cite{Rawat_2023} argue that the cloud has the potential to make an intermediate to high-mass cluster through the hierarchical assembly of both gas and stars, such as those predicated in conveyor-belt type models. 

%In \cloud based on our   observation,  we have observed presence
%of large-scale filamentary structures. 

%The large-scale filaments and hubs can host a number of clumps or protocluster clumps \citep{moral2019,kumar2022}, where high star formation %activity has been reported, leading to the formation of a cluster. 

\cite{Rawat_2023}, based on dust continuum and stellar content analyses, 
proposed that \cloud~has the potential to make a rich cluster, preferentially at the hub location.
%However, to understand the process by which matter accumulates at the hub location, the gas kinematics of the cloud need to be explored. 
% In general, because of turbulence and gravity, the distribution and kinematics of the gas in GMCs is often complicated.
The physical and kinematic structure of gas in GMCs is typically complex due to the interplay of turbulence and gravity. 
%Understanding the role of filaments in the gas assembly processes and the rate at which they are transporting matter from the  large-scale to the small-scale are key for understanding the process and the type of cluster (low vs. high) may emerge from a cloud. 
Gas kinematics provides a diagnostic tool for understanding the physical processes involved in the conversion of gas mass into stellar mass. %Moreover, the Position-position-velocity (PPV) cube data can delineate the filaments more clearly in position-velocity space \citep{shang2022L}, which are often merged together in the continuum data in position-position space.
In this work, using low spatial resolution molecular line data of CO (1--0) isotopologues, we explore 1-square degree area centred around the hub of \cloud, and present the first detailed study of large-scale gas properties and kinematics of the various structures associated with the cloud. We aim to understand the gas assembly processes from cloud-scale to clump-scale; thus, the role of the gaseous structures in the formation of the stars or star clusters as evidenced in the cloud by \cite{Rawat_2023}.
\\

We organize this paper as follows. In Section \ref{data}, we describe the data used in this work. In Section \ref{res}, we present the global gas properties and kinematics of \cloud, and compare our results with the nearby Galactic clouds. In Section \ref{sec:clump}, we discuss the clumps of the cloud and their properties. In Section \ref{filaments}, we discuss filamentary structures, their extraction, properties, and the gas kinematics along the filaments. We also present the measured longitudinal gas mass accretion rate along the filaments. In Section \ref{dis}, we discuss our results in the context of cluster formation scenario in the \cloud~cloud and summarize our findings in Section \ref{summary}.

\section{Data}
\label{data}
The molecular line observations of the \cloud\ complex in \twcos, \thcos, and \eico lines (J = 1--0
transitions) at 115.271, 110.201, and 109.782 GHz, respectively, were 
%The molecular gas is traced by three isotopes of CO emissions, all of J = 1--0
%transitions, 
observed with the 13.7-m radio telescope as part of the Milky Way Imaging Scroll Painting \citep[MWISP;][]{su2019} survey, led by the Purple Mountain Observatory (PMO). The MWISP survey mapping covers the Galactic longitude from l = $9\degree.75$  to $230\degree.25$ and the
Galactic latitude from b = $-5\degree.25$ to $5\degree.25$. The three CO isotopologue line observations were done simultaneously using a 3 $\times$ 3 beam sideband-separating Superconducting Spectroscopic Array Receiver (SSAR) system \citep{shan2012} and using the position-switch on-the-ﬂy mode,  scanning the region at a rate of 50\arcsec per second.  The calibration was done using the standard chopping wheel method that allows switching between the sky and an ambient temperature load. 
%The calibration accuracy is estimated to be within 10\%. 
The calibrated data were then re-gridded to 30\arcsec pixels and mosaicked to a FITS cube using the GILDAS software package \citep{Guill_lucas_2000}. 
The antenna temperature ($T_\mathrm{A}$) has been converted to the main-beam temperature ($T_\mathrm{{MB}}$) using
%the beam filling factor ($f_b$) and main beam efficiency ($\eta_{MB}$). 
%All the intensities throughout the paper are converted to a scale of main beam temperatures with 
the relation $T_\mathrm{{MB}} = T_\mathrm{A}/B_\mathrm{{eff}}$, where B$_\mathrm{{eff}}$ is the beam efficiency, which is 46\% at 115 GHz and 49\% at 110 GHz.
%\citep[for details, see][]{su2019}. 
%All data were reduced using the GILDAS %software4 (Pety 2005). 
%The pointing accuracy is within $\sim$ 5\arcsec. 
The spatial resolutions (Half Power Beam Width; HPBW) of the observations are around $\sim$ 49\arcsec, 52\arcsec, and 52\arcsec for \twcos, \thcos, and \eico, respectively, which correspond to
a spatial resolution of $\sim$ 0.8$-$0.9 pc at the distance of the cloud ($\sim$ 3.4 kpc).
%of $\sim$ 55\arcsec for \twco, \thco, and \eico. 
The spectral resolution of \twco is $\sim$ 0.16 \kms~with a typical rms noise level of the spectral channel is about 0.5 K, and of \thco and \eico is $\sim$ 0.17 \kms~with a rms noise level of 0.3 K \cite[for details, see][]{su2019}.
%according to the status report4 of the PMO-13.7m telescope. 

\begin{figure}
    \centering
    \includegraphics[width=8cm]{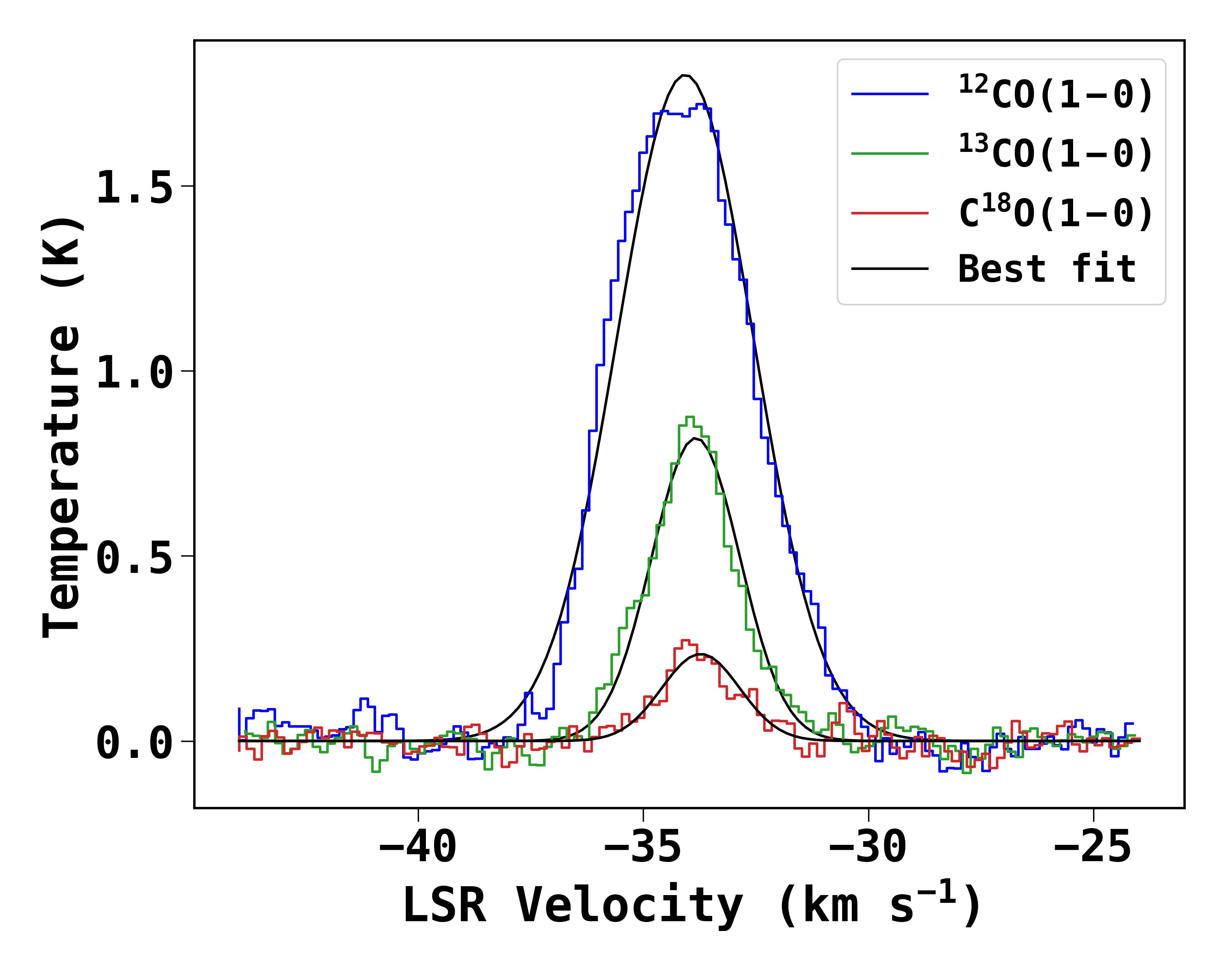}
    \caption{The average \twcos, \thcos, and \eico spectral profiles towards the direction of the \cloud~cloud. The black solid curve shows the Gaussian fit over the spectra.}
    \label{fig_co}
\end{figure}

\section{Results and Analysis}
The advantage of using the CO (1--0) isotopologues is that one can use the \twco emission to trace the enveloping layer (i.e. $\sim$ 10$^2$ cm$^{-3}$) of the molecular cloud to reveal its large-scale low surface brightness structures and dynamics. 
%Relatively diffuse molecular gas also show filamentary features, and such low-density,
%tenuous “striations” were first observed in CO lines (Goldsmith et al. 2008). 
On the other hand, the optically thin \thco and \eico emission (discussed in Section \ref{glo_con}) can trace the denser regions (i.e. $\sim$ 10$^3$–10$^4$ cm$^{-3}$) such as large-scale filamentary structure and dense clumps within the cloud. By combining the CO isotopologues, the overall properties of the diffuse regions of the cloud, as well as the gas properties and physical conditions of the dense structures within it, can be determined. 
\label{res}

\subsection{Global Cloud Morphology, Properties, and Kinematics}
\subsubsection{Gas morphology and kinematics}
\label{glo_kin}
%To see the gas structure, dynamics, and kinematics, we utilise the molecular line data.
\cite{Rawat_2023}, based on \twco~spectrum and comparing the CO gas morphology with the dust continuum images ($\it{Herschel}$ images at 250, 350 and 500 \mum), showed that the \cloud~cloud component mainly lies in the velocity range of $-$37.0 \kms to $-$30.0 \kms\ in agreement with the previous studies \citep[e.g.][]{Urq_2008, md17}. 
%The \twco emission is spread over in a radius of $\sim$ 26 pc ($\sim$ 26\arcm) on the plane of sky.
Fig. \ref{fig_co} shows the average spectrum of all three isotopologues towards the cloud. %As can be seen, among the CO lines, the \eico shows weak emission and is spread over a relatively narrow velocity range compared to \twco and \thco. 
We fitted a Gaussian function to the line profiles and derived the peak velocity, velocity dispersion ($\sigma_{1d}$), and velocity range of each spectrum, which are given in Table \ref{tab:cloud}. %\textbf{The resultant peak velocity for \twco, \thco, and \eico are found to be $-$34.07 \kms, $-$33.83 \kms, and $-$33.72 \kms, respectively, and the corresponding velocity dispersion ($\sigma_{1d}$) are 1.51 \kms, 0.98 \kms, and 0.87 \kms, respectively.} 
The estimated line-width ($\Delta V = 2.35\, \sigma_{1d}$) and 3D velocity dispersion ($\sigma_{3d} =  \sqrt{3} \times \sigma_{1d}$) associated with the \twco~profile are 3.55 and 2.62 \kms, for \thco are 2.30 and 1.70 \kms, and for \eico are 2.04 and 1.51 \kms, respectively. We want to point out that the optical thickness of \twco~may affect the velocity centroid and velocity dispersion of the line profile. Therefore, the \twco data has been used to measure the global properties and distribution of low-density gas, while the kinematics of dense structures or properties of dense clumps have been derived using the \thco~and \eico~data.  
%The \eico shows the presence of two components within the \thco velocity range, however, to compare the morphology (shown below) of the cloud in \eico with respect to the \thco and \thco, we consider both the components together for making an integrated intensity map. 
%We adopt similar procedure for obtaining the aforementioned
%parameters based on \eico data.\\ 

%\href{https://arxiv.org/pdf/1406.3134.pdf}{Sch15}.

\begin{figure}
    \centering
    \includegraphics[width=8.5cm]{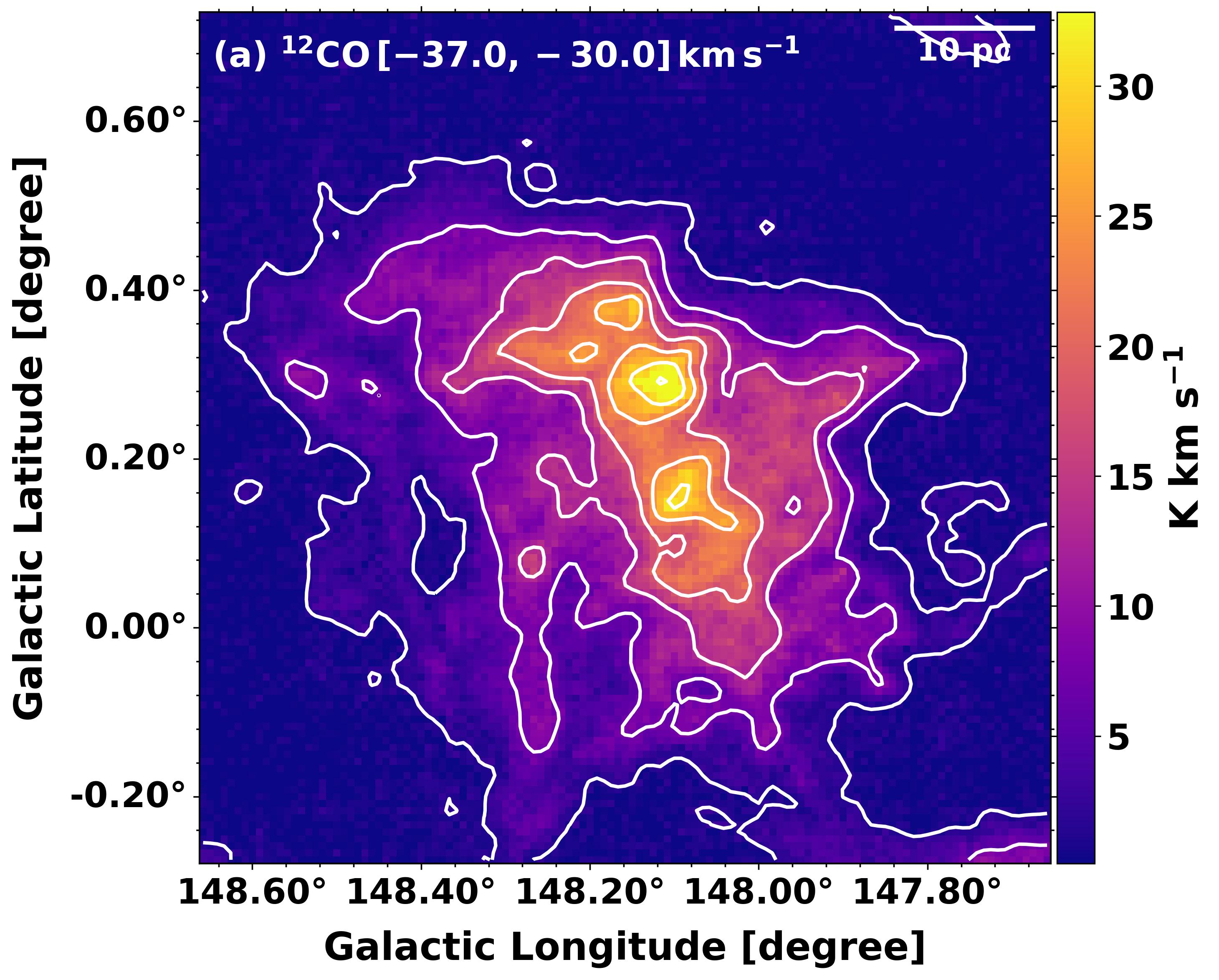}\\
    
    \includegraphics[width=8.5cm]{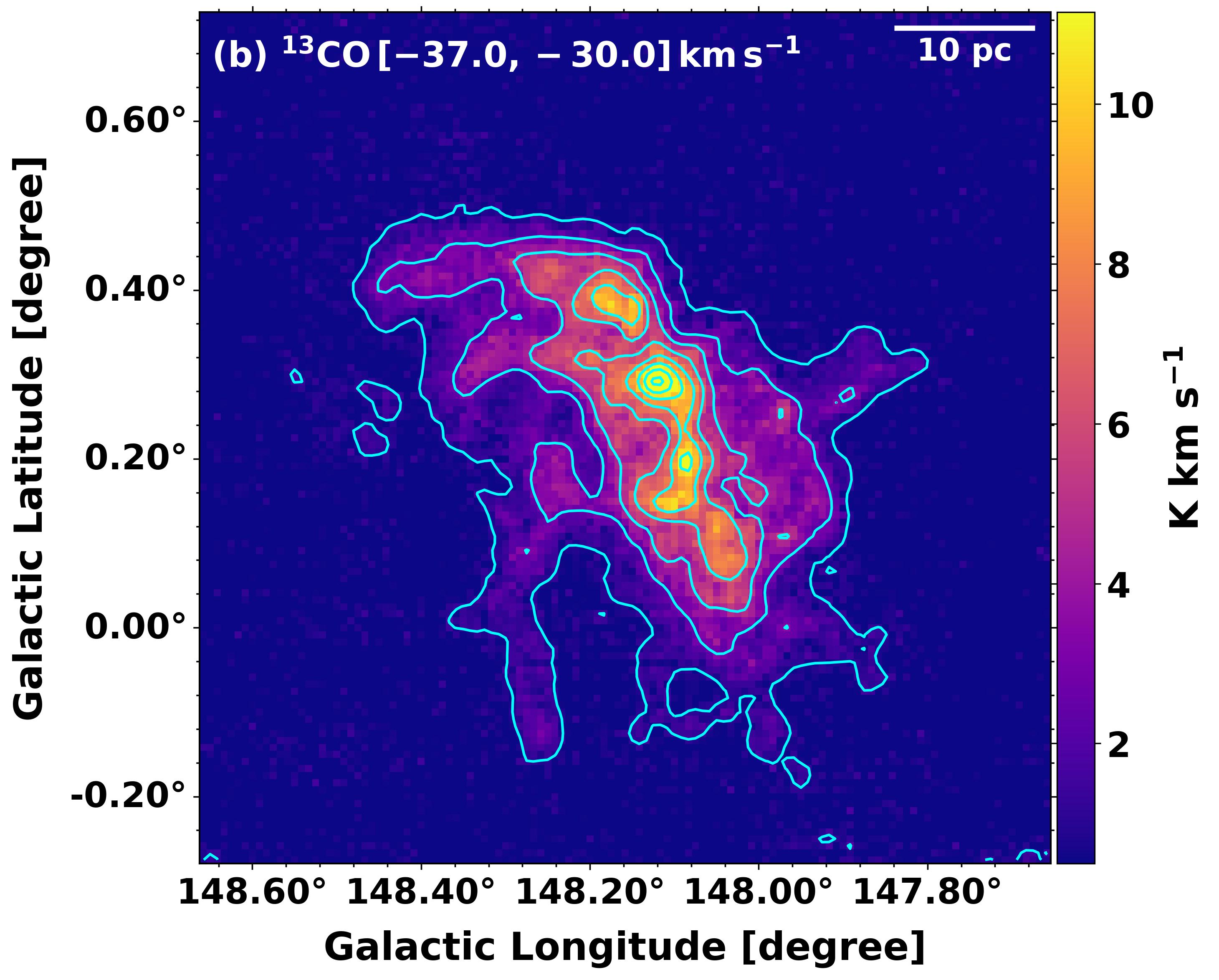}\\
    
    \includegraphics[width=8.5cm]{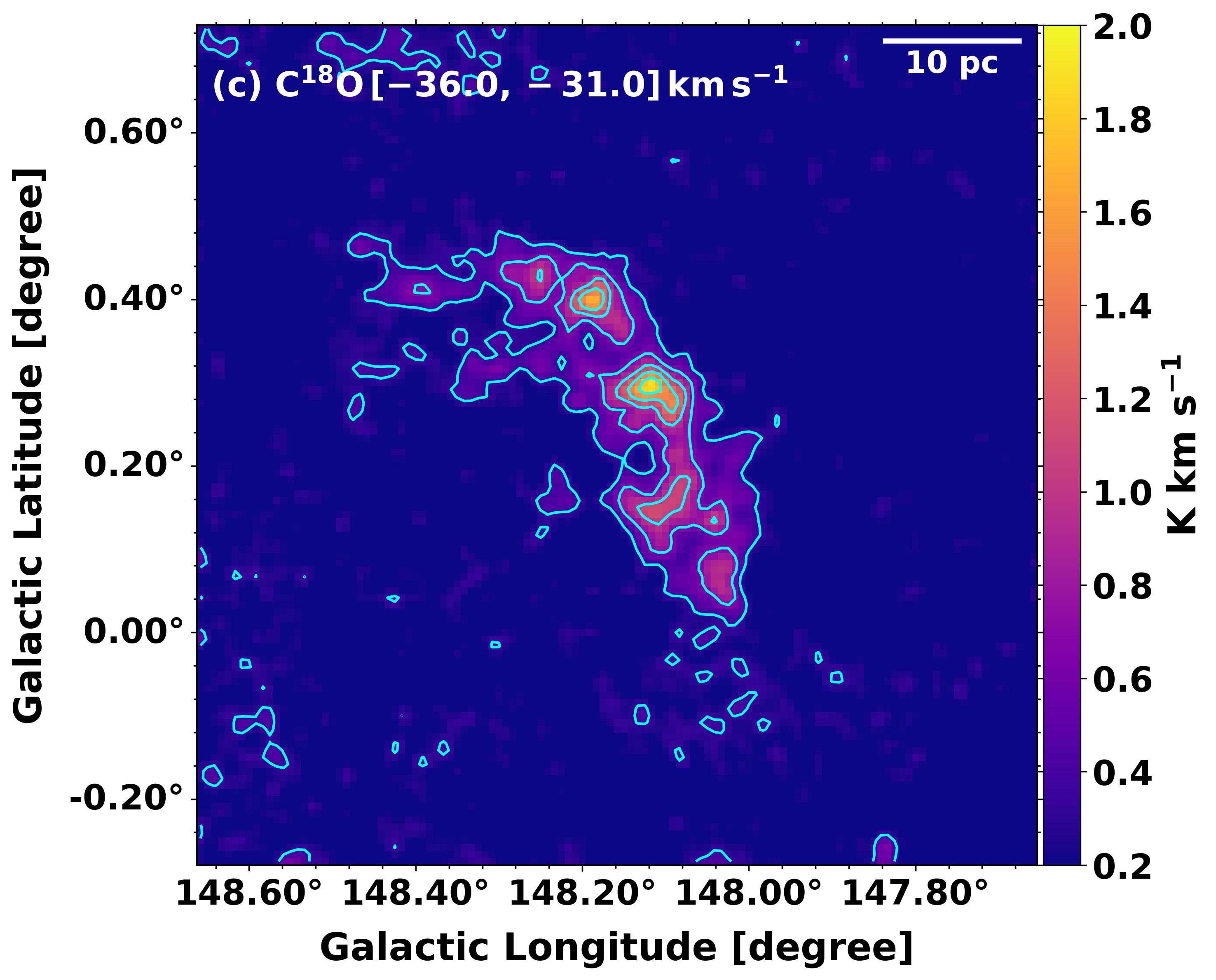}\\
    
    \caption{(a) \twco~integrated intensity (moment-0) map of the cloud with contour levels at 1.5, 7.08, 12.67, 18.25, 23.83, 29.42, and 35 K \kms. 
    %shown above 3$\sigma$ from the background, starting from 1.5 to 35 K \kms, 
    (b) \thco~integrated intensity map of the cloud with contour levels at 0.9, 2.7, 4.5, 6.3, 8.1, 9.9, 11.7, and 13.5 K \kms. (c) \eico~integrated intensity map of the cloud with contour levels at 0.35, 0.68, 1.01, 1.34, 1.67, and 2.0 K \kms. The contours are drawn 3$\sigma$ above the background value of individual maps. We note that the \eico map has been smoothened by 1 pixel to improve the signal.
    }
    \label{fig_int}
\end{figure}

The integrated intensity (moment-0) maps of \twcos, \thcos, and \eico line emissions, integrated in the velocity range given in Table \ref{tab:cloud}, are shown in Fig. \ref{fig_int}. Also shown are the contours above 3$\sigma$ of the background value, where $\sigma$ is the standard deviation of the background emission.
%The molecular emission is integrated in the velocity range [$-$37.0, $-$30.0] km s$^{-1}$ for \twco~and \thco, and [$-$36.0, $-$31.0] km s$^{-1}$ for \eico. 
%The integrated intensity maps reveal the filamentary structure of the cloud and some dense gas regions, which may be are the cluster forming clumps. 
As discussed earlier, in molecular clouds, the \twcos, 
%\textcolor{red}{due to its high abundance and high optical depth}, 
traces better the diffuse emission, while \thco and \eico probe deeper into the cloud and trace higher column density regions. Though the spatial resolution of the data is relatively low, the presence of several filamentary structures can be seen in the \thco map (details are discussed in section \ref{filaments}), while \eico emission seems better at tracing the central area and the dense clumpy structures of the cloud. In \cloud, we find that \thco covers
$\sim$ 87\% of the \twco emission, while \eico covers only 43\%. 
%But \cloud~is devoid of any optically visible cluster, which shows that the cloud is still at the early stages of its evolution.  
%All these aforementioned structures are also apparent in the color-composite image shown in Fig. \ref{fig_int}, which also shows the relative spatial emission area and intensity traced by the isotopologues.

%and DSS2 R-band image, shown in Figure \ref{fig_rgb}. 
%From the Figure also, it is evident that there are at least three-four filaments seem to connect towards the central area of the cl.
%shows the relative spatial emission area and intensity traced by the three isotopologue CO lines

% \begin{figure}
%     \centering
%     %\includegraphics[width=8.5cm]{g148_RGB.png}
%       \includegraphics[width=9.0cm]{ds9_rgb_new.png}
%     \caption{A three-colour composite image of the \cloud~cloud, made using the \eico (red), \thco (green), and \twco (blue) moment-0 maps. The maps have been smoothed by 3-pixel Gaussian smoothing to improve the signal of the diffuse emission.}
%     \label{fig_rgb}
% \end{figure}

\begin{figure*}
     \centering
    \begin{subfigure}[t]{0.49\textwidth}
        \raisebox{-\height}{\includegraphics[width=\textwidth]{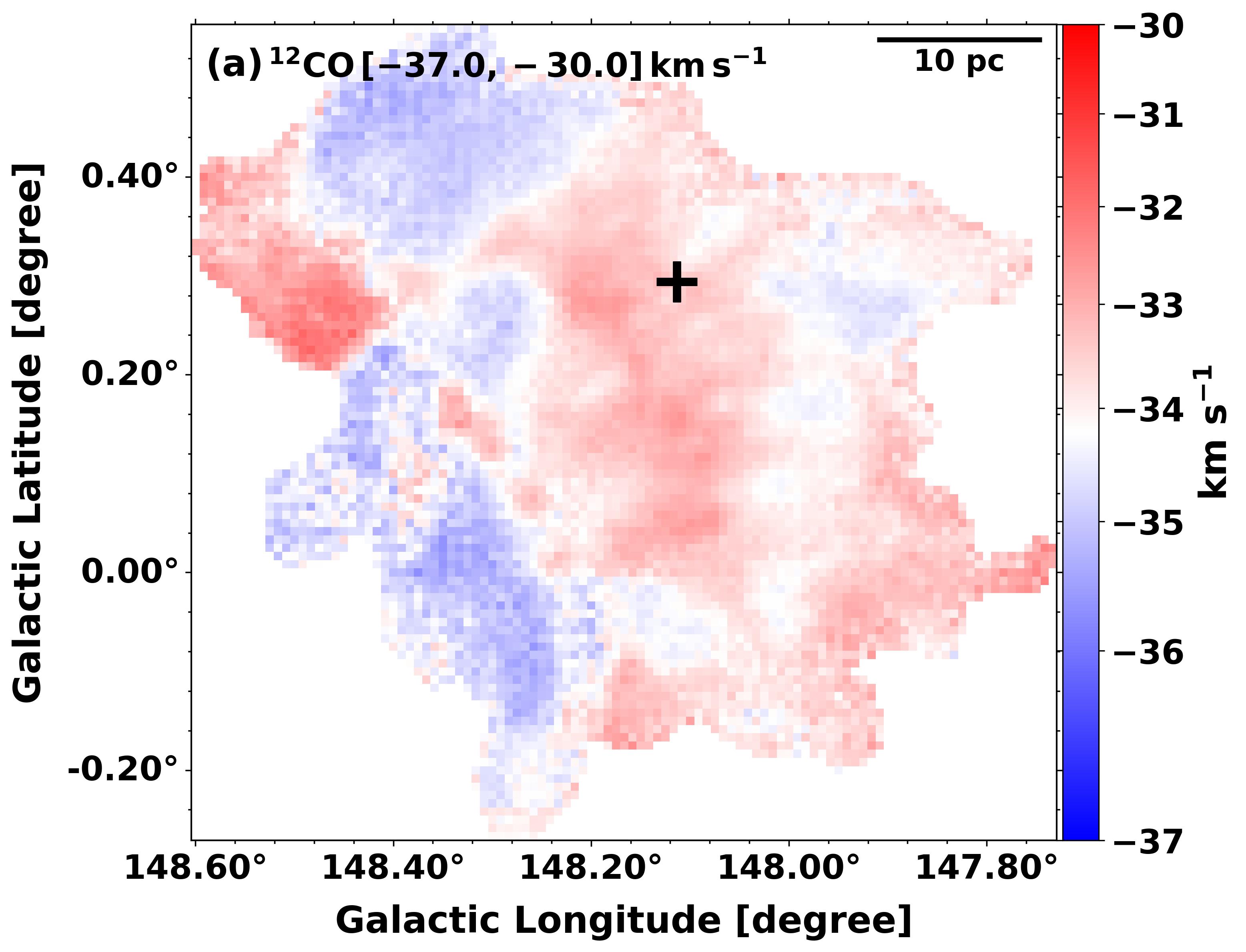}}
        
    \end{subfigure}
    \hfill
    \begin{subfigure}[t]{0.49\textwidth}
        \raisebox{-\height}{\includegraphics[width=\textwidth]{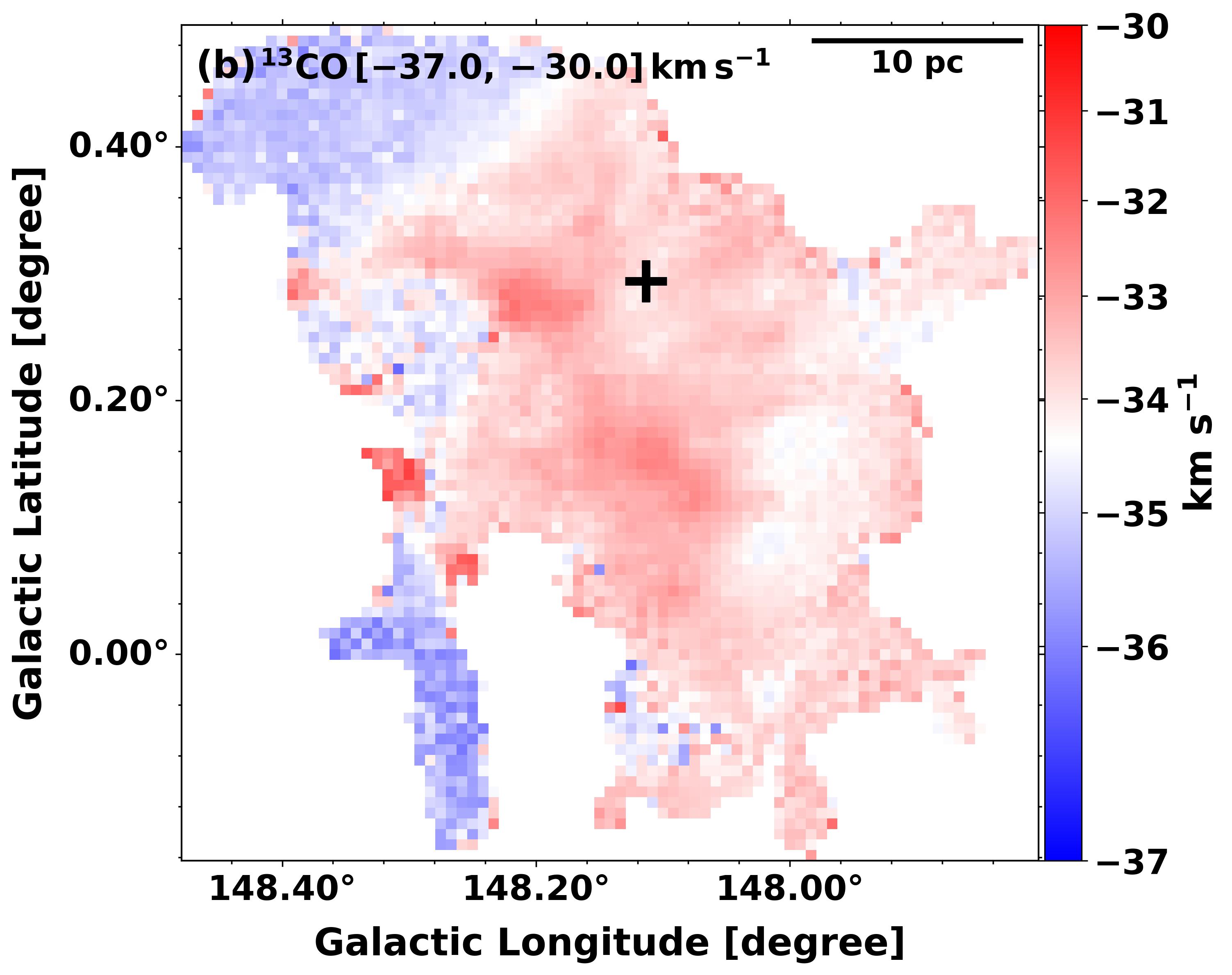}}
    \vspace{0.6cm}    
    \end{subfigure}
    
    \begin{subfigure}[t]{0.49\textwidth}
        \raisebox{-\height}{\includegraphics[width=\textwidth]{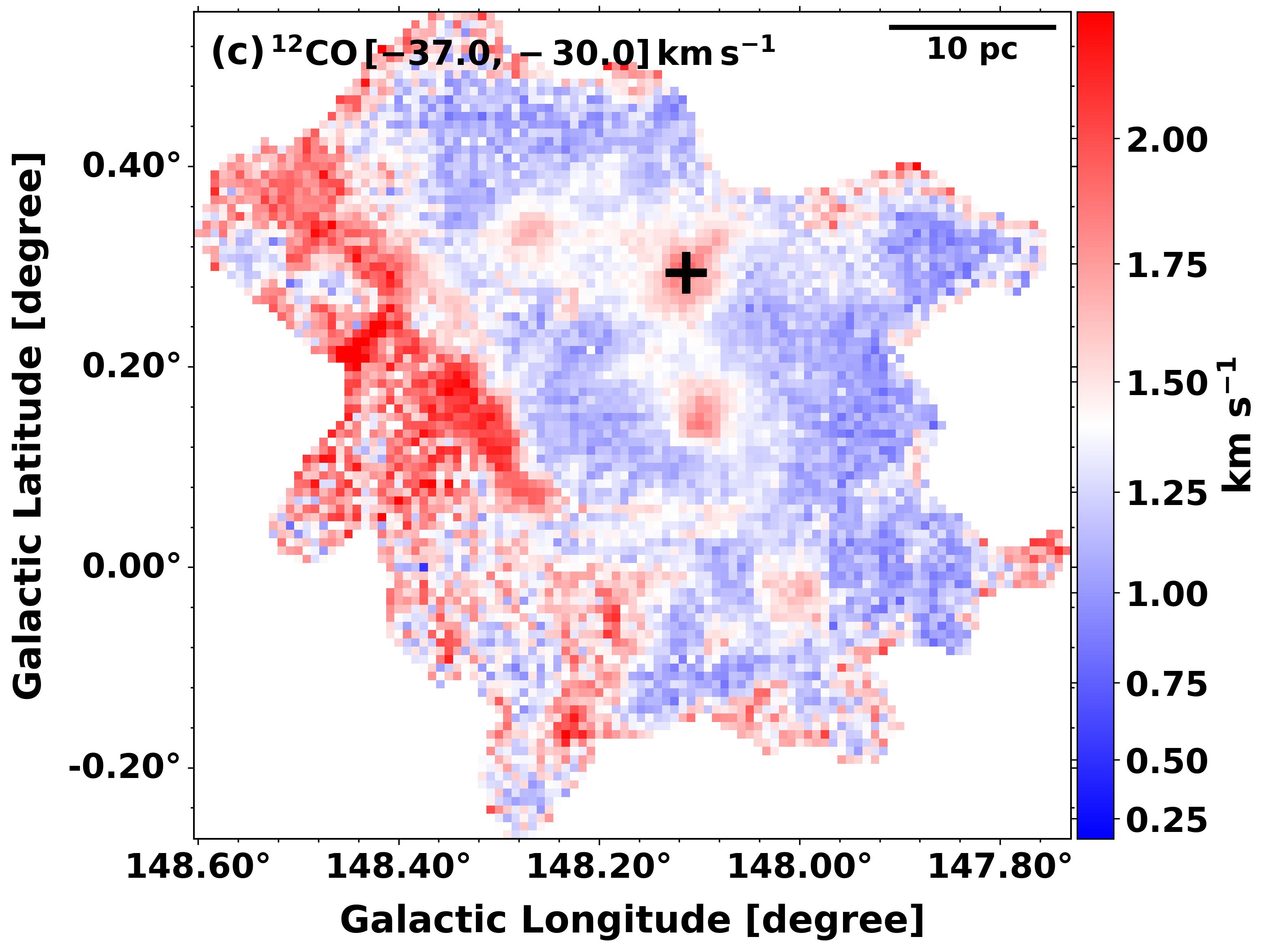}}
        
    \end{subfigure}
    \hfill
    \begin{subfigure}[t]{0.49\textwidth}
        \raisebox{-\height}{\includegraphics[width=\textwidth]{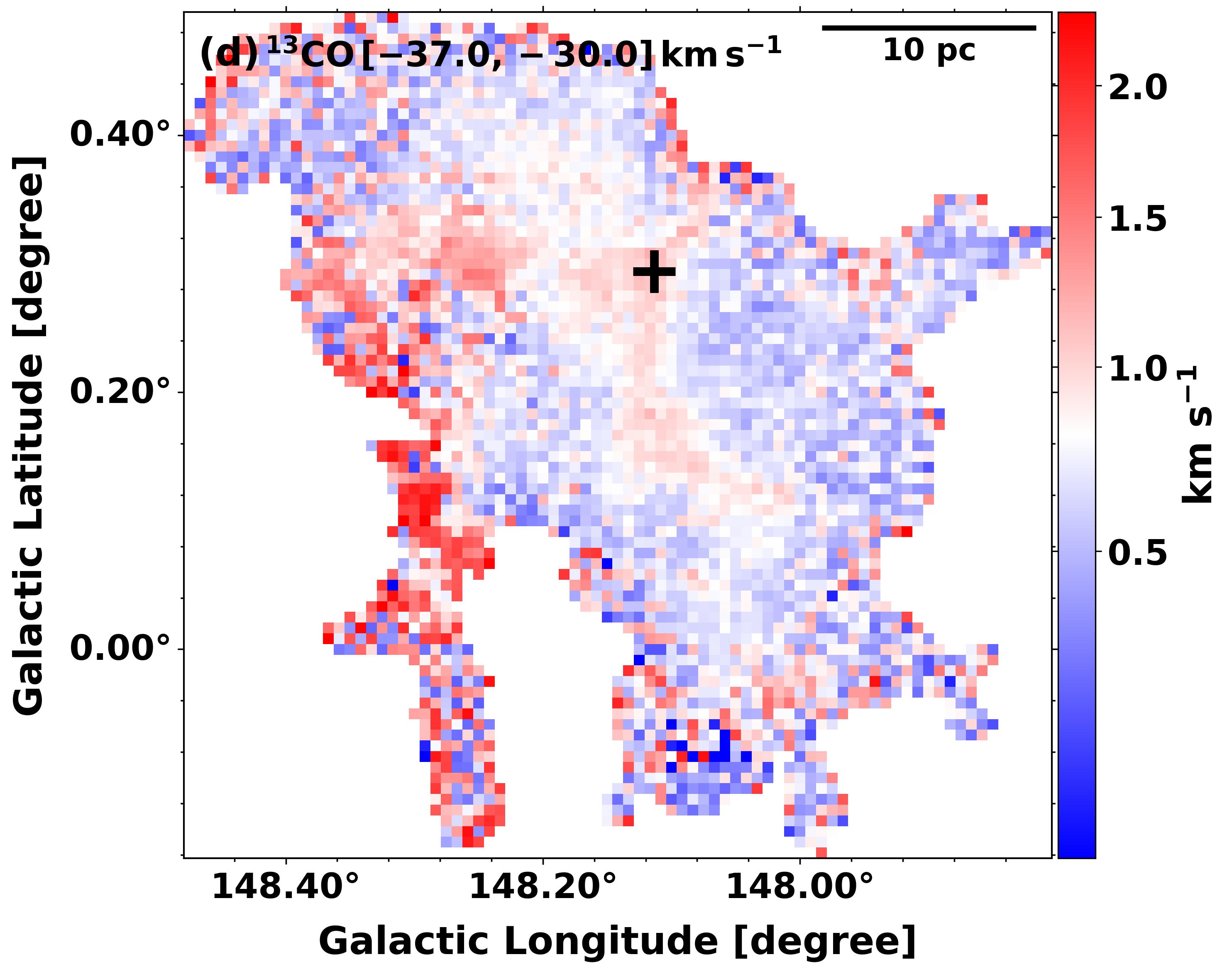}}
        
    \end{subfigure}
    \caption{(a) \twco and (b) \thco velocity maps of \cloud. (c) \twco and (d) \thco velocity dispersion maps of \cloud. The location of the hub is marked with a plus sign.}
    %{\bf We note at the edges of the cloud, due to low-level signals, the pixel values are likely discrepant, thus, should not be taken too literally.}}
    \label{fig_vel}
\end{figure*}

In order to understand the overall velocity distribution and velocity dispersion of the \twco and \thco gas in the cloud, we made intensity-weighted mean velocity (moment-1) and
velocity dispersion (moment-2) maps,  which are shown in  Figs. \ref{fig_vel}a-b and Figs. \ref{fig_vel}c-d, respectively. In general, the velocity distribution maps reveal that the
%the overall velocity distribution is smooth and is in range from $-$36 to $-$34 \kms
outer extent of the cloud exhibits blue-shifted velocities relative to the systematic one, typically ranging from $-$36 to $-$34 \kms, while the central region displays a red-shifted velocity range, from $-$34 to $-$30 \kms. 
We note that, since moment analysis represents the mean velocity of the gas along the line of sight, it is insensitive to the kinematics of the multiple velocity structures, if present in the cloud (more discussion in Section \ref{fil_cha}). Figs. \ref{fig_vel}c-d shows that
the velocity dispersion is not uniform across \cloud, it varies from 0.2 to 2.3 \kms, with a notable increase in the cloud's central area. The velocity dispersion of \twco gas may be on the higher side due to the optical depth effect, but this trend also holds true for the relatively optically thin \thco line. In the central area,  a patchy increase in velocity dispersion can
be seen at several locations. We discuss more
on this in Section \ref{clump_pro}. 
%Given the fact that cloud's excitation temperature is less than 15 K (discussed in Section \ref{glo_con}), one would expect the thermal velocity dispersion 
%of the gas to be less than 0.07 \kms (see Section \ref{clump_pro}). The increased velocity dispersion could thus be due to the presence of non-thermal motions, either due to the gas motions within the cloud 
%the merger of several velocity components or non-thermal motions generated by filamentary accretion flows
%or protostellar feedback due to local star formation activity or a combination of both the processes. %\textbf{(see Section \ref{fil_fil} and \ref{fil_sta})}. 
%For example, the hub location of the cloud (marked with a plus sign in Fig. \ref{fig_vel}c) shows a high velocity dispersion ($\sim$ 2 \kms) in \twcos, where several filaments are expected to interact and merge as found in numerical simulations \citep{Smith_2014, gomez&sema2014}. 
%For example, \citet{Rawat_2023} discussed that the hub is also associated with a massive YSO (Young Stellar Object) with an outflow, thus, its radiation and feedback might have also impacted the dynamics of the surrounding gas. 
Additionally, the \twco map reveals high velocity dispersion
in the north-eastern side of the cloud, whose exact reason is not known to us. External shock compression can result in such high dispersions. Although a young ($\sim$ 4 Myr) \hii region is found to be present in the vicinity of the cloud \citep{rome09}. However, the \hii region is located in the south-western direction of \cloud~and is also at a different distance (i.e. $\sim$ 1 kpc) with respect to it. A  detailed investigation covering wider surroundings of \cloud~is needed to better understand its origin, which is beyond the scope of the present work.

%From these maps, 
%we show the intensity weighted mean velocity (moment--1) map in the velocity range [$-$37.0, $-$30.0] \kms~for \twco~\& \thco~(see Figure \ref{fig_vel}). 
%Figure \ref{fig_vel}a \& b shows that the velocity ranges from $-$ 37.0 \kms~ to $-$ 30.0 \kms~with a median around $\sim$ $-$ 34.0 \kms. 

%The velocity dispersion (moment--2) map is shown in Figure \ref{fig_vel}c-d, %which indicates that there is a high-velocity dispersion near the hub ($\sim$ 3 %\kms; see Section \ref{kine} for details).  

\subsubsection{Physical conditions and gas column density} 
\label{glo_con}
Assuming the molecular cloud is under the local thermodynamic equilibrium (LTE) and 
\twco is optically thick, the excitation temperature (T$_\mathrm{{ex}}$), optical depth ($\tau$), and column density (\nht) of the \cloud~cloud can be calculated with the measured brightness of CO isotopologues. Under LTE, the kinetic temperature of the gas is assumed to correspond to the excitation temperature.

%The hydrogen column density from $^{12}$CO being optically thick 
The brightness temperature under the Rayleigh-Jeans approximation is expressed as $T_\mathrm{{MB}}$ = $f \times [J(T_\mathrm{{ex}}) - J(T_\mathrm{{bg}})](1 - e^{-\tau})$,
%\citep{yan21},
where T$_\mathrm{{MB}}$ is the brightness temperature, $f$ is the beam filling factor, T$_\mathrm{{bg}}$ is the cosmic microwave background temperature, and $J(T) = \frac{h\nu}{k(\exp(h\nu/kT)-1)}$. Taking T$_\mathrm{{bg}}$ $\sim$ 2.7 K and assuming $f$=1,  T$_\mathrm{{ex}}$ can be derived and written in a simplified form \citep{Garden_1991, nishi_2015, xu2018} as :

\begin{equation}
    T^{1 - 0}_\mathrm{{ex}} = \frac{5.53}{\mathrm{ln}\left[1\ \ +\ \ \frac{5.53}{T^{12, 1 - 0}_\mathrm{{MB, peak}}\ \  +\ \ 0.84}\right]},
\end{equation} where $T^{12, 1 - 0}_\mathrm{{MB, peak}}$ is the peak brightness temperature of the \twco emission along the line of sight. 
Based on the above formalism, we derived the excitation temperature at each pixel of the cloud. Fig. \ref{fig_temp}a shows the excitation temperature map, which ranges from 5 K to 21 K with a median around 8 K. The temperature map shows a relatively high temperature in the central region of the
cloud with respect to the outer extent. This is likely due to the fact that the central region is
heated by the protostellar radiation, where it has been found that protostars are actively forming \citep{Rawat_2023}.  The obtained average excitation temperature of the cloud is found to be similar to the \twco based excitation 
temperature of massive GMCs with embedded filamentary dark clouds \citep[e.g. $\sim$ 7.4 K,][and references therein]{Hernan_2015} and also similar to 
other nearby molecular clouds such as Taurus \citep[$\sim$ 7.5 K,][] {Goldsmith_2008} and Perseus \citep[$\sim$ 11 K,][]{Pineda_2008}.
%, and CMa OB1 Complex \citep{Goldsmith_2008, Pineda_2008, Lin_2021}.} 

%The brightest region having T$_{ex}$ around 21 K shows the location of a clump (see section xx.xx). 
%\textcolor{red}{you may discuss the patchy high excitation temperature by
%looking the distribution of 70 micron protostars. You may also briefly discussed the $\it{ Herschel}$ temperature range, and some caveats in getting
%accurate temperature in dense regions using  only CO isotopologues. See Section 3.3 of %\href{https://iopscience.iop.org/article/10.3847/1538-4365/ab5b97}{Sun et al} for various %limitations and caveats on deriving maps}.

%Under local thermodynamic equilibrium, the T$_{ex}$ value is  expected to be the same for all the CO isotopologues; 

Next we derived the optical depth maps of  $^{13}$CO and C$^{18}$O gas using the following relations \citep{Garden_1991, pineda_2010}:

\begin{figure*}
    \centering
    \includegraphics[width=8.5cm]{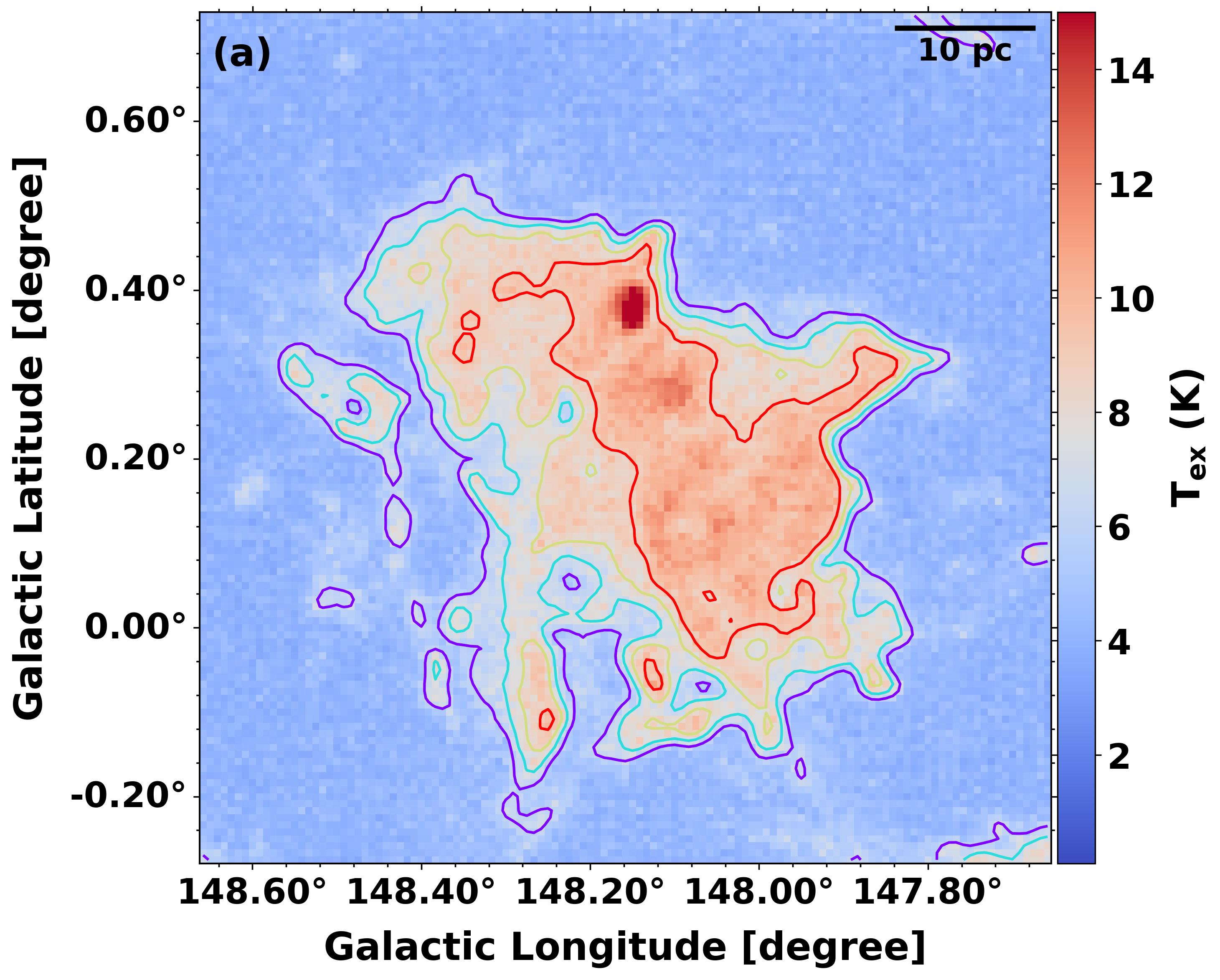}
    \includegraphics[width=8.78cm]{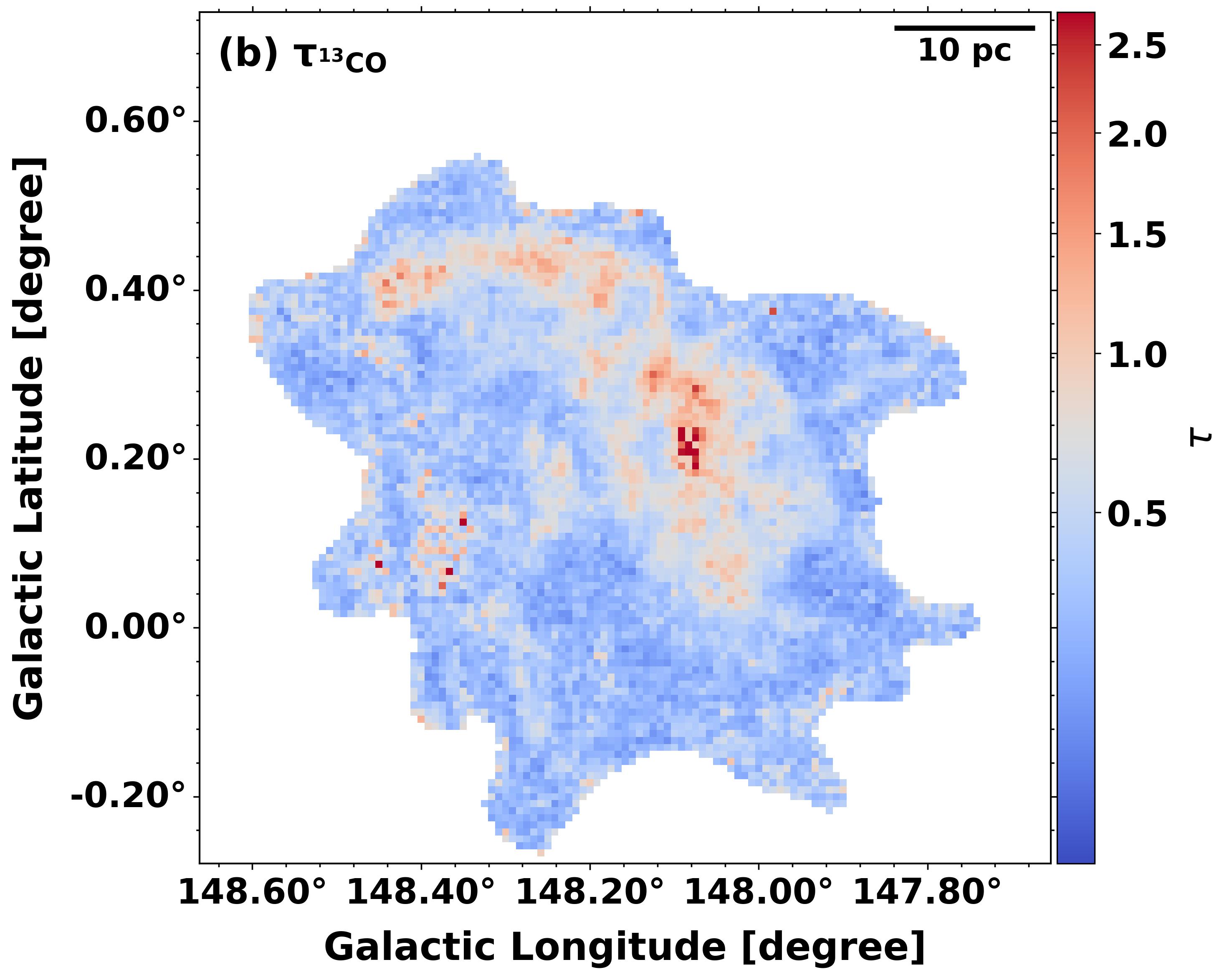}
    \caption{(a) Excitation temperature map overplotted with contours at 6, 7, 8, and 9 K. (b) Optical depth map of \thco. 
    %The location of the hub is marked with a plus sign.
    }
    \label{fig_temp}
\end{figure*}

\begin{equation}
    \tau_{13} = -\mathrm{ln}\left[1 - \frac{T^{13}_\mathrm{{MB, peak}}}{5.29}\left[\frac{1}{\exp(5.29/T_\mathrm{{ex}}) - 1} - 0.164\right]^{-1}\right]
\end{equation}

\begin{equation}
    \tau_{18} = -\mathrm{ln}\left[1 - \frac{T^{18}_\mathrm{{MB, peak}}}{5.27}\left[\frac{1}{\exp(5.27/T_\mathrm{{ex}}) - 1} - 0.166\right]^{-1}\right],
\end{equation} where $T^{13}_\mathrm{{MB, peak}}$ and $T^{18}_\mathrm{{MB, peak}}$ is the peak brightness temperature of \thco and \eico, respectively. 

The optical depths of \thco and \eico lines are estimated to be 0.1 $<$ $\tau(\thco)$ $<$ 3.0 and 0.05 $<$ $\tau(\eico)$ $<$ 0.25, respectively.  Fig. \ref{fig_temp}b shows the optical depth map of the \thco emission. Within the cloud area, we found only a 5\% fraction of the cloud area is of high  ($\tau > 1$) optical depth, implying that most of the observed \thco emission is optically thin.
%the \eico emission is found to be optically thin for most of the observed region. 
 \\ 

%The column density of the linear rigid rotor molecule is related to its optical depth of transition from the J to the J + 1 energy level by the formula (Garden et al. 1991; Bourke et al. 1997)
\noindent We then obtained the column density of $^{13}$CO and C$^{18}$O using the following relations from \citet{bou1997}: 
\begin{multline}
    N(\thco)_\mathrm{{thin}} = 2.42 \times 10^{14} \times \left(\frac{T_\mathrm{{ex}} + 0.88}{1 - \exp(-5.29/T_\mathrm{{ex}})}\right) \\ \times \frac{1}{J(T_\mathrm{{ex}}) - J(T_\mathrm{{bg}})}\ \ \int T_{\mathrm{MB}}(\thco) dv
\label{eq4}
\end{multline}

\begin{multline}
    N(\eico)_\mathrm{{thin}} = 2.42 \times 10^{14} \times \left(\frac{T_\mathrm{{ex}} + 0.88}{1 - \exp(-5.27/T_\mathrm{{ex}})}\right) \\ \times \frac{1}{J(T_\mathrm{{ex}}) - J(T_\mathrm{{bg}})}\ \ \int T_{\mathrm{MB}}(\eico) dv
\label{eq5}
\end{multline}\\

%Figure \ref{fig_cdens} shows the $^{13}$CO based column density map, which is tracing column density
%in the range xx to xx \cms. 
However, the $^{13}$CO based column density may underestimate the column density in the central area
of the cloud, where \thco is optically thick.  
Many studies on the GMCs and Infrared Dark Clouds have accounted for line optical depth while estimating
their physical properties \citep{roma10, hern11}. %\textcolor{red}{For example, \citet{hern11} showed that optical depth correction factors can increase the \thco column density by a factor of $\sim 2$ in the densest regions, such as IRDCs.} 
%However, the optical depth correction factors  are expected to be smaller for the more diffuse GMCs. 
%When the emission is optically thin,  $T_{ex}\int \tau dv  = \frac{\tau_{}}{1 - e^{-\tau_{}}}\int T_{mb} dv$ (Wilson et al. 2009). %Though the \thco is found to be moderately optically thick for only a small fractional area of the cloud, yet 
We, thus applied the following correction to the $N(\thcos)_{\rm{thin}}$ following \citet{pineda_2010, Li_2015}. Since the observed \eico emission is optically thin, no optical depth correction was made to $N(\eicos)_{\rm{thin}}$.

\begin{equation}
    N(\thco)_{\mathrm{corrected}} = N(\thco)_{\mathrm{thin}} \times \frac{\tau_{13}}{1 - e^{-\tau_{13}}}
\label{eq6}
\end{equation}\\

We then convert the \thco and \eico column density to the molecular hydrogen column density ($\nht$) using the relation, N(H$_2$) = 7 $\times$ 10$^5$ N($^{13}$CO) \citep{frek_1982} and N(H$_2$) = 7 $\times$ 10$^6$ N(C$^{18}$O) \citep{Cast_1995}, respectively. 
%Fractional aboundance: \href{https://arxiv.org/pdf/1809.09806.pdf}{HSIEH et al.}
%based on the following abundance ratios:   $n(^{12} \mathrm{CO} ) / n(^{13}\mathrm{CO}) = 59$ and  $n(\mathrm{H}_2)/ n(^{12}\mathrm{CO}) = 1.2 \times 10^{4}$. 
%\begin{equation}
%N(H_2)  = N(^{13}CO)_{thin} \times f(\frac{n^{12}\mathrm{CO}}%{n^{13}\mathrm{CO}}) \times f(\frac{n\mathrm{H}_2}%{n^{12}\mathrm{CO}})
%\end{equation}
The molecular hydrogen column densities from the \thco and \eico gas emission are estimated to be around 0.9 $\times$ 10$^{21}$ \cms $<$
$\mathrm{N(H_2)_{^{13}CO}}$ $<$ 2.4 $\times$ 10$^{22}$ \cms and 1.1 $\times$ 10$^{21}$ \cms $<$ $\mathrm{N(H_2)_{C^{18}O}}$ $<$ 2.0 $\times$ 10$^{22}$ cm$^{-2}$, respectively.
We found for the common area, the column density of both the maps are in agreement with each other
by a factor of 1.5. The observed variation in column density values might be due to the abundance variations of these isotopologues. For example, chemical models and observations suggest that selective photo-dissociation and fractionation can significantly affect the abundance of CO isotopologues \citep[e.g.][]{Shimajiri_2015, liszt_2017}.
%\citep{Dish_1988, visser_2009, Shimajiri_2015, liszt_2017}.

\begin{figure}
    \centering
    \includegraphics[width=8.5cm]{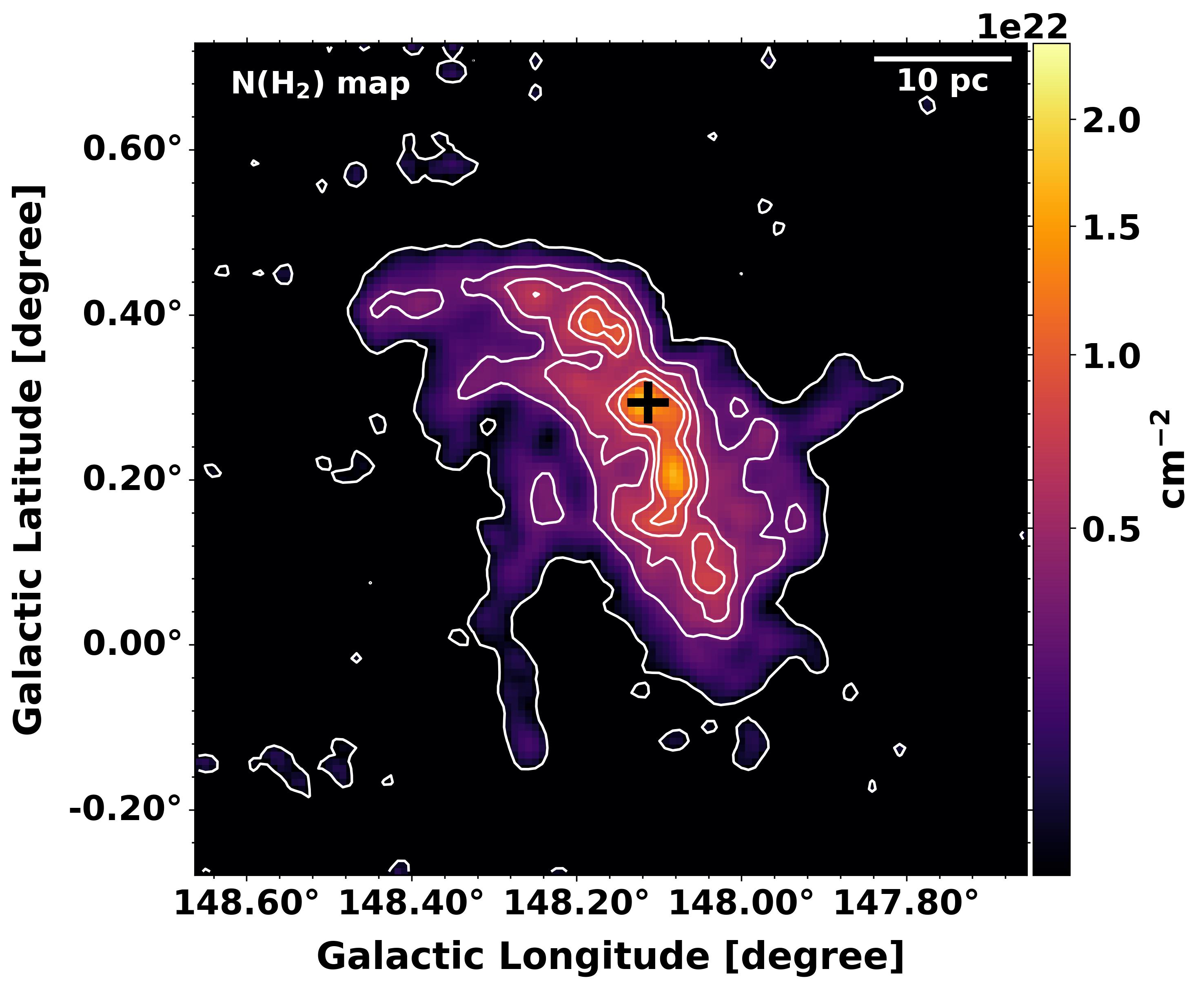}
    \caption{Molecular hydrogen column density map based on \thco. The contour levels are shown above 3$\sigma$ of the background value, starting from 0.9 $\times$ 10$^{21}$ to 9 $\times$ 10$^{21}$ \cms. The location of the hub is marked with a plus sign.
    }
    \label{fig_cdens}
\end{figure}

Since \thco covers a larger area and has a better signal-to-noise ratio compared to \eico, thus, we used \thco based column density map for further analysis, such as in deriving the global properties of the cloud.  Fig. \ref{fig_cdens} shows the \thco based $\nht$ map, tracing well the central dense location of the cloud. 
%seen in the dust column density map shown by \cite{Rawat_2023}. 
We find the peak value of $\nht$ is around 2.4 $\times$ 10$^{22}$ cm$^{-2}$, which corresponds to the location of the hub. 
%We found where both 13CO and C18O emission is detected, the H$_2$ column densities derived by these two isotopologues are not exactly the same. The discrepancy might be due to  the abundance variations of these isotopologues across different environments. For example, chemical models and observations suggest that selective photo-dissociation and fractionation can significantly affect the abundance of CO isotopologues \citep{visser_2009, liszt_2017}. In diffused regions having moderate kinetic temperature \citep[$\lesssim$ 50 K, ][]{liszt_pety_2012}, the chemical fractionation between $\mathrm{^{13}C^+}$ and \twco increases the abundance of \thco. The more abundant CO isotopologues shield themselves better than the less abundant ones from selective photodissociation by UV photons(e.g., van Dishoeck & Black 1988; Shimajiri et al. 2015), . 
%Because of different levels of the self-shielding effect (e.g., van Dishoeck & Black 1988; Shimajiri et al. 2015), C18O is selectively dissociated by far-ultraviolet (FUV) emission more effectively than is 13CO. 
%Since, \thco is optically thin in most part of the cloud and covers larger part of the cloud and also have better SNR compared to \eico, thus, we used \thco based column density map for further analysis.  Figure \ref{fig_cdens} shows the $\nht$ map based on \thco. As can be seen, the peak $\nht$ is around 3 $\times$ 10$^{22}$ \cms~, which corresponds to the location of the filamentary hub found in Paper-I. 

The \twco emission in \cloud~is more extended than \thco emission; thus, for estimating column density of the cloud area located outside the boundary of \thco emission, we also estimated the hydrogen column density of each pixel directly from the \twco intensity, I($^{12}$CO). To do so, we use the relation N(H$_2$) = X$_{\rm{CO}}$  I($^{12}$CO), where X$_{\rm{CO}}$ is the CO-to-H$_2$ conversion factor, whose  typical value is $\sim$ 2.0 $\times$ 10$^{20}$ \cms (K \kms)$^{-1}$ \citep{dame2001, bolatto2013,John_2022} with an uncertainty of around 30\% \citep{bolatto2013}. We also estimated the total \twco column density from the \thco optical depth map, using an average value of \twco/ \thco abundance of $\sim$ 60 \citep{frek_1982} and equation 3 of \cite{Garden_1991}. Doing so, we found that the total molecular hydrogen column density of the cloud based on both approaches is within a factor of 1.3. 

%\citep{lew22} suggested that the canonical X-factor may hold good for cloud scale, but may not hold true on subcloud (parsec level) scales as broad variation in X-factor is seen over subcloud scales. Since, we are
%looking for global properties of the cloud using \thco column density, 
%thus, use of canonical X-factor is not reasonable.  
%Moreover, recent
%observational (e.g., Barnes et al. 2015; Kong et al. 2015; Barnes et al. 2018) studies have suggested an X factor that
%varies significantly with several effects (e.g., excitation, abundance, opacity, and resolution). Obtaining, accurate  X factor for \cloud is
%beyond the scope of the pape r\href{https://iopscience.iop.org/article/10.3847/1538-4365/ab5b97/pdf}{X-fact}. 

%- Have a look to the various aboundance,particulary NH2-NC180 \href{Gou, Section3.3}%%{https://iopscience.iop.org/article/10.3847/1538-4357/ac8933}

%However, one should note that chemical models and observations suggest that selective photo-%dissociation and fractionation can significantly affect the abundance of CO isotopologues %\citep{visser_2009, liszt_2017}. In diffused regions having moderate kinetic temperature %\citep[$\lesssim$ 50 K, ][]{liszt_pety_2012}, the chemical fractionation between %$\mathrm{^{13}C^+}$ and \twco increases the abundance of \thco. The more abundant CO %isotopologues shield themselves better than the less abundant ones from selective %photodissociation by UV photons. For example, in the same regions where the \thco abundance %increases through fractionation, the \eico abundance decreases due to selective %photodissociation. This changes the \thco/\eico abundance ratio in a broad range. 
%\href{https://arxiv.org/pdf/1508.07898.pdf}{gas_pro}

%\paragraph{Composite column density map}

\begin{figure}
    \centering
    \includegraphics[width=8.5cm]{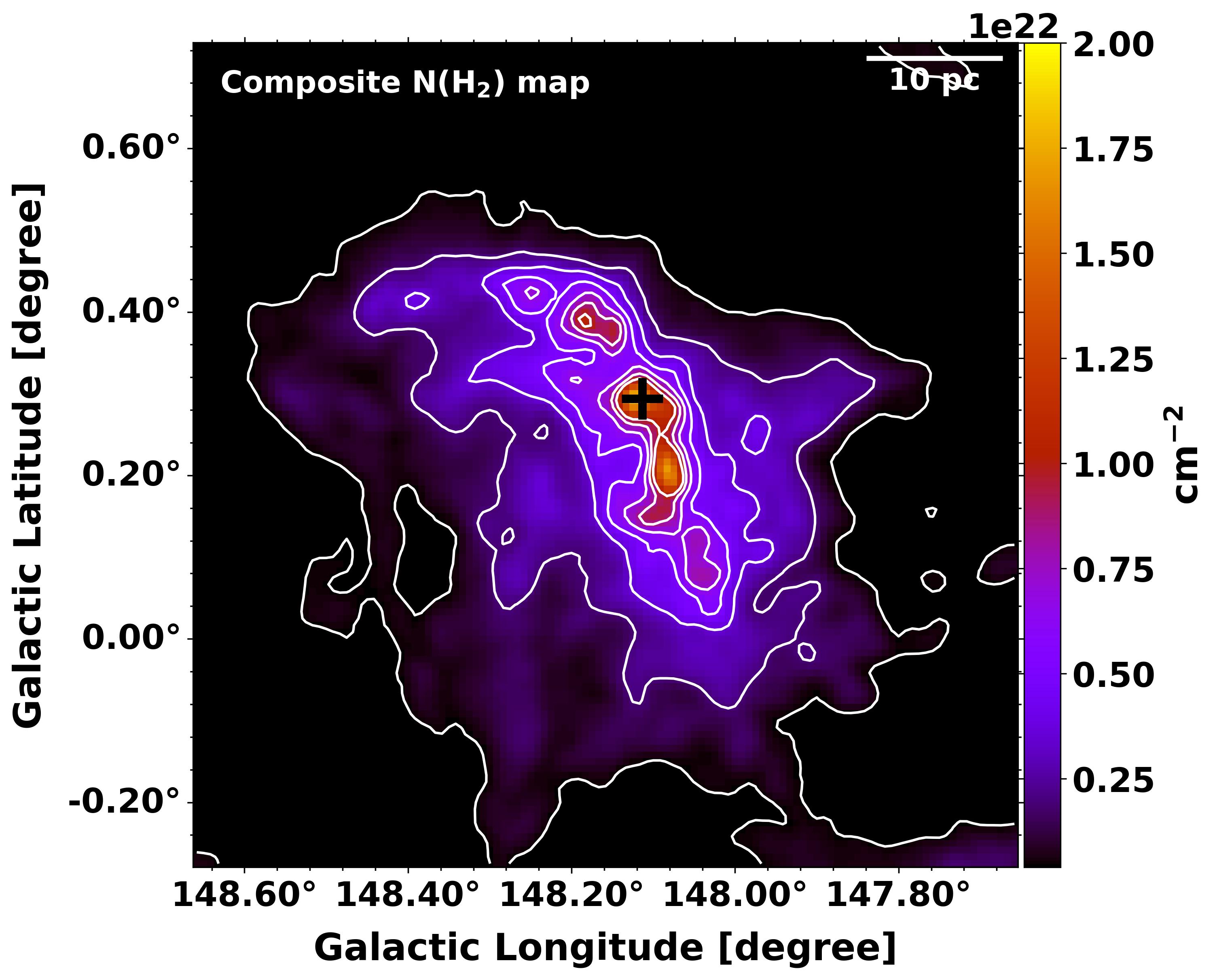}
    \caption{Composite \nht~map based on \twco~and \thco column density maps.  The contour levels are shown above 3$\sigma$ of the background value, starting from 4.3 $\times$ 10$^{20}$ to 1 $\times$ 10$^{22}$ \cms. The location of the hub is marked with a plus sign. 
    }
    \label{fig_com}
\end{figure}

We then combined \twco and \thco column density maps to make a composite molecular hydrogen column density map. For the area lying outside the area of \thco emission, we take the column density values from the \twco emission.
%which roughly corresponds to the column density threshold, $\nht$ $\sim 2 \times 10^{21} \cms$. 
Although we use $\mathrm{N(H_2)_{^{13}CO}}$ column density values within the \thco emission area, we observed that some of the pixels in the central area of the \thco map exhibit lower column density values than the neighbouring pixels. Overall, these outliers do not affect the measured global properties. Nonetheless, in these pixels, if the ratio (R) of $\mathrm{N(H_2)_{^{13}CO}}$ to $\mathrm{N(H_2)_{^{12}CO}}$ is found to be $>$ 1, the pixel values from the $\mathrm{N(H_2)_{^{13}CO}}$ map are considered, otherwise, from the 
$\mathrm{N(H_2)_{^{12}CO}}$ map. 
%As a result, within the \thco area, we adopted the following approach; 
%we defined  a pixel value as  the ratio (R) of $\mathrm{N(H_2)_{^{13}CO}}$ to $\mathrm{N(H_2)_{^{12}CO}}$  and then 
%All the pixels having column density below this threshold acquires $\mathrm{N(H_2)_{^{12}CO}}$ value. 
%For the pixels above the threshold, the R value defined as the ratio of $\mathrm{N(H_2)_{^{13}CO}}$ to $\mathrm{N(H_2)_{^{12}CO}}$ decides the pixel value.
%if R $>$ 1 for a pixel, the pixel values from the $\mathrm{N(H_2)_{^{13}CO}}$ map are considered, otherwise, from the 
%$\mathrm{N(H_2)_{^{12}CO}}$ map. In this way, we made a composite column density map, which 
The combined composite map made in this way is shown in Fig. \ref{fig_com}. The
column density of the composite map lies in the range of 0.2 $\times$ 10$^{21}$ cm$^{-2}$ to 2.4 $\times$ 10$^{22}$ cm$^{-2}$. We checked the difference in column density values at the boundary of \thco emission from both the tracers and found that they are within a factor of 1.2, thus reasonably agreeing with each other.
%We find that gas column density of the central ridge, estimated in the above way, is in close agreement with the dust the values of column densities derived in Paper-I, i.e  the two approaches  are consistent within a factor of 1.5. 
%considering the
%fact that the column density estimated by each tracer is accurate within a factor of two. Despite the uncertainty in excitation temperature, thus, in the column density,  the gas column density of the central ridge, estimated in the above way, is in close agreement with the dust the values of column densities derived in Paper-I, i.e  the two approaches  are consistent within a factor of 1.5. 

\subsubsection{Global cloud properties and comparison
with Galactic clouds}
\label{glo_pro}

We obtained the cloud properties like mass, effective radius, surface density, and volume density,  following the approach described in \cite{Rawat_2023}. Briefly, we estimated the mass of the cloud using the relation.

\begin{equation}
    M_{\mathrm{c}} = \mu_{\mathrm{H_2}} m_\mathrm{H} A_\mathrm{{pixel}} \Sigma N(H_2),
\label{eq7}
\end{equation}

where m$_\mathrm{H}$ is the mass of hydrogen, $A_\mathrm{{pixel}}$ is the area of a pixel in cm$^2$, and $\mu_{\mathrm{H_2}}$ is the mean molecular weight that is assumed to be 2.8 \citep{kauffmann2008}. We define the outer extent (thus the area) of \cloud~for different tracers by considering emission within the 3$\sigma$ contours (see Fig. \ref{fig_int}) and derive its properties within this area.
%outermost contour of the \twco intensity emission above 3$\sigma$, in which the cloud is nearly a circular structure of geometric radius $\sim$ 26 pc (see Fig. \ref{fig_int}a). Similarly, for \thco and \eico, we define the effective area of the cloud based on the outermost contour of \thco and \eico intensity emission above 3$\sigma$.
The cloud mass from \twcos, \thcos, and \eico based $\nht$ column density map, calculated above the 3$\sigma$ emission is $\sim$ 5.8 $\times$ 10$^4$ \Ms, $\sim$ 5.6 $\times$ 10$^4$ \Ms, and $\sim$ 3.5 $\times$ 10$^4$ \Ms, respectively.
%which reflects the larger emission area in \twco compared to \thco and \eico (see Fig. \ref{fig_int}).} 
%This is also evident from their effective radius (i.e., the pixels having \nht $\geq$ 8.4 $\times$ 10$^{20}$ \cms or \ak $\geq$ 0.1 mag), which is 18.8, 17.3, and 12.3 pc for \twco, \thco, and \eico, respectively.
The cloud mass estimated from the composite column density map is found to be $\sim$ 7.2 $\times$ 10$^4$ \Ms. %\textcolor{red}{The cloud masses estimated from \twcos, \thcos, and composite maps are within a factor of 2 of the mass estimated from $\it{Herschel}$ dust emission \citep[1.1 $\times$ 10$^5$, \Ms][]{Rawat_2023}}. 
To check the boundness status of \cloud, we calculated its virial mass using the relation, $M_{vir} = 126 \times 1.33 r_{\rm{eff}} \Delta V^2$, for density index, $\beta = 1.5$ \citep[see Eqn. 2 of][]{Rawat_2023}. Using $r_{\rm{eff}}$ and $\Delta V$ values of \twcos, \thcos, and \eico~(see Table \ref{tab:cloud}), the estimated $M_{\rm{vir}}$ is around $\sim$ 4 $\times$ 10$^4$ \Ms, 1.5 $\times$ 10$^4$ \Ms, and 8.6 $\times$ 10$^3$ \Ms, respectively. Since the virial mass of \cloud, as estimated by \twcos, \thcos, and \eicos, is less than their respective gas mass, it implies that the cloud is bound in all three CO isotopologues. We acknowledge that the optical thickness of \twco line can make the line profile broader, as discussed in Section \ref{glo_kin}, thus, the \twco based virial mass can be an upper limit. Even then, the aforementioned boundness status of the cloud will remain true. %We have only used \twco to obtain the global properties of \cloud to compare it with the nearby molecular clouds that have also been studied in \twco \citep{md17, John_2022}.}     

% which is higher than the virial mass \citep [$\sim$ 5.5 $\times$ 10$^4$ \Ms;][]{Rawat_2023} of the cloud, implying the cloud is bound. 
%The  mass and other properties of the cloud  are tabulated in Table \ref{tab:cloud}. 

The typical uncertainty associated with the estimation of gas mass from the \twcos, \thcos, and \eico emissions is in the range 35$-$44\%. Because the uncertainty in 
%Taking the uncertainty of 30\% 
the assumed $\rm{X_{CO}}$ factor and in the isotopic abundance values of CO molecules, used in converting N(CO) to $\nht$ is around 30 to 40\% \citep{Wilson_rood_1994, savage2002, bolatto2013}. The distance uncertainty associated to the cloud is around 9\%. In addition,
the estimation of N(CO) is also affected by the uncertainty associated with the estimated gas kinematic temperature. In the present case, we find that the average gas kinetic temperature (8 K) is lower than the average dust temperature of the cloud, 14.5 $\pm$ 2 K \citep{Rawat_2023}. When gas and dust are well mixed, the gas kinematic temperature better corresponds to the dust temperature, and this occurs when the density is  $>$ 10$^4$ \cmq. For example, \cite{Goldsmith_2001} found that dust and gas are better coupled at volume densities
above 10$^5$ \cmq, which are typically not traced by \twco and \thco data (n$_\mathrm{{crit}}$ < 10$^4$ cm$^{-3}$). They find a temperature difference of  $\sim$ 4 K at density $\sim$ 10$^5$ \cmq and completely negligible at density $\sim$ 10$^6$ \cmq. 
Moreover, it is also suggested that if the volume density of the gas is lower than the critical density of \twcos %or the filling factor of \twco emission is lower than unity
, this would lead to a lower excitation temperature  \citep{hey09}.
Assuming the true average temperature of the gas is around 14 K,
we estimated that it would change the \thco column density by a factor of 14\%, hence the estimated gas mass would also change by this factor. 

In the present work, though we have derived the masses using canonical values of $\rm{X}$ factor and the CO abundances, however, it is worth mentioning that many studies have suggested that these values increase towards the outer galaxy \citep{naka06, Pineda_2013, hey15, Patra_2022}. Since \cloud~is located in the outer galaxy (i.e. $\sim$ 11.2 kpc from the Galactic centre), the derived masses are likely underestimations. For example,
we find that implementing  $\rm{X_{CO}}$ value from the relation given in \cite{naka06}, would increase the total $\mathrm{N(H_2)_{^{12}CO}}$ column density and thus, the mass by a factor of $\sim$ 2. 

The total gas mass estimated for G148.24+00.41 in the present work using the composite column density map, within uncertainty, agrees with the dust-based gas mass $\sim$ ($1.1 \pm 0.5) \times 10^5$ \Ms, derived by \citet{Rawat_2023}. In Table \ref{tab:cloud}, we have tabulated the cloud mass, mean column density, effective radius, surface density, and volume density of the cloud. 
%The composite map is providing higher mass for \cloud~because it gets contribution both from \twco and \thco, in which \twco covers the outer diffused region of the cloud and \thco traces the inner dense regions of the cloud.
The derived surface mass density from \twcos, \thcos, \eicos, and composite map is $\sim$ 52, 59, 72, and 63 \Ms~pc$^{-2}$, respectively. The surface mass density from \twco is similar to the value which we have obtained for \cloud~from the dust continuum and dust extinction-based column density maps
for the same area, \citep[i.e. $\Sigma_{\rm{gas}}$ = 54 \Ms~pc$^{-2}$, see][] {Rawat_2023}. Since the isotopologues trace different areas of the cloud, their estimated surface densities are different, with a gradual increase from low-density to high-density tracer.
%Table \ref{mass} shows the cloud properties calculated from \twco, \thco, \eico, and composite-based column density maps. 
%The difference in the surface densities derived from \twco and \thco is within a factor of 1.2.  

Comparing the properties of \cloud~with other Galactic clouds, we find that the \twco surface density of \cloud~is significantly higher than the average surface density ($\sim$ 10 \Ms~pc$^{-2}$) of the outer Galaxy molecular clouds of our own Milkyway \citep{md17}. \citet{md17} studied Galactic
plane clouds using  \twco and found that the average mass surface density of clouds is higher in the inner Galaxy, with a mean value of 41.9 \Ms~pc$^{-2}$, compared to 10.4  \Ms~pc$^{-2}$ in the outer Galaxy.  Similarly, we also find that the derived \thco surface density of \cloud~is on the higher side of the surface densities of the Milkyway GMCs studied by \cite{hey09}. \citet{hey09} found an average surface density value $\sim$ 42 \Ms~pc$^{-2}$ using \thco data, assuming LTE conditions and a constant H$_2$ to \thco abundance, similar to the approach used in this work. 
Recently \citet{John_2022}  investigated nearby star or star-cluster forming GMCs, including Orion-A, using \twco emission and similar X$_{\rm{CO}}$ factor used in this work. Comparing the surface densities of these clouds, we find that the 
surface density of \cloud~is higher than most of their studied  GMCs (average $\sim$ 37.3 $\pm$ 10 \Ms~pc$^{-2}$) and comparable to the surface density of Orion-A \citep[see Fig. 8 of][]{John_2022}. All the aforementioned comparisons support the inference drawn by \cite{Rawat_2023} on \cloud~based on the dust continuum analysis, i.e. \cloud~is indeed a massive GMC like Orion-A.
\\

%Taking the uncertainty of 30\% in the assumed $X_{CO}$ factor \citep{bolatto2013}, and including the distance uncertainty of 9\%, the uncertainty in the \twco based cloud mass is around 35\%. However, as discussed, the $X_{CO}$ factor is expected to increase in the outer galaxy, so assuming a constant value for $X_{CO}$ throughout the galaxy will systematically underestimate the estimated mass and surface density of the cloud. \cite{hey15} estimated that for the outer galaxy, this effect can be significant, even up to $\sim$ 50\%. Since this cloud is in the outer galaxy (11.2 kpc from the galactic centre), therefore the mass of the cloud from the \twco column density is likely underestimated. 
%Moreover, the \thco abundance w.r.t to \nht is expected to increase towards the outer galaxy, as the metallicity of gas decreases with the galactic distance \citep{bal2011}.

\begin{table*}

\caption{\cloud~properties from CO emission. The mass of the cloud from $\rm{^{12}CO}$, $\rm{^{13}CO}$, and $\rm{C^{18}O}$ is calculated above 3$\sigma$ from the mean background emission. The FWHM is the line-width of the spectra, calculated as $\Delta V = 2.35\sigma_{1D}$.} %The errors quoted here are due to uncertainty in the estimated distance.}
  
\begin{tabular}{|p{2.0cm}|p{2.0cm}|p{1.3cm}|p{1.8cm}|p{1.8cm}|p{1.5cm}|p{1.0cm}|p{1.0cm}|p{1.5cm}|} 
\hline
\hline
Emission & Velocity interval & V$_{\rm{peak}}$ & FWHM ($\Delta V$) & Mean \nht & Mass & n(H$_2$) & r$_{\mathrm{eff}}$ & $\Sigma_{\mathrm{gas}}$\\

    & (\kms) & (\kms) & (\kms) & ($\times$ 10$^{21}$ \cms) & (\Ms) & (\cmq) & (pc) & (\Ms~pc$^{-2}$)\\
\hline
\hline
\twco & ($-$37.0, $-$30.0) & $-$34.07 & 3.55 & 2.3 & 5.8 $\times$ 10$^4$ & 30 & 18.8 & 52 \\ 
\thco & ($-$37.0, $-$30.0) & $-$33.83 & 2.30 & 2.6 & 5.6 $\times$ 10$^4$ & 37 & 17.3 & 59 \\ 
\eico & ($-$36.0, $-$31.0) & $-$33.72 & 2.04 & 3.2 & 3.5 $\times$ 10$^4$ & 63 & 12.4 & 72 \\
Composite-map  & ($-$37.0, $-$30.0) & $ \ \ \ \ \ $--- & $ \ \ $--- & 2.8 & 7.2 $\times$ 10$^4$ & 37 & 19.0 & 63 \\
(\twco \& \thco) &  &  &  &  &  &  &  &\\

\hline
\hline

\end{tabular}

\label{tab:cloud}

\end{table*}

%Although the mean column density is same, the mass estimated from the %composite column density map is almost 1.2 times the mass from \thco. %Also, including \eico, the mass reaches upto $\sim$ 8.0 $\times$ 10$^4$ %\Ms, which can be the upper limit of cloud mass. The mass estimate from %CO emission is within the factor of 2 from the dust mass (paper I). The %median column density is also similar, around 3 $\times$ 10$^{21}$ \cms.\\

%\end{document}

\subsection{Filaments and Filamentary Structures}
\label{filaments}
As mentioned in Section \ref{int}, based on dust continuum maps, \cite{Rawat_2023} suggested that the central cloud region likely consists of six filaments, forming a hub filamentary system (HFS) with the hub being located at the nexus or junction of these filaments. 
%And based on the  $\it{Spitzer}$ images, we find that a cluster is deeply embedded in the hub. 
%Although the formation mechanism of filaments in molecular clouds is still not clear \cite[e.g.][]{pine22}, however, based on various dust continuum and molecular observations, it is now certain that filaments are ubiquitous in the interstellar medium \citep{Andre2010, Andre2014}, and they play a crucial role in the  star-formation process \citep[e.g.]{hacar22,pine22}. In this regard, 
Molecular clouds with HFSs are of particular interest because these are the sites where cluster
formation would take place, as advocated in many simulations \citep[e.g.][]{Naranjo_2012, gomez&sema2014, Gomez_2018, sema2019}.
%also supported by observations \citep{liu2019, kumar2020}.
Massive and elongated hub regions are sometimes referred to as “ridges” \citep[e.g.][]{Hennemann_2012, Tige2017, Motte_2018}.

%Hubs or ridges are located close to the gravity center and thus may aid in global collapse or large-scale gas inflow along the filaments \citep{myers2009, Andre2014, kumar2020}. Simulations suggest that the HFS could also occur due to the Global Hierarchical Collapse model \href{https://par.nsf.gov/servlets/purl/10166284}{(GHC)}
%Regardless of the origin of the HFS, 

%Simulations suggest, the kind of star cluster that would emerge from a hub depends on the properties of the associated filaments, like their mass reservoir, mass-inflow rate, and dispersal time.  Therefore, the identification and characterization of filaments in molecular clouds are becoming crucial for understanding their role in the formation and growth
%of the star cluster in the hub. 
The line-of-sight (LOS) velocity gradient, traced by molecular lines, is commonly interpreted as a proxy for the plane-of-sky (POS) gas motion. Recent molecular line observations have revealed the kinematic structures of several HFSs in nearby clouds, and significant velocity gradients are observed along several filaments that are attached to HFSs \citep[e.g.][]{liu2012, Friesen_2013, Hacar_2018, Dewangan_2020, Chen_2020, Yang_2023, Hong_Liu_2023}. These gas motions are thought to represent dynamical gas flows which are fuelling the hub. 
%The comparison between the density structures and the direction of the velocity gradient further suggests that these gas flows are possibly driven by gravity \citep[e.g.][]{Friesen_2013, Hacar_2018}. Local velocity gradients within HFSs have been measured in a few systems \citep[e.g.][]{Peretto_2014, Williams_2018, Chen_2020}.
In the following, we identify and characterize the filamentary structures in the cloud and discuss their role in star and cluster formation observed in the cloud. 
%to describe the identification method of velocity coherent filaments. We then, discuss the %method to %extract filament profiles and their properties like length, width, mass, and %longitudinal accretion. %Finally, we present the position-velocity map along the filamentary %structures. 

\subsubsection{Identification of global filamentary structures}
\label{fil_glo}

\begin{figure}
    \centering
    \includegraphics[width=8.5cm]%{13CO_moment_0__filaments.jpe}
    {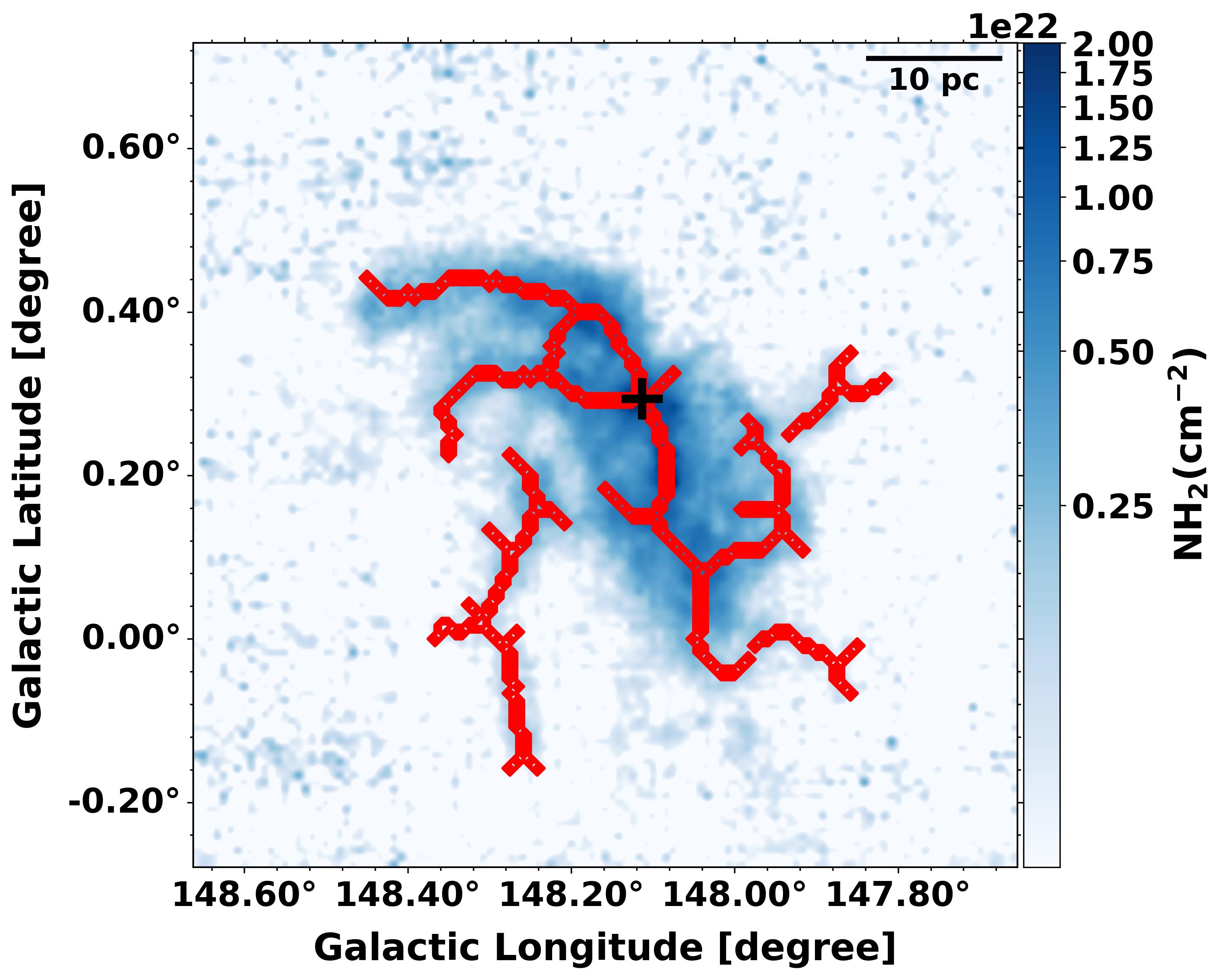}
    \caption{Global skeletons of \cloud~showing the filamentary structures, main ridge, nodes, and central hub location, over the \thco based $\nht$ map. The location of the hub is marked with a plus sign.
    }
    \label{fig_ske}
\end{figure}

We used a python-based package - $\it{FilFinder}$\footnote{\href {https://github.com/e-koch/FilFinder} {https://github.com/e-koch/FilFinder}} \citep{koch2015} to identify the filamentary structures of the cloud (details of FilFinder is given in Appendix A) using the \thco based molecular hydrogen column density map. 
Fig. \ref{fig_ske} shows the extracted skeletons of the \cloud~cloud. It can be seen that the $\it{Filfinder}$ algorithm reveals several filamentary structures, including the main central filament that runs from north-east to south-west, and also several nodes where the filaments are intersecting.  
%Like any filament-finding algorithm, it is not always possible to detect faint small-scale structures, particularly those that are close to bright structures or confused with the nearby surroundings.  This is also possible in case of \cloud, as we see some of the structures appear to be isolated and not connected to the main filament through the filamentary network.
%However, it is worth mentioning that such cases are commonly seen in molecular clouds \textcolor{red}{(e.g. ).} 
%We want to stress that due to low-resolution data, we are possibly missing some of the small-scale filamentary structures of the cloud that are seen in the dust continuum image (e.g. see Fig. \ref{fig_hub}).

%\textcolor{red}{The $\it{FilFinder}$ connects all the branches to the main skeleton (see Fig. \ref{fig_fil1}). Although the filament extraction with column density map made from the integrated intensity map, integrated
%over the entire velocity range reveals the presence of global large-scale structures.} 

% Suggestion......................
The column density map is created from the integrated intensity map. So it's important to acknowledge that in the integrated intensity map, multiple individual velocity features may blend together and appear as a single one, as observed in nearby filamentary clouds \citep[e.g.][]{hacar2013, hacar2017} or in distant ridge \cite[e.g.][]{hu21,cao22}. Velocity sub-structures of gas in a cloud can be inspected using channel maps, in which the emission integrated over a narrow velocity range is examined. 
%Similarly, position-position-velocity (PPV) maps can separate sub-structures  that are spatially confused in the plane of the sky but are separated in velocity.
It is then possible to identify individual velocity coherent structures that are very likely to correspond to the physically distinct structures of the cloud.

%It is possible to identify the likely distinct features of a given cloud by examining different velocity intervals of the data cube using
%channel maps. 
%However, in dust continuum and integrated intensity maps, the small velocity coherent features can get merged together to give an impression of single large-scale filamentary features \citep{hacar2013}.

\subsubsection{Small-scale gas motion and velocity coherent structures}
\label{fil_cha}

Fig. \ref{fig_chan} shows the \thco velocity channel maps with a step of 0.34 \kms. As can be seen from the channel maps, along with several compact emissions,  multiple spatially elongated velocity structures are also present. These elongated structures 
are marked as St-1, St-2, St-3, St-4, St-5, and St-6 on the map. The
location of these structures in the map corresponds to either the maximum intensity feature or has relatively the longest distinct visible structure, or both.  These structures have noticeable differences
in velocity because they emerge in different velocity channels. 
%The channel maps suggest, the structure, St-1, emerges around $\leq$ $-$36.5 \kms and joins with the structure, St-2,  at around $-$35.0 \kms. St-2 is also joined by the structure, St-3, which is prominent between $-$35.5 and $-$34.7 \kms. The structures, St-4 and St-5 become prominent between $-$34.5 \kms and $-$33.7 \kms. The structure, St-6, is better apparent at $-$33.4 \kms, while  St-7 emerges around $-$33.4 \kms and becomes prominent at -32.4 \kms.

The majority of these structures appear to move towards the hub location, marked by a plus sign
on the map. The merger and convergence
of these structures form a nearly continuous structure in the central area of the cloud that we refer to as the "ridge" in which the hub is located.
The ridge is marked by a solid green line on the channel map. 
%The interface between St-1 and St-2 shows much larger velocity dispersion, as revealed in the second-moment map of \thco emission. 
The ridge also seems to be attached to several small-scale strand-like, nearly perpendicular elongated structures (shown by arrows in Fig. \ref{fig_chan}). The kinematic association of such perpendicular structures 
with the main filament/ridge indicates the possible direct role of the surrounding gas 
on the formation and growth of the main filament/ridge \citep[e.g.][]{cox16}. Besides, one can see that the structure, St-2, is composed of 2-3 small-scale 
filamentary structures that are seen in the channel maps at around $-$35.1 to $-$ 34.4 \kms. These structures are indistinguishable in the integrated intensity map shown in Fig. \ref{fig_int}b, emphasizing
that some of the elongated filamentary structures that we are seeing in the integrated intensity map could be the sum of multiple velocity coherent structures. 
%Simulations also predict that large-scale filaments may actually be composed of a series of small-scale filaments. For example, \cite{Smith_2016} proposed a "fray and gather" model, in which the sub-filaments are formed first and then gathered together by large-scale motions within the cloud, initially by large-scale turbulent modes and afterward gravitationally. 

%especially in the channel-[$-$34.780, $-$34.436] \kms, that are not distinguishable in integrated intensity maps. 
%Also, it is evident from the channel map that 2--3 filamentary features on the north-eastern side are showing significant emission only in the velocity range $-$37.0 \kms~to $-$34.0 \kms. In
%the velcoity rnage xx
%In the PPV space \href{https://www.aanda.org/articles/aa/pdf/2019/11/aa34903-18.pdf}{ngc2264}, it is possible to identify velocity coherent structures using similar method adopted in  Hacar et al. (2013), where the velocity components are grouped based on how closely they are linked in both position and velocity simultaneously.  Fig. xx shows the PPV representation of various structures present in the cloud. 

\subsubsection{Likely velocity coherent filaments}
\label{fil_fil}
In molecular clouds, small-scale velocity coherent filaments (VCF) have been identified using position-position-velocity (PPV) maps, where the velocity components are grouped based on how closely they are linked in both position and velocity simultaneously \citep[e.g.][]{hacar2013}. In the literature, identification of VCFs is primarily done using high-resolution and high-density tracer (e.g. NH$_3$, N$_2$H$^+$, \eico)  data cubes and preferentially on the nearby clouds, where structures are well resolved  \citep[e.g.][]{hacar2017,Hacar_2018,shima2019}. However, as witnessed from the channel maps, the gas kinematics of the cloud is quite complicated with overlapping structures. Disentangling and identifying individual velocity coherent structures is challenging with the present data. None the less, to identify the likely VCF of \cloud, we followed an approach similar to that of the nearby molecular clouds. We visually inspected the \thco data cube and identified the velocity coherent structures that are continuous in position as well as velocity in the data cube. We then made the integrated intensity map of the structure by integrating the emission in the velocity range that encompasses the majority of its emission. In this way, we identified six likely velocity coherent filamentary structures %(shown in Fig. \ref{fig_fil_cut}) 
in \cloud.

\begin{figure*}
    \centering
    \includegraphics[width=17cm]{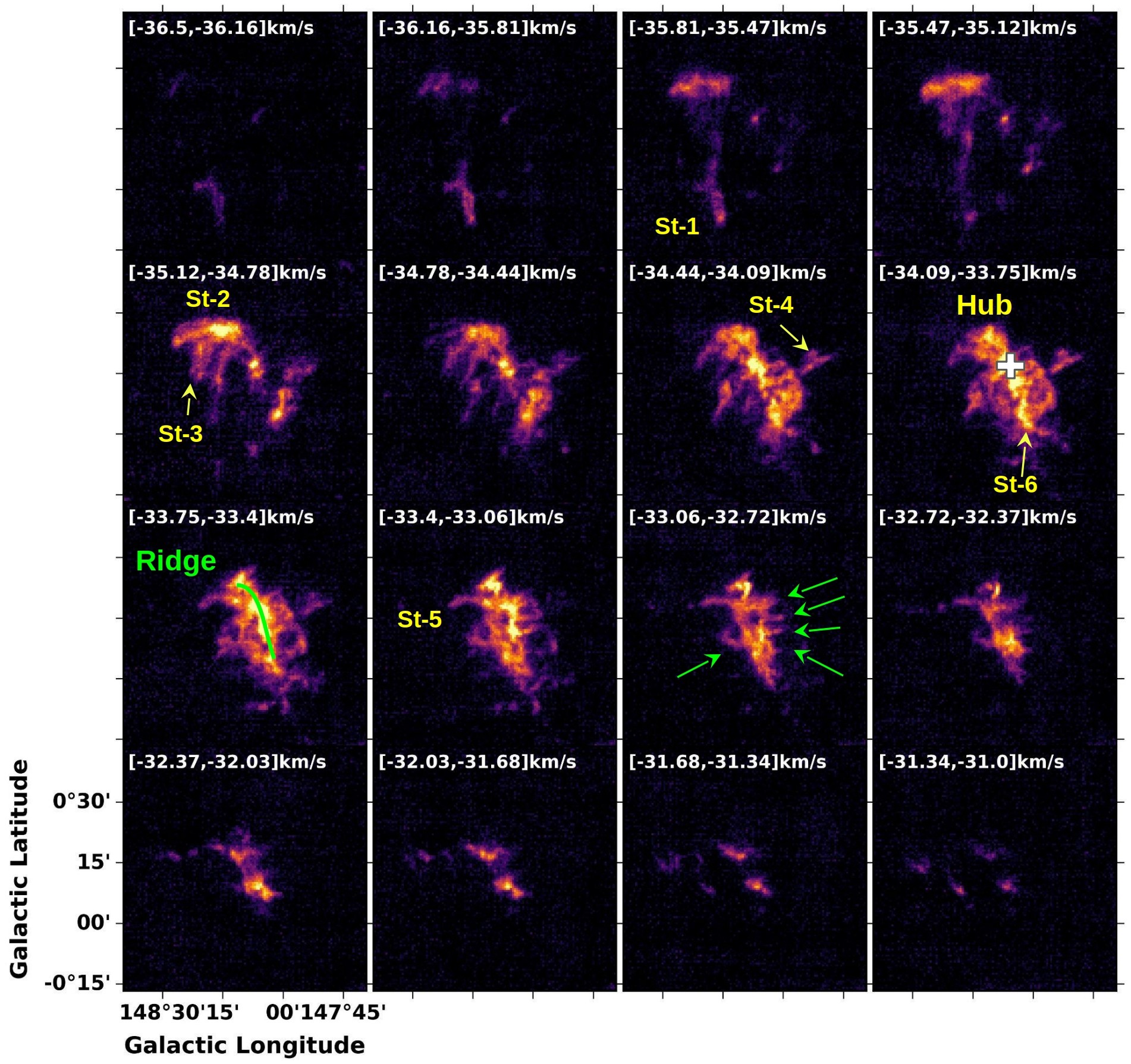}
    \caption{ Velocity channel maps in units of K \kms for the \thco emission.  The velocity ranges of the channel maps are indicated at the top left of each panel. The ridge (green curve), strands (green arrows), structures (yellow arrows), and the hub location (plus) are marked in the channel maps. 
    }
    \label{fig_chan}
\end{figure*}

\begin{figure}
    \centering
    \includegraphics[width=8.5cm]{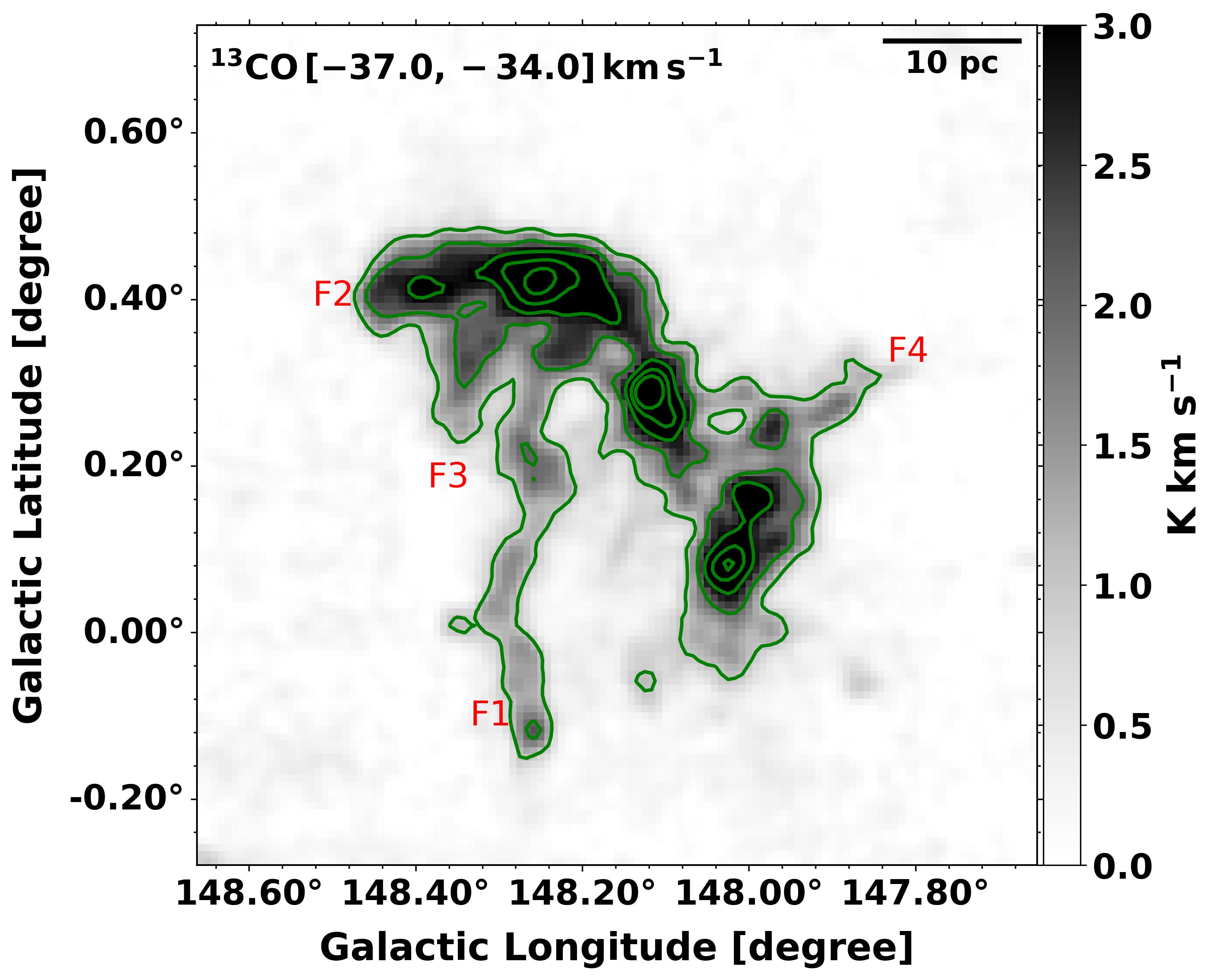}
    \caption{\thco integrated intensity map in the sub-velocity range, $-$37.0 \kms to $-$34.0 \kms. The filamentary features- F1, F2, and F3 are prominent in this velocity range. Filament-F4 is also visible here.   
    }
    \label{fig_velcut}
\end{figure}

Fig. \ref{fig_velcut} shows an example of an intensity map, integrated in the sub-velocity range, [$-$37.0, $-$34.0] \kms, where filament F1, F2, and F3 are identified, while F4, F5 and F6, are identified in the full velocity integrated intensity map (see Fig. \ref{fig_ske}). Compared to Fig. \ref{fig_velcut}, the identification and delineation of F3 filament is confusing and difficult in Fig. \ref{fig_ske}, whereas in Fig. \ref{fig_velcut}, the structure of  F3 is better apparent,  and seems to connect to F2. We note that although our approach is subject to the choice of velocity range, we find that except for one, the majority of the identified structures matched well with the structures shown in Fig. \ref{fig_ske}, but separated into different velocity coherent filaments. 

All the identified  structures are marked in Fig. \ref{fig_fil_cut} as F1, F2, F3, F4, F5, and F6. Most of these filaments correspond to the structures marked in Figure \ref{fig_chan}. The length of the filaments lies in the range of 15$-$40 pc.
%while the velocity range used to make their integrated intensity maps are tabulated in Table \ref{tab:fil_prop}. 
It is important to emphasize that while we have identified six probable filaments within the cloud based on our data, we acknowledge that the identification of such structures is also subject to the resolution of the data. %For example, we could not clearly distinguish St-5 and St-7 marked in the channel map, thus, tentatively considered them as a single structure (i.e. the filament F6). 
%The structure, St-6, forms the central dense
%area of the cloud, the majority part of which coincides with the high-column density elongated structure ($\nht $>$ 8 \times 10^{21} \cms$) seen in the $\it{ Herschel}$ map (i.e the  south-eastern filament of the Fig. \ref{fig_hub}).
Comparing the morphology of the \thco integrated intensity based filamentary structures identified here with the dust-based filamentary structures visually identified by \cite{Rawat_2023} in the central region of the cloud, we find that the filaments F2, F5, and F6 reasonably agree with the major filaments of \cite{Rawat_2023} marked in Fig. \ref{fig_hub}. While the smaller $\it{Herschel}$ filaments attached to the hub are not identifiable in our low-resolution data. Future high-resolution observations may resolve the filaments into multiple sub-filaments \cite[e.g.][]{hu21}. We proceed with the present available data to characterize the identified filamentary structures to get a  sense of their role in the cluster formation of the cloud. 
%Fig. \ref{fig_fil_cut} shows the integrated intensity emission of individual filament constructed 
%5 times above the local background value. 

%In Fig. \ref{fig_velcut}, we have also marked
%the intermediate regions  between  C1 (i.e. the central clump in the hub) and C2 and between C1 and C3 as RF1 and RF2, respectively, which we used to discuss the gas motion towards the central hub between these clumps (discussed
%in Section \ref{fil_kin}).

%is used to investigate the gas flow velocity towards the central hub.}
%in this velocity range. In this figure, the SF1, SF2, and SF3 filaments correspond to the sub-velocity range [$-$37.0, $-$34.0] \kms, while the filament F4 corresponds to the full velocity range in which the cloud lies, i.e. [$-$37.0, $-$30.0] \kms. Therefore, to separate it out, hereafter, we denote the SF2 filament as F2 for the velocity range [$-$37.0, $-$30.0] \kms, and both of them are marked in Figure \ref{fig_velcut}. Figure \ref{fig_velcut} also shows the RF1 and RF2 marked regions, which connect the upper and lower clumps with the central hub/clump, and it is used to investigate the gas flow velocity towards the central hub. The contours in figure \ref{fig_velcut} also clearly reveal the clumpy structures in the cloud (more discussion in section \ref{clumps}). \\

%The velocity coherent features of \cloud~are complicated from both morphological and kinematical points of view, and the spatial resolution of 55\arcsec is not sufficient to resolve the individual features. 

%\textbf{To study the kinematics and properties  of the individual filaments}, we cut down the filament surrounding regions depending upon the contour levels above 3$\sigma$ of
%the local background.  

%We then explored marked in Fig. \ref{fig_velcut}, whose spines, profiles, and properties are then explored separately. Fig. \ref{fig_fil_cut} shows the integrated intensity emission of individual filament cut regions.
%The filament F1, F2, and F3 are analysed in the velocity range [$-$37.0, $-$34.0] \kms, while the filament F4, F5, and F6 are analysed in the velocity range [$-$37.0, $-$30.0] \kms. 
%The regions around the central hub location marked by RF1 and RF2 are analyzed in different velocity ranges: [$-$37.0, $-$30.0] \kms, [$-$37.0, $-$33.0] \kms, and [$-$37.0, $-$34.0] \kms.\\ 
%The filament F2 and F6 form the main central filament/ridge, which is visible in the \eico intensity map. 
\subsubsection{Properties of the filaments}
\label{fil_pro}

\begin{figure*}
    \centering
    \includegraphics[width=17.5cm]{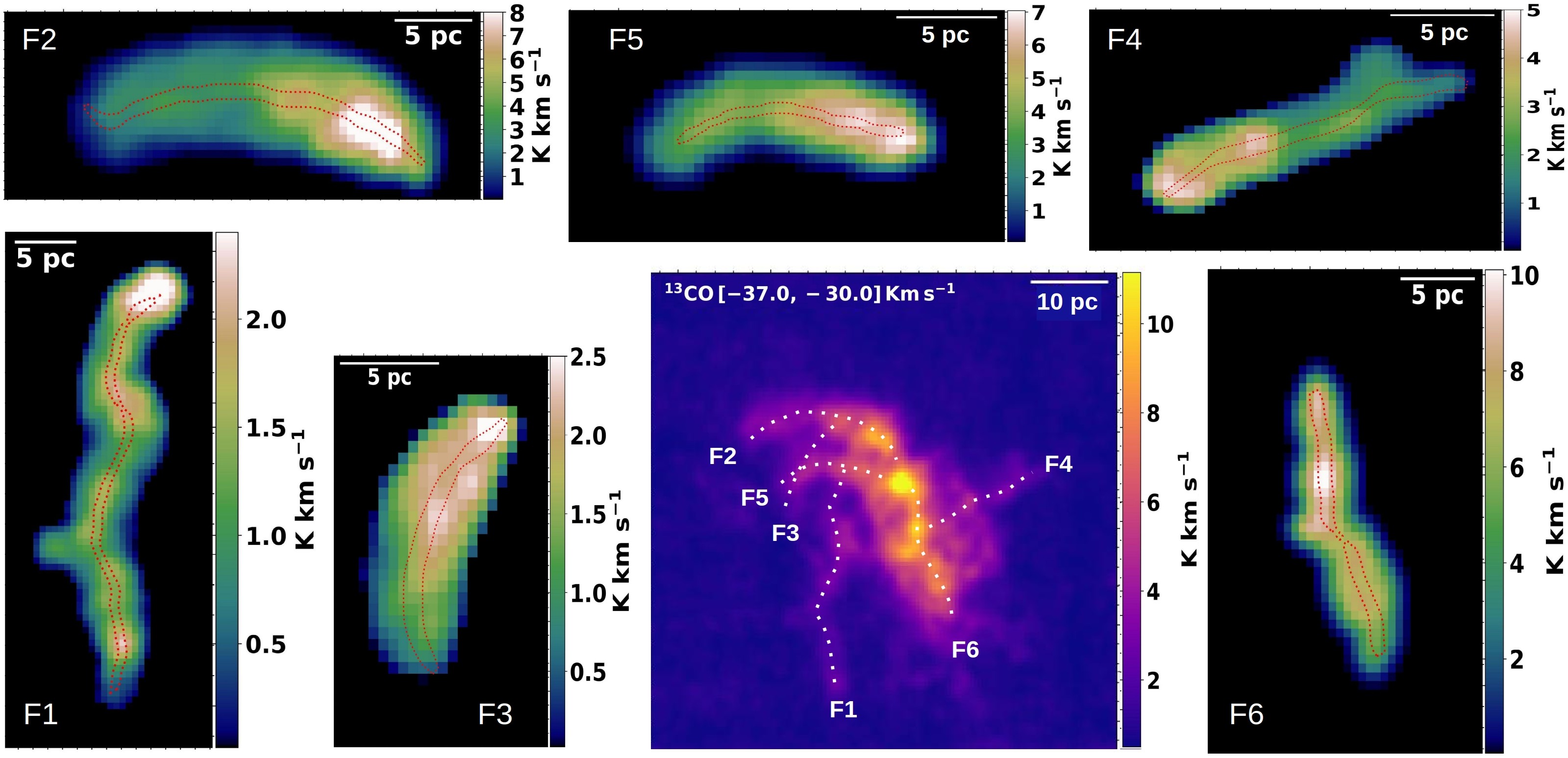}
    \caption{ \thco integrated intensity maps of individual filaments. The red-dotted curve in each filament map shows the filament spine extracted from $\it{Filfinder}$. \\ 
    }
    \label{fig_fil_cut}
\end{figure*}

\begin{figure}
    \centering
    \includegraphics[width=8.0cm]{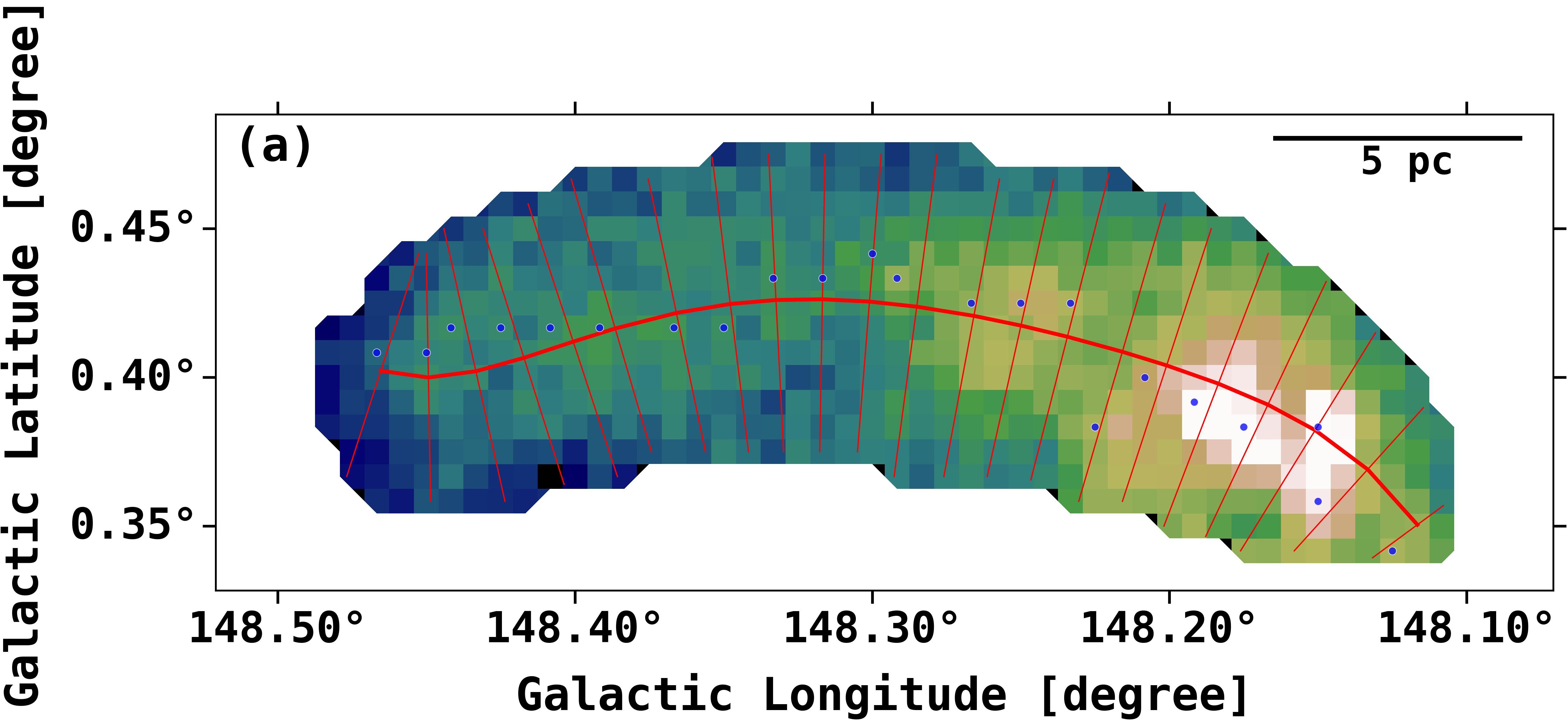}
    \includegraphics[width=8.0cm]{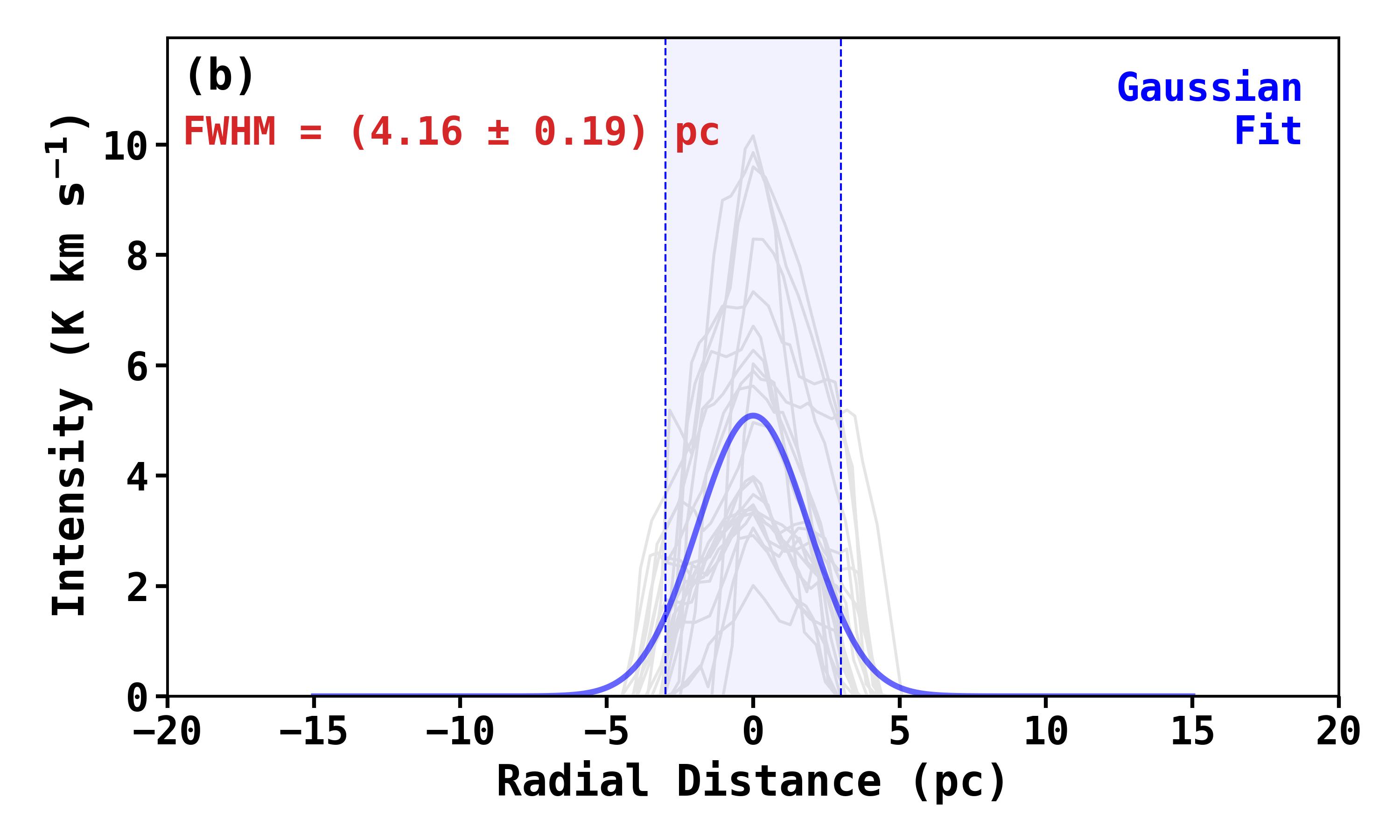}
    \caption{(a) The filament spine of F2 (red solid curve) shown over the \thco integrated intensity emission, integrated in the velocity range, [$-$37.0, $-$34.0] \kms. (b) The radial profile of filament F2, built by sampling radial cuts (red solid lines perpendicular to filament spine shown in panel a) at every 2 pixels (roughly 1 beam size $\sim$ 52\arcsec or 0.9 pc). The radial distance at a given cut is the projected distance from the peak emission pixel, shown by blue dots in panel a. The grey dots trace the profile of each perpendicular cut, and the blue solid curve shows the Gaussian fit over these filament profiles. The light-blue shaded region shows the range of radial distance taken for the Gaussian fit. 
    }
    \label{fig_spine_cut}
\end{figure}

We make use of $\it{RadFil}$ \footnote{\href {https://github.com/catherinezucker/radfil} {https://github.com/catherinezucker/radfil}} \citep{zuck-chen2018}, a python-based tool to obtain the radial profile and width of the filament. $\it{RadFil}$ also uses the $\it{FilFinder}$ to generate the filament spines. It requires two inputs, image data, and filament mask, which we provided from the output of $\it{FilFinder}$ for the individual filament.  %We again employed $\it{FilFinder}$ to identify the spines of the filaments. 
%The mask is determined for every filament based on their 5$\sigma$ boundary.
Fig. \ref{fig_fil_cut} shows the extracted filament spines over their \thco integrated intensity emission. 
 $\it{RadFil}$ makes radial profiles of perpendicular cuts drawn along the filament spine as shown in Fig. \ref{fig_spine_cut}a for filament F2  and then obtains filament width (Full Width Half Maxima, FWHM) by fitting a Gaussian function considering all
the radial profiles (the detailed description of the procedures is given in Appendix B) as shown in  Fig. \ref{fig_spine_cut}b.

%$\it{RadFil}$ also uses the $\it{FilFinder}$ to generate the filament spines. It requires two inputs, image data, and filament mask, which we provided from the output of $\it{FilFinder}$ for the individual filament.  %We again employed $\it{FilFinder}$ to identify the spines of the filaments. 
%The mask is determined for every filament based on their 5$\sigma$ boundary and is shown in Fig. \ref{fig_fil_cut}. 
%Fig. \ref{fig_fil_cut} also shows the extracted filament spines over the integrated intensity emission. 

We obtained the deconvolved FWHM by taking into account the beam size (52\arcsec) as: \\
$\mathrm{FWHM_{decon} = \sqrt{FWHM^2 - FWHM_{bm}^2}}$ \citep{kon2015}, where FWHM$_{\rm{bm}}$ is the beam size. The obtained FWHM$_{\rm{decon}}$ for all the filaments are listed in Table \ref{tab:fil_prop}. 
%An example of the Gaussian fit to
%the profiles of the F2 filament is shown in Fig. \ref{fig_spine_cut}b. 

In our case, the filament widths turn out to be in
the range of 2.5$-$4.2 pc, with a mean $\sim$ 3.7 pc. The obtained
widths are found to be higher than the typical width of $\sim$ 0.1 pc obtained from $\it{ Herschel}$-based dust emission analysis of nearby clouds \cite[e.g.][]{Andre2010,Andre2014,Arzou2019}. 
%For example, for eight nearby clouds (distance $<$ 500 pc), \citet{Arzou2019}, found  filament widths to be around 0.1 $\pm$ 0.06 pc. 
However, it is worth noting that many observations and simulations have also argued that the width of filaments depends on  many factors such as the fitted area, used tracer, resolution of the data, distance, evolutionary status of
the filaments, and magnetic field \citep[e.g. see][]{Smith_2014, Sch_2014, Fed_2016, Pan_2017, Suri_2019, Pan_2022}.
%although it still debated 
%But the robustness of these results and whether they indicate a characteristic scale for filament widths continued to be debatable \citep{Pan_2017, Pan_2022} as  it is advocated that filament width is possibly affected by many factors such as the fitted area, distance, tracer, resolution, evolutionary status, and magnetic field \citep[for e.g. see][]{Smith_2014, Schisano_2014, Fed_2016, Pan_2017, Suri_2019}
%\cite{Pan_2017} found no indication of this characteristic filament width and reported that this discrepancy could be due to the adopted methodology for filament width calculation.
For example, \cite{Pan_2022} found that the mean filament width for the nearby clouds is different from that of far away clouds. They also found that the mean per cloud filament width scales with the distance approximately as 4$-$5 times the beam size. Although the debate on the characteristic filament width of 0.1 pc is yet to be settled \citep[see discussion in][]{Pan_2017, Pan_2022}; we want to emphasize
that 
%They have also shown that the filament width determination is highly dependent on the resolution of the data set used, as the resolution can affect the shape of the radial profiles of the filaments.
%The mean filament width determined using the \thco~molecular data for \cloud~is 3.66 pc. 
our extracted filament widths might be on the higher side because \cloud~is located at a distance of $\sim$ 3.4 kpc and analysed with the low-resolution ($\sim$ 0.9 pc) and low-density tracer CO data. 
%The physical beam size of our data at the distance
%of the cloud is about 0.9 pc and therefore, we are limited by the resolution.
Moreover,
%whereas the filamentary clouds for which a small characteristic filament width ($\sim$ 0.1 pc) is found, is either at a relatively small distance and analyzed with the high-resolution data \citep[for e.g. see][]{Hacar_2018, Arzou2019, Li_2022}. 
 some of the filaments (e.g. F2) could be the sum of a series of sub-filaments, whereas filaments in nearby clouds are well resolved. In addition, \thco is tracing better the enveloping layer of the filaments. None the less, it is worth mentioning that using the PMO \thco data, \cite{Liu_2021} and \cite{guo22}, found similar mean filament widths of $\sim$ 3.8 pc and $\sim$ 2.9 pc, respectively, for Galactic plane filamentary clouds located at 2.4 kpc and 4.5 kpc, respectively.
 %We hypothesized that, the high-width obtained using \thco data could be due to the fact that it is tracing better the enveloping layer of the filaments. 
%\textcolor{red}{Also, \cite{gou2021}, found a filament width of 0.6 pc for California molecular filament using the PMO data. 
%Similarly, based on dust continum analys \cite{wang_2015} gives a census of large-scale filaments (L $\sim$ 37--99 pc) having filament width in the range 0.6--3.0 pc at distances of 2.8–-5.4 kpc.}

Future, high-resolution molecular data may be able to better characterize the filaments of \cloud. However, with the available data, we proceed to derive the properties of the filaments, such as mean line mass and column density, as well as the kinematics and dynamics
of the filaments along their spines. %Additionally, we investigate their mass reservoir, stability, and gas kinematics, as these factors are crucial in understanding the role of filaments in star formation and star cluster formation within the cloud.
%\subsubsection{Filament Mass}

We estimated the total mass of the filaments within their widths using the same procedure 
discussed in Section \ref{clump_pro}, and then divide
the mass by the length of the filaments to obtain mass per unit length, M$_{\rm{line}}$, of the filaments. The properties of the filament, such as total mass, mean \nht, aspect ratio (i.e. length/width), and M$_{\rm{line}}$ are tabulated in Table \ref{tab:fil_prop}. The aspect ratios of the filaments are in the range of 4$-$10. Generally, a filament is characterized by an elongated
structure with an aspect ratio greater than $\sim$ 3$-$5 \citep{Andre2014}.  
%We note, 
%Before obtaining Mass$_{line}$, we
%The filament F2 consists of three clumps (C3, C4, \& C5, see Section \ref{clump} for details on clump detection), therefore to find the mass of the filament only,
%we subtract projected clumps' mass from the mass of
%the individual filaments for obtaining the  mass 
%and M$_{line}$ of the filaments. %found the mass of the filament F2 to be around 4.4 $\times$ 10$^3$ \Ms.
%The mass in the filaments ranges from $\sim$ 1.3 $\times$ 10$^3$ to  4.4 $\times$ 10$^3$ with Filament F2 having the highest mass, and the total mass contained in the filaments is around 1.4 $\times$ 10$^4$, which is around 21\% of the total cloud mass.
%\textcolor{red}{The mass of St-6 or filament F6 is complicated to determine as it consists of clumps, and its boundary is not well defined.}
%Using filament mass and length, we can derive the filament line mass, $M_{line}$ (\Ms pc$^{-1}$). 
The M$_{\rm{line}}$ of the filaments F1, F2, F3, F4, F5, and F6 is found to be 92, 171, 93, 138, 233,  and 396 \Ms~pc$^{-1}$, respectively, with a mean around 187 \Ms~pc$^{-1}$.
%\textcolor{red}{The  M$_{line}$ estimation of filament F6 is complicated to determine as it consists of overlapping structures, none the less, from its present shape shown in Fig. \ref{fig_fil_cut}, we crudely  estimated its
%$M_{line}$ to be $\sim$ 396 \Ms pc$^{-1}$. }
M$_{\rm{line}}$ is a critical parameter for assessing the dynamical stability of the filaments, which we discuss in Section \ref{fil_sta}.

%The Mass$_{line}$ might be the lower limit by taking the larger filament lengths, e.g., for filament - F3, most of the filament mass is contained within the cores, while the filament has a long tail having diffused emission. So, total Mass$_{line}$ may not be a good choice to define the stability of such filaments \citep{kim2022}.    
%This is consistent with filaments found in the Galactic high-mass star-forming regions (Hill et al. 2012 for Vela C; André et al. 2016 for NGC 6334). Such filaments with very large line-mass are supposed to be quite unstable against the gravitational collapse unless we assume that there is an additional support force, such as strong magnetic fields (see also discussions in André et al. 2016).
  
\begin{table*}
\caption{Filament properties determined from \thco. The filament F1, F2, and F3 are extracted in the velocity range, [$-$37.0, $-$34.0] \kms, while the filament F4, F5, and F6 are extracted in the velocity range, [$-$37.0, $-$30.0] \kms. } 

\begin{tabular}{|p{1.2cm}|p{1.2cm}|p{1.2cm}|p{1.5cm}|p{1.5cm}|p{1.5cm}|p{2.0cm}|p{1.2cm}} 
\hline
\hline
Filament  & Length & Width & Mass & Mass$_{line}$ & Mass$_{crit}$ & Mean N(H$_2$) & $\mathrm{\frac{\sigma_{nt}}{c_s}}$\\

&  (pc) & (pc) & ($\times$ 10$^3$ \Ms) & (\Ms~pc$^{-1}$) & (\Ms~pc$^{-1}$) & ($\times$ 10$^{21}$ \cms) \\
\hline
\hline
F1 &  38.0 & 3.61  & 3.5  & 92 & 90 & 1.1 & 2.4 \\
F2 &  25.7 & 4.16  & 4.4  & 171 & 94 & 3.0 & 2.4 \\ 
F3 &  14.0 & 3.53  & 1.3  & 93 & 90 & 1.5 & 2.4 \\ 
F4 &  18.1 & 3.28  & 2.5  & 138 & 215 & 2.3 & 3.7 \\
F5 &  14.6 & 2.69  & 3.4  & 233 & 495 & 4.4 & 5.8\\
F6 &  17.4 & 2.49  & 6.9  & 396 & 283 & 9.0 & 4.0\\

\hline
\hline

\end{tabular}

\label{tab:fil_prop}

\end{table*}

\begin{figure*}
    \centering
    \includegraphics[width=16.8cm]{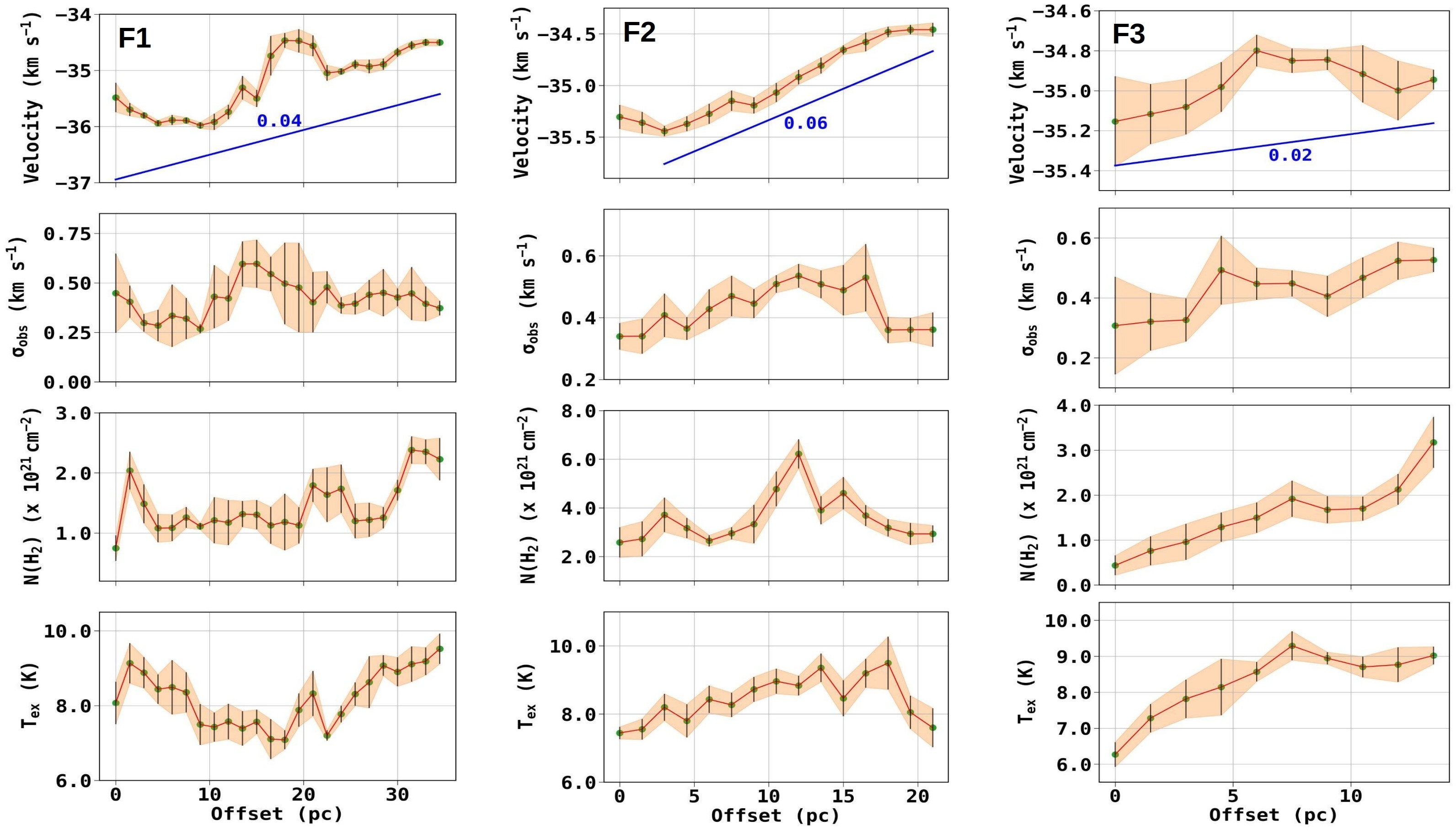}
    \includegraphics[width=16.8cm]{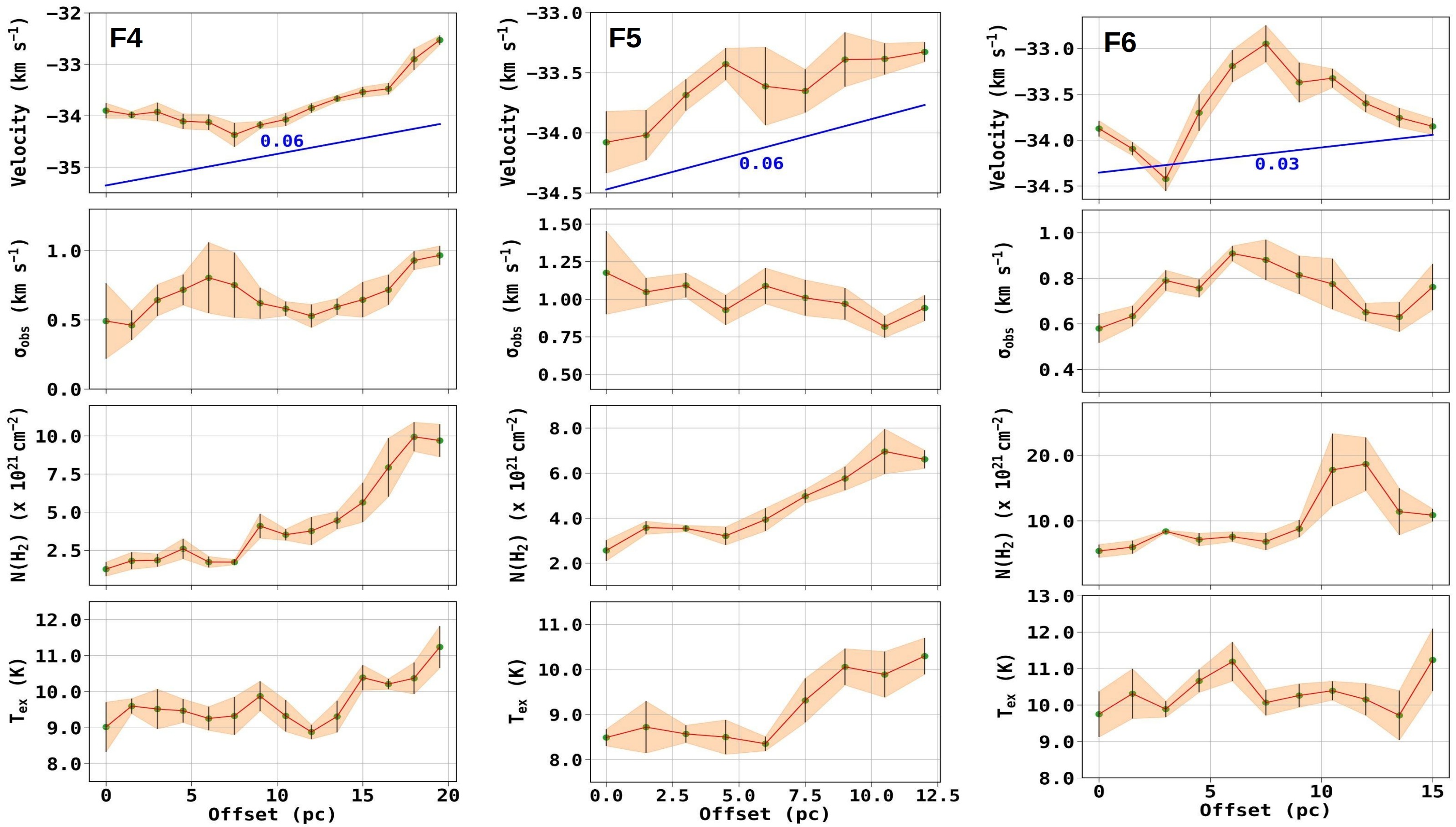}
    \caption{The average velocity, velocity dispersion, column density, and excitation temperature as a function of distance from the filament tail to the head, determined using \thco. The offset 0 pc is at the filament tail. The error bars show the statistical standard deviation at each point. The blue solid line in the top panel of each filament plot shows the linear fit to the data points, whose slope (marked in the plot) gives the velocity gradient along the filament.
    }
    \label{fig_fil_pro}
\end{figure*}

\subsubsection{Kinematics and  dynamics of the gas along the spine }
\label{fil_kin}

%The velocity flow structure along the filaments is evident from the channel map, velocity map, and velocity dispersion map (see Fig. \ref{fig_chan} and \ref{fig_vel}). Position-velocity diagrams of the filaments are  shown in Appendix C. The position-velocity plots are extracted along the paths, which roughly trace the spines of the filaments, shown in \ref{fig_fil_cut}. A hint of velocity gradient from the tail of the filaments towards the hub can be seen for filaments F1, F2, and F5. For filament F3, there is no clear evidence of velocity gradient. Filament F4 also shows a velocity gradient towards the clump, C2. Fig. \ref{fig_hub_prof}a shows the position-velocity diagram of the ridge, in which there is a velocity gradient from the north-east side up to the hub. After the hub, there is a relatively small variation in the velocity up to south-west. Although, if we look into the small path within the ridge, RF1 and RF2 regions (shown by dashed-dotted arrows in Figure \ref{fig_hub_prof}a same as in Fig. \ref{fig_velcut}), which connects the upper junction point or clump C3 and clump C2 with the hub, a gradient seems to be present towards the hub. Fig. \ref{fig_hub_prof}a also shows the positions of the clumps identified from \eico integrated intensity map (see Section \ref{clump}.\\

To examine the kinematics, physical conditions, and dynamics of the gas along the filaments, we used \thco molecular line data and estimated the parameters within the filament width. In Fig. \ref{fig_fil_pro}, we show the variation of velocity, velocity dispersion, column density, and excitation temperature along the filament
spines from their tail to head. We refer to the tail as the farthest point of the filament spine from the hub, while
the head is referred to as the tip of the filaments near the hub. 
%and gradient along the filaments towards the main ridge of the \cloud~cloud, 

%We sample the filament spines by circles of size around 3 $\times$ pixel size ($\sim$ 90\arcsec) and calculated the mean quantities within them. 

%A mean velocity from each circle is taken as a single point and examined as a function of distance along the filament spine. 
%The distance of each point along the filament is taken as an offset in position (centre of the circle) measured along the filament spine from the starting point (eastern end for SF1 and SF2, southern end for SF3, and western end for F4) of the filament spine. 
%As discussed previously, the velocity range used to for deriving
%these parameters' for filaments - SF1, SF2, and SF3 is [$-$37.0, $-$34.0] \kms, and for filament -F2 and F4 is [$-$37.0, $-$30.0] \kms. 
%Figure \ref{fig_kine}, in the upper panels, shows the variation of velocity as a function of distance along the filament. In the same way, the variation of velocity dispersion, column density, and excitation temperature along the filament is also examined and displayed in Figure \ref{fig_kine}.
%\textbf{The velocity ranges used for different filaments are mentioned in Table \ref{tab:fil_prop}, and the same velocity ranges are also used to see the variation of the above-mentioned filament properties along the filament spine.}

In filamentary clouds, the observed velocity gradient along the long axis of the filaments is referred to as the longitudinal in-fall motion of the gas. To assess the amplitude of the longitudinal flow 
along the filaments' long axis, we estimate the velocity gradient of each filament by doing the linear fit to the observed velocity profile along their spines.
%An exemplary fit is shown by the blue-solid line  
%in the upper panels of Fig. \ref{fig_fil_pro}.
In some filaments (e.g. F6), noticeable fluctuations in the velocity profiles are seen. Similarly, for filaments F1 and F4, we see a negative gradient towards the tail of the filaments. 
% These filaments are host to relatively massive clumps whose local
% gravitational potential could have affected the overall velocity profile of the filament.
This could be due to the local gravitational effect of the compact structures and associated star formation activity \citep[e.g.][]{Peretto_2014, Yang_2023}. For example, in F1, a noticeable dense compact gas is seen in the tail (see Fig. \ref{fig_fil_cut}), which might have reversed the flow of direction due to local gravity. Similar situations have also been seen
in other filaments as well, for example, see Filament Fi-NW of the SDC 13 hub filamentary system \citep{Peretto_2014}.

\begin{figure}
    \centering
    \includegraphics[width=8.5cm]{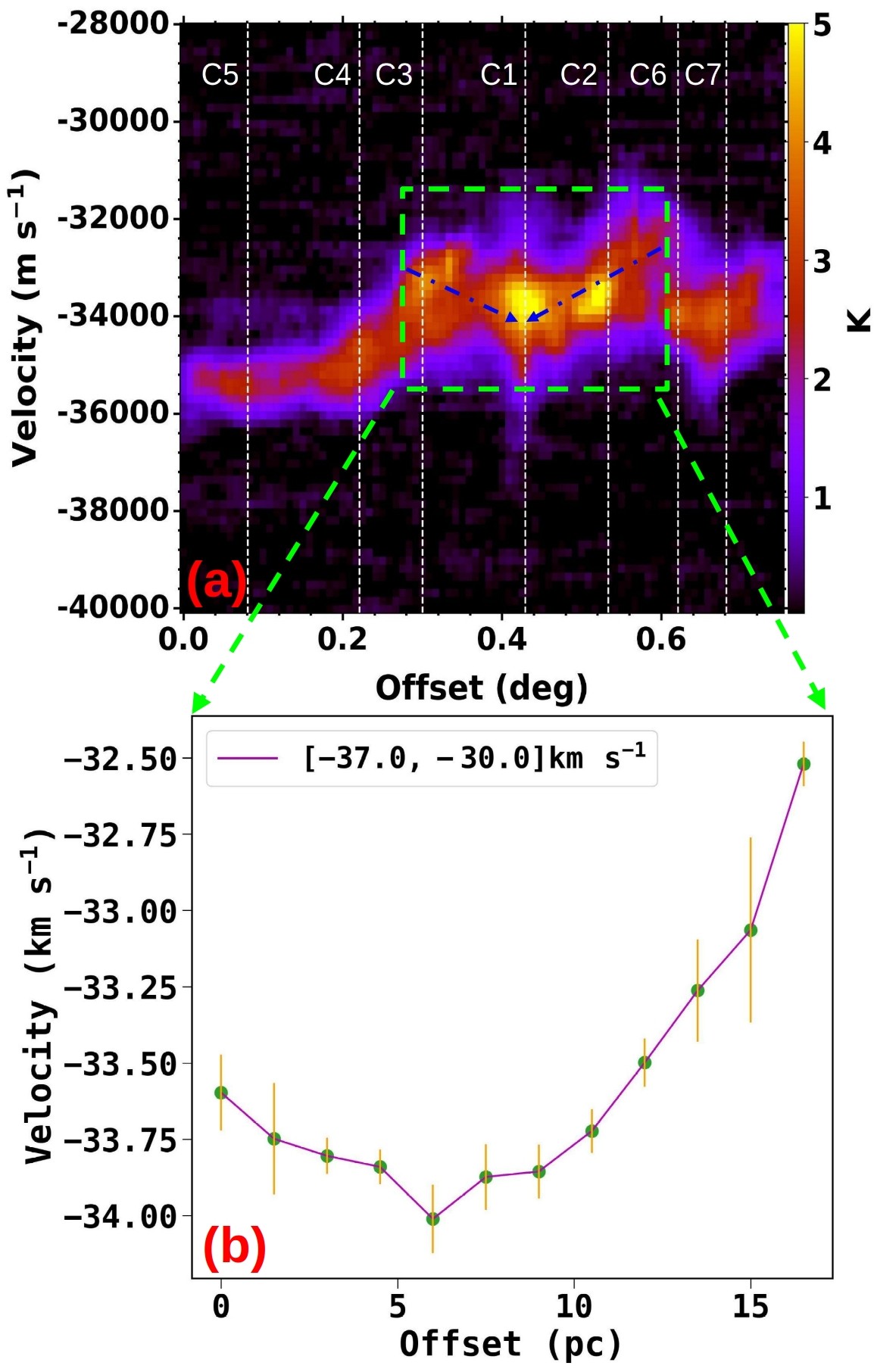}
    %{central_connection_vel_profile.jpeg}
    %\includegraphics[width=8.5cm]{}
    \caption{(a) The position-velocity (PV) diagram of the full ridge based on \thco, which is shown in Fig. \ref{fig_chan}. The green-dashed box shows the region that is used to see the gas flow structure along the blue dashed-dotted arrows, toward the central hub/clump. The vertical dashed lines show the location of identified clumps, marked with their names (see Section \ref{sec:clump}). (b) The variation of average velocity with distance along the arrows (shown in panel a), which shows the velocity gradient towards the central hub/clump. 
    }
    \label{fig_hub_prof}
\end{figure}

From the linear fit, the overall velocity gradient along the filament - F1, F2, F3, F4, F5, and F6 are found to be 0.04, 0.06, 0.02, 0.06, 0.06, and 0.03 \kms pc$^{-1}$, respectively.
%\textbf{Overall, the observed velocity gradients along the filaments though low, among which the filaments F3 and F6 show the lowest or weak velocity gradient. }\\ 
%\textcolor{red}{For a few filaments we see no strong gradient from north to south could be due either to barely any streaming motions along the filament.} 
We note that the observed velocities are line-of-sight projected velocities, and thus, small velocity gradients in some filaments could be due to the filament orientation close to the plane-of-sky. Filaments with low inclination angles would make any identification of gas flows along the filaments very difficult.
%which may be the case for filament-F6/ridge.
None the less, the observed velocity gradient for most of the filaments is close to the velocity gradient observed in large-scale giant molecular filaments (GMFs), i.e. filaments with lengths $>$ 10 pc. \citep[e.g.][]{Ragan_2014, wang_2015, Zhang_2019}. For example, \cite{Ragan_2014} found 0.06 \kms pc$^{-1}$, as an average of the 7 filaments in their sample. Similarly, \cite{wang_2015} find velocity gradient in the range 0.07$-$0.16 \kms pc$^{-1}$ in their sample of GMFs. Similar gradients have also been seen in
some large-scale individual filaments \citep[e.g.][]{Hernan_2015, Zernickel_2015, Wang_2016}. Higher velocity gradients have been observed in filaments at parsec and sub-parsec scales with high-resolution data, particularly in those filaments/elongated structures that are close to the hub or massive clumps \cite[e.g.][]{liu2012,Chen_2020,zhou22,zhou23}.
%Fig. \ref{fig_hub_prof}a shows the Position-Velocity (PV) diagram of the central filamentary area from north-east to south-west (see Fig. \ref{fig_ske}) that covers the spines of filaments F2 and F6 (marked in Fig. \ref{fig_fil_cut}). The positions of the identified clumps are also marked in the figure. An increase in the velocity at the positions of the clumps can be seen from the figure, and there is a dip in the velocity towards the hub/C1. This dip in velocity is seen from the positions of C3 and C2 towards C1, and this region is marked by RF1 and RF2 and shown by arrows in the figure. Figure  \ref{fig_hub_prof}b shows the variation of average velocity with distance along the arrows shown in Fig. \ref{fig_hub_prof}a.
The general finding is that the velocity differences ($\delta V$) between the filaments and central clump/hub become larger as they approach the
central clump \citep[i.e. $\delta V \propto  \delta R^{-1}$, where $\delta R$  is the distance to the clump; e.g. see][]{hacar22}. This is also observed in \cloud~as in the proximity of hub (i.e. within the distance of 3 pc), we find that the associated filaments F2 and F6 show higher velocity gradients,  $\sim$ 0.2 \kms pc$^{-1}$, towards their respective heads, which can be seen from Fig. \ref{fig_hub_prof}. Fig. \ref{fig_hub_prof}a shows the Position-Velocity (PV) diagram of the central filamentary area covering (see Fig. \ref{fig_ske}) spines of the filaments F2 and F6 (marked in Fig. \ref{fig_fil_cut}). The figure also shows the positions of the clumps identified in Section \ref{sec:clump}. Fig \ref{fig_hub_prof}b 
shows the gas velocity variation along the arrows marked in Fig. \ref{fig_hub_prof}a. The gas profile shows a dip in the PV diagram, like the V-shaped structure found in other filaments, which is considered as a signature of gas inflow along the filaments towards a hub/clump \cite[e.g.][]{zhou22}.

%An increase in the velocity at the positions of the clumps can be seen from the figure,
%This is also observed in \cloud~as in the proximity of hub (i.e. within the distance of 3 pc), we find the associated filaments F2 and F6 show higher velocity gradient towards their heads, $\sim$ 0.2 \kms pc$^{-1}$ (see Fig. \ref{fig_hub_prof}b), implying that the inflows along these filaments towards the hub are likely dominated by the gravity of the central clump/hub.

To understand the level of turbulence in the filaments, we also calculated the non-thermal velocity dispersion ($\sigma_{\rm{nt}}$) and Mach number ($M = \sigma_{\rm{nt}}/c_{\rm{s}}$) from the observed velocity dispersion.
%, using the same procedures outlined in Section \ref{clump_pro}. 
We calculate the non-thermal velocity dispersion ($\sigma_{\mathrm{nt}}$) from the total observed velocity dispersion ($\sigma_{\mathrm{obs}}$) using the relation,
\begin{equation}
    \sigma_{\mathrm{nt}} = \sqrt{\sigma_{\mathrm{obs}}^2 - \sigma_{\mathrm{th}}^2},
\label{sig_nt}
\end{equation}
where $\sigma_{\mathrm{th}}$ = $\sqrt{k_\mathrm{B} T_\mathrm{K}/\mu_\mathrm{i} m_\mathrm{H}}$ is the thermal velocity dispersion. $T_\mathrm{K}$ is the gas kinetic temperature, $k_\mathrm{B}$ is the Boltzmann constant, and $\mu_\mathrm{i}$ is the mean molecular weight of the observed tracer (e.g. $\mu (\thco) = 29$ \& $\mu (\eico) = 30$). The mean $\sigma_{\mathrm{obs}}$ is obtained from the velocity dispersion (moment-II) map within the filament region. Using the average $T_{\mathrm{ex}}$ of the filaments as $T_{\mathrm{kin}}$, we calculated the $\sigma_{\mathrm{th}}$ and $\sigma_{\mathrm{nt}}$ of the filaments. Using $\sigma_{\mathrm{nt}}$ and thermal sound speed, $c_\mathrm{s}$ = $\sqrt{k_\mathrm{B} T_\mathrm{K}/\mu m_\mathrm{H}}$ with mean molecular weight per free particle, $\mu$ = 2.37 \citep{kauffmann2008}, we can also calculate the total effective velocity 
dispersion

\begin{equation}
    \sigma_{\mathrm{eff}} = \sqrt{\sigma_{\mathrm{nt}}^2 + c_{\mathrm{s}}^2},
\label{sig_eff}
\end{equation}
The Mach number for the filaments is tabulated in Table \ref{tab:fil_prop}.
%Using $\sigma_{NT}$ and $c_s$ values in equation \ref{sig_eff}, we calculated the total effective velocity dispersion for all the filaments. The Mach number, $M = \sigma_{NT} / c_s$ is given in Table \ref{tab:fil_prop}.  
The gas in the individual molecular filaments of \cloud~is found to be supersonic with sonic Mach number $\sim$ 2$-$6\footnote{
The velocity dispersion may be overestimated if the
molecular lines are optically thick \citep{Gold_1999, Hacar_2016}. 
Along the spine, the optical depth of the \thco emission for most of the filaments is close to 1. Following the suggestion made by \cite{Hacar_2016}, this would increase the line width only by 15\%, suggesting that even after
applying the optical depth correction to the line width, the filaments would remain supersonic.}. This is in agreement with the results of \cite{wang_2015} and \cite{matt18} toward a sample of large-scale filaments measured with the low-resolution (30$-$46\arcs) \thco data using Galactic Ring Survey and SEDIGISM survey data \citep[for details, see Table. 1 of][]{Schuller_2021}. However, we want to stress that the derived properties are from the medium-density tracers such as \thco, but the high-density tracers that would trace very central regions of the filament may give different results. For example, \cite{pineda_2010} comparing high-density and low-density tracers suggested that the sub-sonic turbulence is surrounded by supersonic turbulence in the filaments of the Perseus cloud. Results from high-resolution observations also show that the velocity dispersions of resolved nearby filaments and fibres are close to the sonic or sub-sonic speed \cite[e.g.][]{hacar2013, Friesen_2016, hacar2017, saha22}. All these results tend to suggest that the level of turbulence is scale-dependent, and subsonic velocity coherent filaments possibly condense out of the more turbulent ambient cloud/filament. 

From Fig. \ref{fig_fil_pro}, we also noticed that the majority of filaments exhibit an increasing velocity dispersion as they approach the hub or ridge. The figure also shows that in the majority of the filaments, the increase in velocity dispersion is proportional to the column density of the gas as we move from the tail to the head of the filaments, which is
also evident in the integrated intensity maps shown in Fig. \ref{fig_fil_cut}.
%Besides, we also observed that the filaments with larger velocity gradients tend to  show larger velocity dispersion. 
%Although, we can not complete rule out that the increase in line width
%at the heads of filaments due to richer star-formation activity in the central area of the cloud. 
%It is suggested that turbulent energy is expected to dissipate in
%stagnation regions generated by shock compression, contrary
%to the observed trends, the growing velocity
%dispersion more likely originates from gravity instead of turbulence. 
%In addition to velocity gradient, these signatures supports the  flow-driven simulations of molecular clouds (e.g. Heitsch et al. 2008; Carroll-Nellenback et al. 2013) and unresolved infall motion as proposed by Vázquez-Semadeni et al. (2019). 
%Other  studies  also  mentioned  that  gravitational  collapse  could
%generate  additional  turbulence  motion  (Klessen  \&  Hennebelle
%2010; Peretto et al. 2014; Hacar et al. 2017)
%Supersonic velocity dispersions are expected in the case of gravitationally accelerated infalling material that enters the filaments through shocks at the boundaries (Heitsch et al. 2009).  
In filaments, strong velocity gradients due to rotation have also 
been observed, but primarily at smaller scales, such as close to the dense clump or along the minor axis of the filaments. 
%The latter is interpreted as the merger of sub-filaments of different velocities. 
In the present case, the velocity gradients along the long-axis of the filaments over large scale ($>$ 5 pc), as well as the increase in velocity dispersion and column density as they approach the bottom of the potential well of the cloud, suggest for longitudinal flow of gas along the filaments toward the hub/ridge as found in numerical simulations \citep[e.g.][]{Heitsch_2008, Carroll_2014, sema2019}. 

%As discussed previously, the filament SF1, SF2, and SF3 seem to first move towards the upper junction point (see Figure \ref{fig_chan} and \ref{fig_velcut}), and filament F4 seems to first move towards the lower clump (C2, see Figure \ref{fig_chan} and \ref{fig_dendro}), therefore to observe the inflow of gas from upper junction point and clump C2 towards the central hub, we also explore the velocity gradient between RF1 and RF2 region (see Figure \ref{fig_velcut}). Since we have divided the filamentary analysis in the full-velocity ($-$37.0 to $-$30.0 \kms) and sub-velocity ($-$37.0 to $-$34.0 \kms) range, therefore, we analyze the gas flow structure in both the velocity ranges. Figure \ref{fig_hub_prof} shows the velocity variation from region RF1 to RF2, and it is evident from the figure that there is a gradient from both sides of the hub. 

\subsection{Dense Clumps and Properties}
\label{sec:clump}
Fig. \ref{fig_int} suggests that the cloud has fragmented into several clumpy structures. These are the clumps of the cloud where
star formation could take place. In order to understand the properties and dynamics of these clumps, we utilized \eico data, as it is a better tracer of denser gas and found to be optically thin in \cloud.
%to identify the clumps and determine their properties, such as mass, size, velocity dispersion, and stability, because it is an optically thin tracer of dense gas and is hence a better choice than \thco, which is comparatively optically thick.

\subsubsection{Identification of clumps}
\label{clump}
\begin{figure*}
    \centering
    \includegraphics[width=17.5cm]{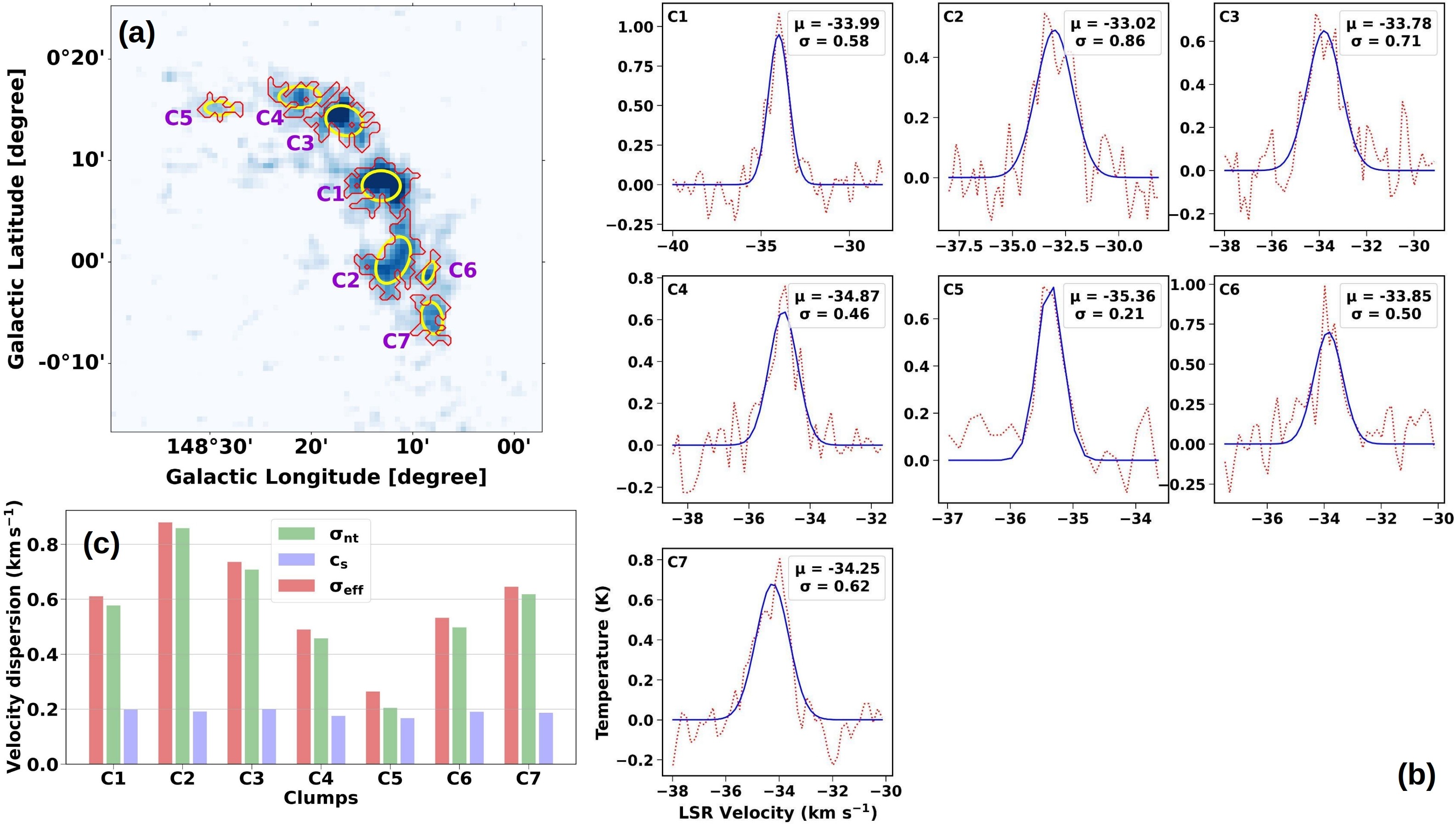}
    \caption{ (a) The location of the clumps identified using ASTRODENDRO over \eico intensity map. The red contours show the leaf structures identified using dendrograms, and the ellipses show the clump within them. (b) The average \eico spectral profile of the clumps over which the solid blue curve denotes the best-fit Gaussian profile, and their respective mean and standard deviation are given in each panel. (c) The histogram plot of non-thermal ($\sigma_{\mathrm{nt}}$), thermal sound speed ($c_\mathrm{s}$), and the total effective ($\sigma_{\mathrm{eff}}$) velocity dispersion of the clumps.}
    \label{fig_dendro}
\end{figure*}

%The clump features are visually evident from the \thco and \eico integrated intensity maps. To clearly identify the clumps in the \cloud~cloud, 
For identifying clumps of \cloud, we implemented the dendrogram \citep{Roso_2008} method using ASTRODENDRO python package\footnote{\href{http://www.dendrograms.org/}{http://www.dendrograms.org/}}. The dendrogram is a structure-finding algorithm that identifies hierarchical structures in the input two- or three-dimensional array. 
%It identifies the nested structures, like clumps and cores, based on intensity or column density in the molecular cloud.
%\citep[see previous work; ][]{Roso_2008, good_2009, Burk_2013, koch2017, and2021}.
The output of the dendrogram depends on three parameters: the \emph{minimum value} that defines the background threshold, the \emph{minimum delta or difference} that defines the separation between two substructures, and the \emph{minimum pixels} that defines the minimum number of pixels or size needed for the structure to be called an independent entity. We ran the dendrogram over the \eico integrated intensity map to find the clumps. We carefully investigate and set the following optimum extraction parameters to detect parsec scale clumpy structures while avoiding faint noisy structures. We set the minimum value to be 3$\sigma$ above the mean background emission, the minimum delta to be 1$\sigma$, and the minimum size to be 12 pixels. Doing so, we identified seven clumps in the cloud, which are marked in Fig. \ref{fig_dendro}a as C1 to C7. The ID, size, and position angle of the clumps are tabulated in Table \ref{tab:clumps}. 

%and their properties are listed in Table \ref{tab:clumps}. 
%The central clump (C2) is the most massive clump having a mass of around 2700 \Ms, which is accreting mass from the filaments. 

\begin{table*}

\caption{Clump properties. The mass, line-width ($\Delta V = 2.35 \sigma_{\rm{obs}}$), virial parameter ($\alpha$), and the ratio of non-thermal velocity dispersion ($\sigma_{\rm{nt}}$) to thermal sound speed ($c_{\rm{s}}$) are calculated using the \eico molecular line data.} 

%The errors quoted here are due to uncertainty in the estimated distance. 

\begin{tabular}{|p{1cm}|p{1.2cm}|p{1.2cm}|p{1.7cm}|p{1.7cm}|p{1.4cm}|p{1.2cm}|p{1.4cm}|p{1.2cm}|p{1.2cm}} 
\hline
\hline
Clump & Major axis & Minor axis & Position angle & Mean \nht & Mass & T$_{\mathrm{ex}}$ & $\Delta V$ & $\alpha$ & $\frac{\sigma_{\mathrm{nt}}}{c_\mathrm{s}}$\\

    & (pc) & (pc) & (deg) & ($\times$ 10$^{21}$ \cms) & (\Ms) & (K) & (\kms) &  &  \\
\hline
\hline
C1 & 2.2 & 1.4 & 219.16  & 11.0 & 2100 & 11.3 & 1.36 & 0.2 & 3.0 \\ 
C2 & 2.5 & 1.5 & 104.45  & 7.0 & 1800  & 10.5 & 2.02 & 0.6 & 4.5 \\ 
C3 & 1.9 & 1.4 & 200.56  & 8.1 & 1600 & 11.5 & 1.67 & 0.4 & 3.5 \\
C4 & 2.0 & 1.0 & 220.56  & 5.6 & 820 & 8.8 & 1.08 & 0.3 & 2.6 \\
C5 & 1.4 & 0.7 & 218.06  & 4.2 & 260 & 8.0 & 0.49 & 0.1 & 1.2 \\
C6 & 1.1 & 0.5 & 108.68  & 7.3 & 280 & 10.4 & 1.18 & 0.5 & 2.6 \\
C7 & 1.6 & 1.0 & 142.86  & 6.7 & 730 & 10.0 & 1.46 & 0.5 & 3.3 \\

\hline
\hline

\end{tabular}

\label{tab:clumps}

\end{table*}

\subsubsection{Properties of the clumps}
\label{clump_pro}
%In section \ref{gas_prop}, we derived the excitation temperature (T$_{ex}$) using the \twco peak main beam temperature, assuming it to be the same for all the CO isotoplogues. Using the same excitation temperature map, we can get the mean value of T$_{ex}$ for the clumps. The mean T$_{ex}$ of the clumps lies in the range $\sim$ 8 to 11.5 K, with the mean around 10.1 K. 
%\textbf{We determined the clump properties like mass, virial parameter, line-width, and velocity dispersion using both the \thco and \eico molecular line data.}
We estimated the mass of the clumps using the integrated intensity emission within the clump boundary, the average excitation temperature from
the excitation temperature map shown in Fig. \ref{fig_temp}, and equations \ref{eq4}, \ref{eq5}, and \ref{eq7} described in Section \ref{glo_con}. The clumps are found to be massive with masses in the range 260$-$2100 \Ms, with the most massive being the central clump, C1, associated with the hub of the cloud. The second most massive clump (C2) is of mass $\sim$ 1800 \Ms.
%however, from visual
%inspection of the intensity map, we find that it actually consists of two close-by dense compact structures. It is plausible that these structures could be two clumps that we were unable to resolve with the present
%data. Consequently,
The mass of C2 is likely an upper limit, as the clump is possibly tracing the part of the filament that connects C1 and C2. 
The effective radius ($R_{\mathrm{eff}}$) of the clump is calculated as $\sqrt{ab}$, where a and b are the semi-major and semi-minor axes of the clump (given in Table \ref{tab:clumps}). The $R_{\mathrm{eff}}$ of the clumps are found to be in the range 0.8$-$1.9 pc, with a mean value of $\sim$ 1.4 pc. %The mass and size of the identified clumps are consistent with the definition of a clump, which has typical sizes between 0.3 and 3 pc \citep{Berg_Tafalla2007}.
%Within the effective radius, we also estimated the clump properties (also given in Table \ref{tab:clumps}) based on \thco emission and found that the clump masses are in agreement with each other within 30\%.
%And if we choose, the  minimum  pixel value corresponding to half beam size, then it resolves 
%the clump into two sub-clumps with mass xx and xx \musn. Although it is recommended to no of pixels, corresponding to 1 beam size for identifying clumps, however, based on the crowding of the structures, half beam size has also been adopted for identifying clumps and cores in molecular clouds (e.g Williams 2022)
%The central clump (C2) is the most massive clump having a mass of around 2700 \Ms, which is accreting mass from the filaments. 
%All the clumps except C6 have effective {\bf sizes $\geq$ 1 pc}.
%The mass and size of the identified clumps are consistent with the definition of a clump, which has typical sizes between 0.3 and 3 pc \citep{Berg_Tafalla2007}.

Velocity dispersion can reflect the level of turbulence in clumps, and the mean line-width, $\Delta V$ of the clump is related to the velocity dispersion ($\sigma_{\mathrm{obs}}$) as 2.35$\sigma_{\mathrm{obs}}$. We get the observed velocity dispersion by fitting a Gaussian profile over the \eico spectrum of the clumps. Fig. \ref{fig_dendro}b shows the average spectral profile of all the clumps. The velocity dispersion of the clumps is in the range of 0.21 to 0.86 \kms, with a mean value of 0.56 \kms.  As determined for the whole cloud, one can also infer whether the clumps are bound or not by calculating the virial parameter, $\alpha = \frac{M_{\mathrm{vir}}}{M_\mathrm{c}}$, where $M_{\mathrm{vir}}$ and $M_\mathrm{c}$ are the virial mass and gas mass of the clumps, respectively. We calculated $M_{\mathrm{vir}}$ %using equation, $M_{\mathrm{vir}} = 126\, R_{\mathrm{eff}} \Delta V^2$ 
using density index, $\beta$ = 2, by assuming a spherical density profile for the clumps.
The $R_{\mathrm{eff}}$, mean T$_{\mathrm{ex}}$, $\Delta V$, $M_{\mathrm{c}}$, and $\alpha$  values of the clumps are tabulated in Table \ref{tab:clumps}. 
The $\alpha$ value of all the clumps is found to be less than 2, suggesting that they are gravitationally bound and, thus, would form or are in the process of forming stars. This will also remain true even if we take $\beta$ =1.5.\\

To determine the contribution of non-thermal (turbulent) support against gravity in the clumps, we calculate the non-thermal velocity dispersion and total effective velocity dispersion from the total observed velocity dispersion, using the same procedures outlined in Section \ref{fil_kin}. Fig. \ref{fig_dendro}c shows the $\sigma_{\mathrm{nt}}$, $c_\mathrm{s}$, and $\sigma_{\mathrm{eff}}$ values of the clumps based on \eico data. From the figure, it can be seen that for all the clumps, the non-thermal velocity dispersion or the turbulence contribution is more dominant than the thermal component. Using the ratio $\sigma_{\mathrm{nt}}/c_\mathrm{s}$, we calculate the Mach number, which is given in Table \ref{tab:clumps}. The Mach number 
%from \thco ranges from 2.6 to 5.5, with a mean around 4, while from \eico, it 
lies in the range of 1.2 to 4.5, with a mean of around 3. Thus, the clumps have supersonic non-thermal motions. The non-thermal motions could be either due to small-scale gas motions within the clump 
%the merger of several velocity components or non-thermal motions generated by filamentary accretion flows
or protostellar feedback due to local star formation activity or a combination of both the processes. For example, \citet{Rawat_2023} discussed that the hub is also associated with a massive YSO (Young Stellar Object) with an outflow, thus, its radiation and feedback might have also impacted the dynamics of the surrounding gas.

\section{Discussion}
\label{dis}
\subsection{Stability of the Filaments}
\label{fil_sta}

The stability of the filament can be evaluated by comparing its observed line mass, $M_{\mathrm{line}}$, with the critical line mass, $M_{\mathrm{crit}}$. 
Assuming filaments are 
in cylindrical hydrostatic equilibrium, $M_{\mathrm{crit}}$, is expressed as \citep{Fiege_Pud_2000}:
%\citep{Ostriker_1964}: 

\begin{equation}
    M_{\mathrm{crit}} = \frac{2\sigma_{\mathrm{eff}}^2}{G} \sim 464\, \sigma_{\mathrm{eff}}^2\, (\Ms\, pc^{-1}),
\end{equation}

where $\sigma_{\mathrm{eff}}$ is the effective velocity dispersion in \kms and G is the gravitational constant. The filament is unstable to axisymmetric perturbation if its line mass exceeds its critical line mass \citep{Inut_1992}. In the case of isothermal filament, $\sigma_{\mathrm{eff}}$ = c$_\mathrm{s}$, where c$_\mathrm{s}$ is the sound speed of the medium \citep[e.g.][]{Ostriker_1964}. In this scenario, the critical line mass only depends on the gas temperature. The average temperature ($T_{\rm{ex}}$) of the filaments estimated within their widths lies in the range 8$-$10 K, which corresponds to $M_{\mathrm{crit}}$ $\sim$ 13$-$17 \Ms~pc$^{-1}$. The line masses for the filaments that we estimated with our data are significantly above the critical thermal line mass. This suggests that either the filaments are collapsing radially or they are supported by additional mechanisms such as non-thermal turbulent motions. These turbulent motions can be generated either due to already formed stars within the filaments or by the radial accretion/infall of
the surrounding gas onto the filaments \citep{Hen_2013, clarke_2016}. The presence of non-thermal motions would increase the effective sound speed, thereby would increase the effective velocity dispersion ($\sigma_{\mathrm{eff}} = \sqrt{c_\mathrm{s}^2 + \sigma_{\mathrm{nt}}^2}$) of the filament, and thus, the critical line mass. At present, the observed velocity dispersion along the filament is higher than that one would expect for a cloud with a temperature in the range 10$-$15 K. Therefore, to understand the present dynamical status of the filaments, we compute the $M_{\mathrm{crit}}$ for the filaments assuming that they are supported by thermal as well as non-thermal motions. Using the mean 
%total or
effective velocity dispersion of the filaments (0.44, 0.45, 0.44, 0.68, 1.03, and 0.78 \kms), we calculated the $M_{\mathrm{crit}}$ values as 90, 94, 90, 215, 496, and 283 \Ms~pc$^{-1}$ for  F1, F2, F3, F4, F5, and F6, respectively, with a mean value around $\sim$ 211 \Ms~pc$^{-1}$. 

%The $M_{crit}$ values obtained from this assumption can provide insights into the stability of the filaments and the formation of stars within them. However, it is important to note that external pressure and magnetic fields can also play a significant role in supporting the filaments and affecting their fragmentation. 

\begin{figure}
    \centering
    \includegraphics[width=8.5cm]{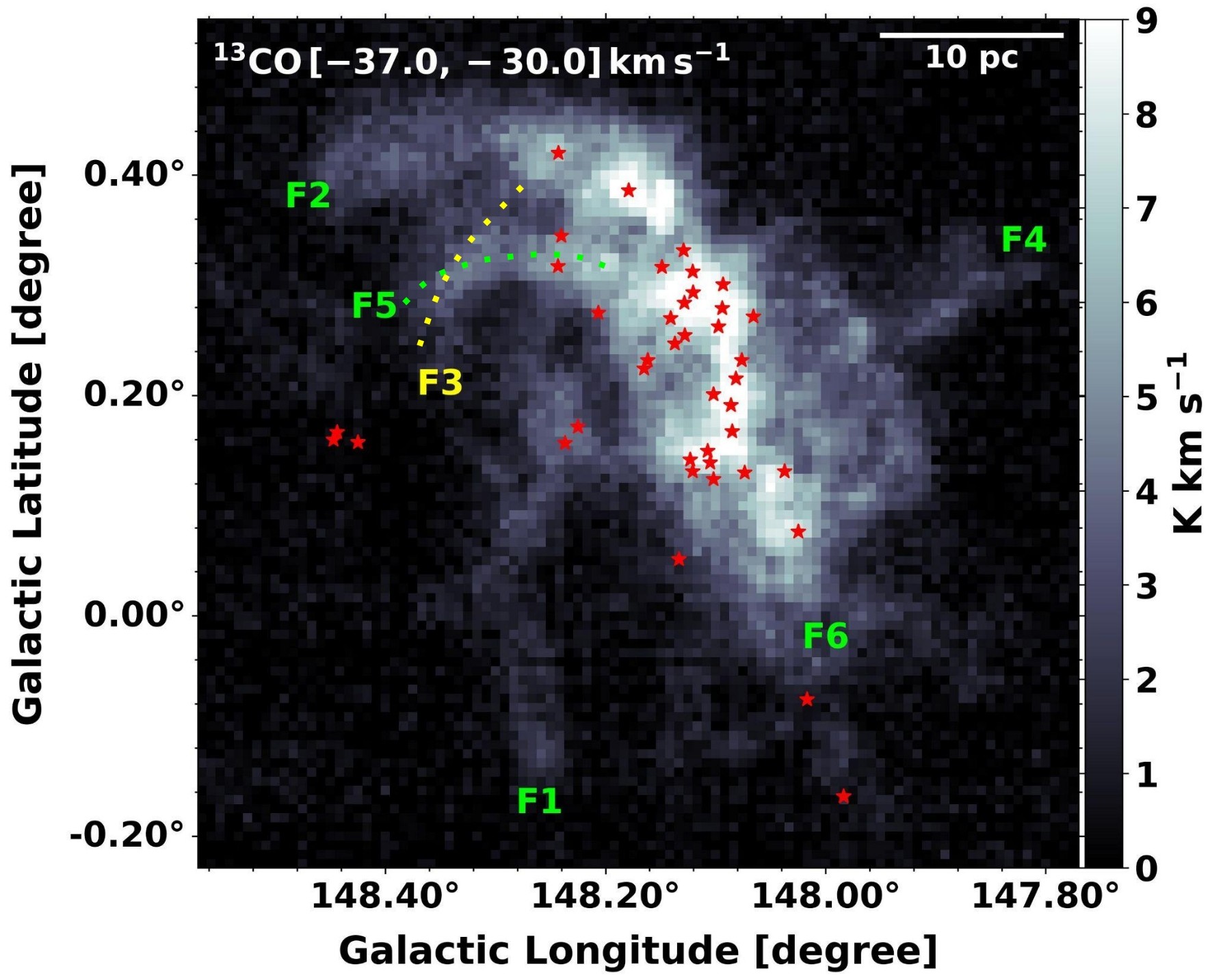}
    %{central_connection_vel_profile.jpeg}
    %\includegraphics[width=8.5cm]{}
    \caption{The distribution of protostars from $\it{Herschel}$ 70 micron point source catalogue \citep{herschel_2020} on the \thco integrated intensity map. 
    }
    \label{fig_yso}
\end{figure}

 \cite{Arzou2019} based on $\it{ Herschel}$ analysis and considering thermal line mass
 as the critical mass, categorised the filaments of the nearby clouds as supercritical filaments ($M_{\mathrm{line}}$ $\geq$ 2 $M_{\mathrm{crit}}$,), transcritical filaments (0.5 $M_{\mathrm{crit}}$ $\leq$ $M_{\mathrm{line}}$ $\leq$ 2 $M_{\mathrm{crit}}$ ), and subcritical filaments ($M_{\mathrm{line}}$ $\leq$ 0.5 $M_{\mathrm{crit}}$). They suggested that thermally subcritical filaments
 are gravitationally unbound entities, while transcritical and supercritical filaments are the preferable sites for gravitational collapse and core formation. Based
 on the thermal line mass, all of our filaments are super-critical, thus, might have undergone collapse and sub-sequence fragmentation to form cores. This fact is evident from the distribution of protostars on the filaments, shown in Fig. \ref{fig_yso}. The figure shows that most of the protostars have been formed in the ridge/F6 of the \cloud~cloud, and a few protostars seem to be formed at the head of the filaments F1, F2, and F5. Taking the contribution of non-thermal
 motion, we find that the line mass of F1, F2, F3, and F6 is larger than their $M_{\mathrm{crit}}$ values, suggesting that they are still gravitationally unstable, whereas for F4 and F5, the $M_{\mathrm{line}}$ is smaller than the $M_{\mathrm{crit}}$ value, suggesting that they are possibly stable against collapse. However, we note that the line masses are estimated with canonical
 values of $\mathrm{^{12}C}$ to $\mathrm{^{13}C}$ isotope ratio and can be higher by a factor of 1.3, if isotopic ratio at the galactocentric distance of the cloud is considered \citep[e.g.][]{Pineda_2013}.
 
 %\textcolor{red}{However, we want to remind that, the masses of the filaments are likely underestimated by a factor of $\sim$ 1.3 as the $\mathrm{^{12}C}$ to $\mathrm{^{13}C}$ isotope ratio at the location of the cloud is likely higher by a factor of $\sim$ 1.3 than what adopted in this work \citep[see][for $\mathrm{^{12}C}$ to $\mathrm{^{13}C}$ isotope ratio variation with galactocentric distance]{Pineda_2013}.}
 
 %This fact is evident from the distribution of young stellar objects (YSOs) (see figure 16 in paper I), which shows that most of the stars have been formed along the filament SF1 and in the ridge of the \cloud~cloud. 
In the above discussion, we have investigated the dynamical status of the filaments, however, we note that in the dynamical scenario of cloud formation and evolution, filaments are very likely to deviate from true equilibrium structures.
%And also, the assumption that filaments are smooth cylindrical structures is an approximation because real clouds are likely to have much more density inhomogeneities than what we have assumed here. 
Because in the dynamical scenario of cloud collapse, filaments are described as dynamical structures that continuously accrete from the ambient gas while feeding dense cores within them. 
% \textbf{where the filament formation can happen either due to gravity-driven gas flow, as advocated in the GHC model \citep{gomez&sema2014, sema2019} or turbulent-driven gas flow discussed in the I2 model \citep{Padoan_2020}.} %with gravity being the main driving force of filament formation \citep[see][]{gomez&sema2014, Naranjo_2015, sema_2017}. 
Moreover, it has also been found that due to the gravitational focusing effect, finite filaments are more prone to collapse at the ends of their long axis \citep{Burkert_2004, Pon_2011}, even when such filaments are subcritical. Thus, though the average properties of some of the filaments are sub-critical, they have a higher concentration of column density at their heads due to longitudinal flow along their axis, where filaments can transit from sub-critical to super-critical.

\subsection{Mass flow Along the Filament Axis}
\label{fil_acc}

%From channel maps and filament profiles, we can qualitatively see that the matter is flowing toward the clumps, but we can also quantitatively estimate the mass accretion rate of the matter flowing through the filaments. 

Assuming that the observed velocity gradient in filaments is due to gas accretion flow, we estimate the mass accretion rate, $\dot{M_\parallel}$ along the filaments using a simple cylindrical model and relation given in \cite{kirk_2013},

\begin{equation}
    \dot{M_\parallel} = V_\parallel \times \rho (\pi r^2) \ \ = \ \  V_\parallel \left(\frac{M}{L}\right),
\end{equation}
where $V_\parallel$ is the velocity along the filament, which is multiplied by the density,  $\rho = \left(\frac{M}{\pi r^2 L}\right)$, and the perpendicular area ($\pi r^2$) of the flow. The r, M, and L are the radius, mass content, and length of the cylinder, respectively. By taking the plane of sky projection with an inclination angle, $\alpha$, the observed parameters of the cylinder are: $L_{\rm{obs}} = Lcos(\alpha)$, $V_{\parallel,\rm{obs}} = V_\parallel sin(\alpha)$, and $V_{\parallel,\rm{obs}} = \Delta V_{\parallel,\rm{obs}} L_{\rm{obs}}$. After simplification, the $\dot{M_\parallel}$ expression reduces to    
\begin{equation}
    \dot{M_\parallel} = \frac{\Delta V_{\parallel,\rm{obs}} M}{tan(\alpha)},
\end{equation} where, $\Delta V_{\parallel,\rm{obs}}$ is the observed velocity gradient along the filament. Taking the obtained mass and velocity gradient of the filaments (see Table \ref{tab:fil_prop}), and $\alpha$ = 45$\degree$ \citep{kirk_2013}, the estimated mass accretion rate for filament F1, F2, F3, F4, F5, and F6 is around $\sim$ 140, 264, 26, 150, 204, and 207 \Ms~Myr$^{-1}$, respectively. Among which, the filaments F2, F5, and F6 are directly tied to the hub (see Fig. \ref{fig_fil_cut}), whose combined accretion rate is around  $\sim$ 675 \Ms~Myr$^{-1}$. We note that the combined mass-accretion rate to the hub is an upper limit as the F6 filament will not transfer its mass entirely to the central hub due to the presence of an additional clump competing with it in the filament. However, we have not accounted for the contribution of small-scale filaments attached to the hub, as seen in the $\it{ Herschel}$ dust continuum image \citep [Fig 14 of][]{Rawat_2023}, which would conversely add to the combined accretion rate.

Taking the above-measured accretion rate as a face value, we find that it is either comparable or higher than some of the well-known cluster-forming hubs found in the literature, such as  Mon R2 \citep[400$-$700 \Ms~Myr$^{-1}$,][]{Morales_2019}, Serpens \citep[100$-$300 \Ms~Myr$^{-1}$,][]{kirk_2013}, Orion \citep[385 \Ms~Myr$^{-1}$,][]{Rodri_1992, hacar2017}, the DR 21 ridge \citep[1000 \Ms~Myr$^{-1}$,][]{sch2010}, G326.27-0.49 \citep[970 \Ms~Myr$^{-1}$,][]{mook_2023}, and G310.142+0.758 \citep[700 \Ms~Myr$^{-1}$,][]{Yang_2023}. %the DR 21 ridge \citep[1000 \Ms~Myr$^{-1}$,][]{sch2010}, SDC 13 \citep[20$-$50 \Ms~Myr$^{-1}$,][]{Peretto_2014}, and G326.27-0.49 \citep[970 \Ms~Myr$^{-1}$,][]{mook_2023}.
%W33 \citep[17000 \Ms Myr$^{-1}$,][]{Liu_2021}, G326.27-0.49 \citep[970 \Ms Myr$^{-1}$,][]{mook_2023}.
This comparison, however, should be treated with caution because all these measurements have been done with different tracers having different resolutions that cover different scales around the hubs. Measuring the accretion rate for massive clouds that have hub filamentary systems, such as those mentioned above, in a uniform way, would give more valuable insight into the accretion rate and the mass assembly time scales of such systems.

\subsection{Overview of Cluster Formation Processes in G148.24+00.41}

\begin{figure}
    \centering \includegraphics[width=8.5cm]{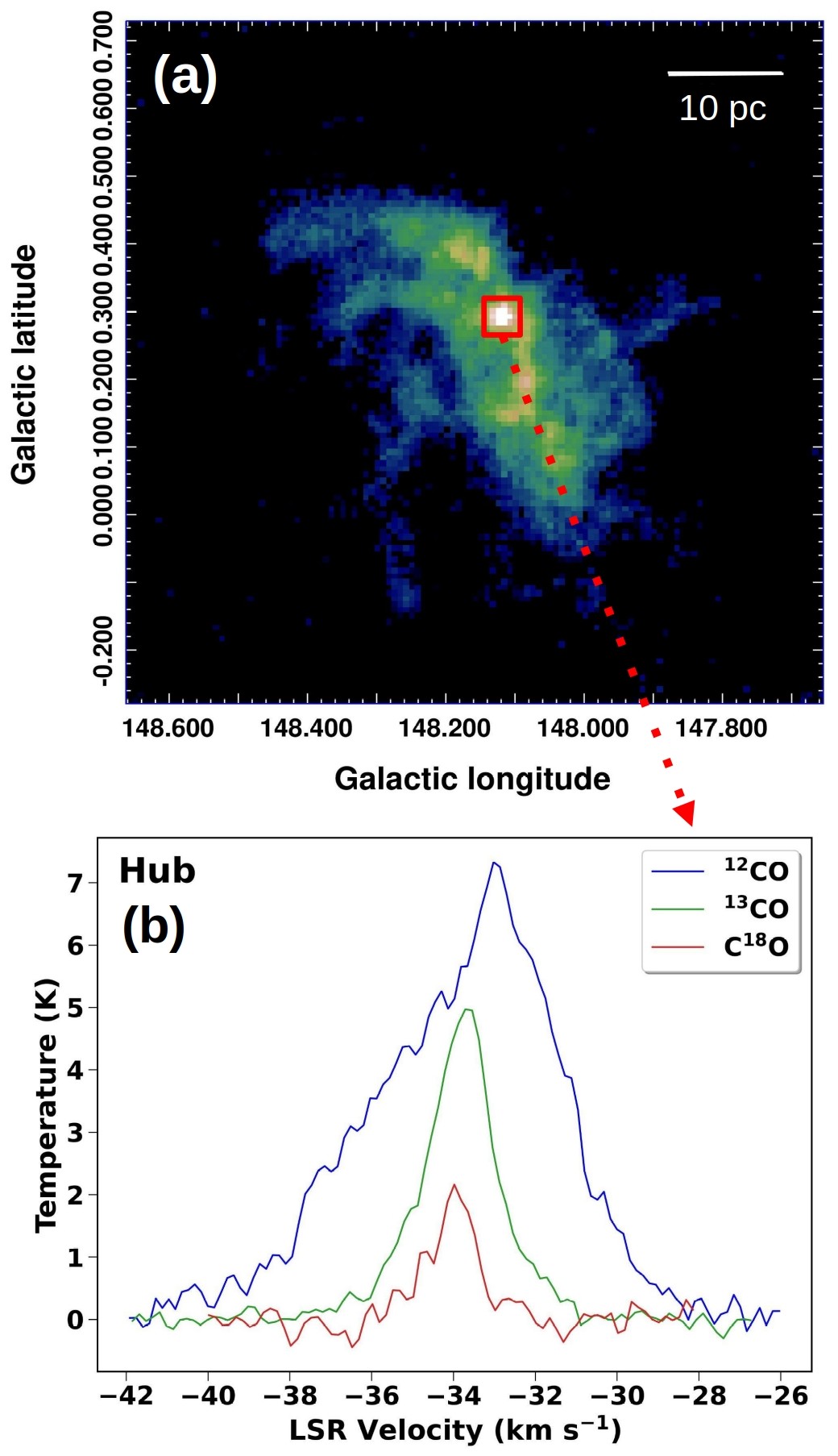}
    \caption{(a) The \thco integrated intensity map showing the location of the hub by a red box, having size $\sim$ 3.5 $\times$ 3.0 pc. (b) The average $\rm{^{12}CO}$, $\rm{^{13}CO}$, and \eico spectral profile of the hub region (shown in panel a). %(c) The \twco (red) and \thco (black) spectra of each pixel at the hub location, along with the contour levels of \thco integrated intensity emission at 10.8, 12.9, and 15.0 K \kms.
    }    
    \label{fig_clumps_disp}
\end{figure}

\cite{Rawat_2023} studied the cloud's profile, structure, and fractalness, as well as the spatial, temporal, and luminosity distribution of the protostars with respect to the cloud's central potential, and suggested that the cloud is likely in a state of hierarchical collapse.

\citet{sema2019} suggested that due to non-homologous collapse in molecular clouds, a classical signature of spherical collapse is not expected over a larger scale. However, at the clump scale, a global velocity offset between peripheral \twco and internal \thco, as found by \cite{barnes22}, is a signature of collapse. According to \cite{Barnes_2019}, if the average \twco profile is red-shifted with respect to the average \thco profile, the motion of the enveloping \twco gas is inwards, while if it is blue-shifted, then the motion is outwards.  Fig. \ref{fig_clumps_disp}b shows the line profiles of the CO-molecules within the 3 pc area (marked by the red rectangle in  Fig. \ref{fig_clumps_disp}a) around the hub. The figure shows that the \twco profile is redshifted with respect to \thco profile, inferring the net inward motion of \twco envelope \citep{barnes22}. However, to get the conclusive signature of infall motion at the clump scale, we need high-density tracer data \citep{Yuan_2018, Liu_2020, Yang_2023}. %\textcolor{red}{Another signature of the inward motion of gas caused by gravity is the self-absorbed emission line profile ("double-peak" or "P-Cygni profile") of an optically thick line \citep{sch2015a}. Since the  \twco is optically thick compared to \thco line in dense regions of molecular clouds,  in Fig. \ref{fig_clumps_disp}c, we also show the relative line profiles of both the lines around the hub area in a $\sim$ 0.5 $\times$ 0.5 pc grid. As can be seen, we do see double-peak profiles towards the centre of the hub, however, with the present data, we can not exclude the contribution of other small-scale dynamical processes, such as outflows and rotation to the shape of the profiles. Future high-resolution and high-dense tracer molecular observation would shed more light on this issue. These results also indicate the collapse of the cloud in the vicinity of the hub.} 

In \cloud, there are six filaments with converging flows heading towards the hub of the cloud. For the filaments having aspect ratio, $\rm{A_0 = Z_0/R_0} \gtrsim 2$, one can calculate the longitudinal collapse timescale using a single equation, $t_{\rm{COL}} \sim (0.49 + 0.26A_0) (G\rho_0)^{-1/2}$ \citep{cla15}, where $\rm{Z_0}$, $\rm{R_0}$, and $\rho_0 = M_{\rm{line}} / \pi \rm{R_0^2}$ is the half-length, radius, and density of the filament, respectively. In our case, the aspect ratio of all the filaments is greater than 2. Using the aforementioned formalism,  we find that the 
longitudinal collapse timescale of these filaments is in the range of 5$-$15 Myr, while the free-fall time of the central clump ($t_{ff} = \sqrt{3\pi/32G\rho\rm{_c}}$, where $\rho\rm{_c}$ is the density of the clump) is found to be $\sim$1 Myr. Since in dynamical hierarchical collapse, each scale accretes from a larger scale, implying that the filaments may continue to fuel the clump for a longer time, provided that they remain bound. Taking the upper limit of combined inflow rate to the C1-clump as $\sim$ 675 \Ms~Myr$^{-1}$, we estimate that to assemble the current mass of the clump, i.e. $\sim$ 2100 \Ms, a minimum time of $\sim$3 Myr would be needed, while the age of the cloud based on formed young stellar objects is around 0.5$-$1 Myr \citep[for details, see][]{Rawat_2023}. This implies that while the mass assembly is ongoing towards the clump, the star formation in the cloud might have initiated around 0.5$-$1 Myr ago. However, we acknowledge that our estimated mass assembly time scale to the C1-clump can be an upper limit due to the following reasons: i) the accretion rate was higher during the early phase of cloud evolution, ii) overestimation of the clump mass due to low-resolution data, iii) missing the contribution of other small-scale filaments such as those seen in $\it{Herschel}$ images, or a combination of all. Future high-resolution observations focusing on the clump area would shed more light on the latter two hypotheses. None the less, the derived accretion rate is close to those found in some of the well-known cluster-forming hub-filamentary systems (discussed in Section \ref{fil_acc}) and also to the prediction of massive cluster-forming simulations \citep[e.g.][]{sema_2009,how18}. For example, \cite{sema_2009} using numerical simulations, suggest that the formation of massive stars or clusters is associated with large-scale collapse involving thousands of solar masses and accretion rates of $\sim$ 10$^{-3}$ \Ms~yr$^{-1}$. 
%In contrast, low- and intermediate-mass stars or clusters in their simulation are associated with isolated accretion flows that are a factor of 10 smaller. 

\begin{figure}
    \centering \includegraphics[width=8.5cm]{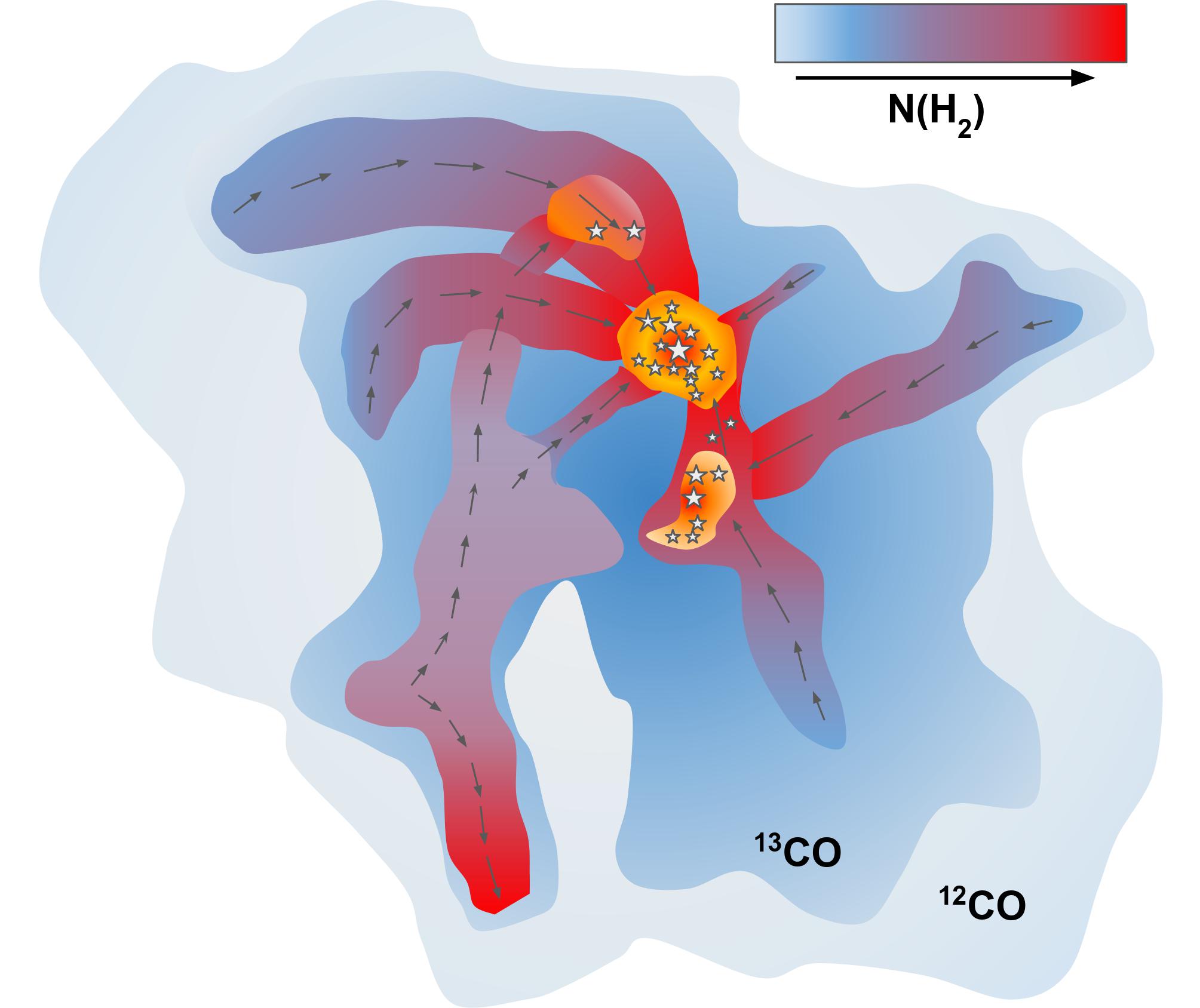}
    \caption{Cartoon illustrating the observed structures in \cloud. The black arrows represent the directions of the overall gas flow. The background colour displays the local density of \twco and \thco.}    
    \label{fig_cartoon}
\end{figure}

The \cloud~cloud has fragmented into seven massive clumps in the range of 260$-$2100 \Ms, and the majority of them have the potential to form an independent group of stars or cluster \citep[e.g. to form a massive star and associated cluster, a minimum mass  $\geq$ 300 \Ms~is needed; see Appendix A in][]{san19}. However, our search for the presence of embedded sources within the clumps using mid-infrared data (i.e. using 3.6 $\mu$m  $\it{Spitzer}$ images) resulted that the massive clumps are associated with stellar sources, and the hub (i.e. the clump C1) hosts the most compact and richer stellar group. \cite{Rawat_2023} also found that the most luminous ($\sim$ 1900 \lsun) protostar of the complex is located within the hub. Thus, we hypothesize that in the \cloud~cloud, the cluster formation in the hub is facilitated by filamentary accretion flows from large-scale cloud to small-scale clumps/hub, which can either be gravity-driven \citep[GHC;][]{gomez&sema2014, sema2019} or turbulence-driven \citep[I2;][]{Padoan_2020}. The cluster in the hub has the potential to grow into a richer cluster by gradually accumulating additional cold gas. \cite{Rawat_2023}, based on the spatial and temporal distribution and fractal subclustering of the stellar sources in \cloud, suggest that GHC may be the dominant mechanism responsible for the formation of the stellar cluster in this cloud. Based on the low-resolution CO data, used in this work, it is difficult to distinguish between the aforementioned two models. %In I2, the kinetic energy is expected to dominate at the clump scale because of its turbulent origin, whereas in GHC, the gravitation energy should dominate up to the cloud scale \citep{Yang_2023}. In our case, we found that the virial parameter is less than one, both for clumps and cloud, implying the dominance of gravitational energy. To check the possibility of the I2 model, we
Future shock tracer observational data would be helpful in this regard, as the I2 model suggests
the formation of filaments due to shocks, while in GHC, the filaments form due to large-scale gravity flow \citep{Yang_2023}. 
% It is, therefore, difficult to favour one model over the other based
% on the results of this study, which opens the scope for future
% observations of this region using high-density tracers and shock
% tracers.} 
Figure \ref{fig_cartoon} illustrates the potential structure and overall gas kinematics of the cloud, forming clusters at the nodes
of the filamentary flows, with the richest cluster being located at the bottom of the cloud's potential.

\section{Summary and Conclusion}
\label{summary}

In the present work, we studied the gas properties and kinematics of the cloud. Based on CO analysis, we confirm that the cloud is massive ($\sim$ 10$^5$ \Ms), bound, and hosts a massive clump of mass $\sim$ 2100 \Ms~nearly at its geometric centre. %\cite{Rawat_2023}, from the $\it{ Herschel}$ dust continuum image, found that the clump is situated at the nexus of several filamentary structures, mimicking the morphology of the hub-filamentary system. 
Based on the low-resolution \thco data, in the present work, we identified six likely velocity coherent, large-scale (length $>$ 10 pc and aspect ratio $>$ 4) filamentary structures in the cloud. Out of which, three filaments (namely F2, F5, and F6) are directly tied to the clump located in the hub. We could not identify and characterize three relatively small-scale filaments that are attached to the hub as seen in the $\it{ Herschel}$ images, thus, their role and properties are not investigated in this work. 
Among the studied filaments, we find that most of them are massive, with high mass per unit length, $M_{\rm{line}}$. We estimated that each filament has the potential to fuel the cold gaseous matter at a rate ranging from 26 to 264 \Ms~Myr$^{-1}$ to the centre of the cloud. 
%From the \thco gas 
%analysis, which is likely tracing the enveloping layer of the filaments, we find that the filaments are turbulent with sonic Mach number between 2$-$6.}

The filaments have undergone fragmentation as several protostars (age $\leq$ 5 $\times$ 10$^5$ yr) that are identified
using 70 $\mu$m and 160 $\mu$m images by \cite{Rawat_2023} are found to be associated with the filaments. Particularly, the filament F6 seems to be associated with a chain of protostars along its spine. The filament F6 has a high line mass, thus possibly actively forming protostars. In the case of the other filaments, the protostars are located close to their respective head, where
strong density enhancement is seen in their respective integrated intensity map. These density enhancements could be due to the filamentary accretion flows along their long axis towards the bottom of the potential well, i.e. towards the hub location. In fact, the velocity profile of the filaments suggests
that each filament is possibly undergoing longitudinal
collapse, as the majority of them tend to show a velocity gradient in the range 0.03$-$0.06 \kms pc$^{-1}$. 
%We want to %stress that, 
% Though the derived gradients are relatively
% low compared to gradients found in the case of hub-filamentary systems (discussed in Section \ref{fil_acc}). 
These gradients are derived over the entire length of the filaments, i.e. for the length scale of 10$-$15 pc %(e.g. gravity-induced free-fall velocity gradient $\propto$ $\sqrt{M_{hub}/R^3}$) 
and found to be comparable to
the gradients of the large-scale filaments. We find that the increase in velocity along the filaments is also correlated with the increase in column density and velocity dispersion. We have also found higher velocity gradients near the hub location (see Fig. \ref{fig_hub_prof}b), implying the acceleration of gas motion towards the hub. We note that, though the kinematic  
features are suggestive of large-scale flows toward the hub, but due to the presence of other clumps in the ridge, the kinematics of the filaments are found to be complicated. Future high-resolution observations will be essential to better understand the kinematics and dynamics of the gas in the filaments and the hub, and unveil the multi-scale process of massive cluster formation.  

The cloud has fragmented into seven massive clumps having mass in the range 260$-$2100 \Ms. We found that the clump located at the hub of the cloud is the most massive one, associated with a massive YSO and a stellar cluster. All these evidence suggests that within the cloud, the hub is the dominant place where a prominent cluster is in the process of emerging. %Also, \cite{Rawat_2023}, based on fractal analysis of the stellar sources in the complex, found that the cloud is moderately fractal. 
Overall, our results are consistent with the flow-driven gas assembly, leading to the formation of the dense clump in the hub and the subsequent emergence of a stellar cluster. 

\section*{ACKNOWLEDGEMENT}

We thank the anonymous referee for the comments and suggestions that helped to improve the paper. The research work at the Physical Research Laboratory is funded by the Department of Space, Government of India. This research made use of the data from the Milky Way Imaging Scroll Painting (MWISP) project, which is a northern galactic plane CO survey with the PMO-13.7m telescope. We are grateful to all the members of the MWISP working group, particularly the staff members at the PMO-13.7m telescope, for their long-term support. MWISP was sponsored by the National Key R\&D Program of China with grant 2017YFA0402701 and CAS Key Research Program of Frontier Sciences with grant QYZDJ-SSW-SLH047. D.K.O. acknowledges the support of the Department of Atomic Energy, Government of India, under project identification No. RTI 4002.
%DKO acknowledges the support of the Department of Atomic Energy, Government of India, under Project Identification No. RTI 4002. 
We thank Eric Koch and Catherine Zucker, for the discussion on using the $\it{FilFinder}$ and $\it{RadFil}$ packages, respectively, to extract the filamentary features and their profiles.

\section*{Data Availability}
We used the CO molecular data from PMO, which can be shared by the PMO database on reasonable request. We have also used {\it Herschel} 250 $\mum$ image and 70 micron point source catalogue. The Herschel images are publicly available from Herschel Science Archive (\hyperlink{http://archi ves.esac.esa.int/hsa/whsa}{http://archi ves.esac.esa.int/hsa/whsa}) and the {\it Herschel} point source catalogue is available on Vizier.

\bibliographystyle{mnras}
\bibliography{G148_cloud.bib} % if your bibtex file is called example.bib

%\appendix
\newpage
\begin{appendix}
\section{FilFinder Algorithm and Large-Scale Filaments}
\begin{figure*}
    \centering
    \includegraphics[width=16cm]{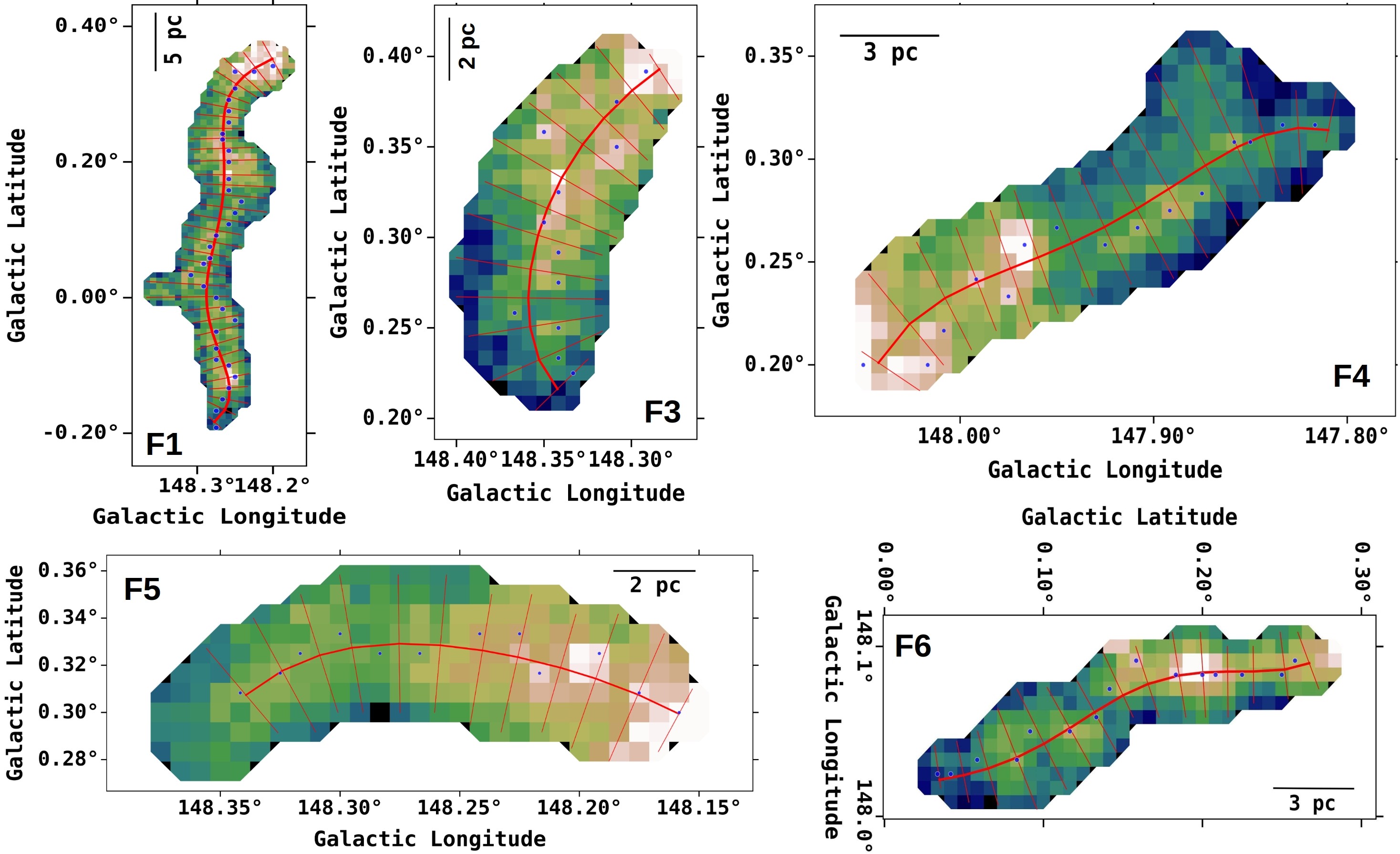}
    \caption{The filaments spines (red solid curve) of F1, F3, F4, F5, and F6 shown over there \thco integrated intensity emission. The blue dots and perpendicular cuts (red solid lines) are the same as in Fig. \ref{fig_spine_cut}a.
    }
    \label{fig_B1}
\end{figure*}

The $\it{FilFinder}$ package picks out the structures within a given mask by comparing each pixel to those in the surrounding neighbourhood using adapting thresholding. The algorithm then reduces the filament mask to skeletons using the Medial Axis Transform (MAT) method. 
%The filament mask can be modified by different parameters like flattening, smoothing, global threshold, size threshold, and other parameters \citep[see][for details]{koch2015}. 
Finally, it prunes down the skeleton structure to a filamentary network. $\it{Filfinder}$ not only extracts bright filaments but also reliably extracts fainter structures such as striations.
%The output of Filfinder strongly depends on
%the input masks that define  the boundaries of the regions of interest. 
%Given the distant nature of the cloud and the low-resolution image being used to extract the filamenatry strcutures, here we aim to extract the likely large-scale filamentary structures of
%the cloud.
We  set the following optimum values in $\it{Filfinder}$ for creating mask and applying adapting thresholding, whose output matches better with the elongated structures visually seen in the column density map: i) global threshold, the intensities below this value are cut off from being included in the mask, as two times the background column density value, ii)
adaptive threshold, the expected full width of filaments for
adaptive thresholding, as 10 pixels (i.e. $\sim$ 5 times beam-width), iii) smooth size, used to smoothen the image to minimize the extraneous branches on the skeletons as 2 pixels (i.e. $\sim$ 1.0 times beam-width), iv) size threshold, the minimum dimensions expected for a filament as 100 pixels$^2$ (i.e. $\sim$ 20 times beam area). The emission structures were first flattened to 95 percentiles before applying the adapting thresholding to suppress the significantly brighter objects than filamentary structures such as dense cores. 
%(improve filfinder write-up, e.g. see \href{https://arxiv.org/pdf/2304.01757.pdf}{see apendix of ASHES paper}) \href{https://hal.science/hal-02428498v1/file/1912.11515.pdf}{30dor}

\section{RadFil and Filament profiles}
\begin{figure*}
    \centering
    \includegraphics[width=6.7cm, height=3.9cm]{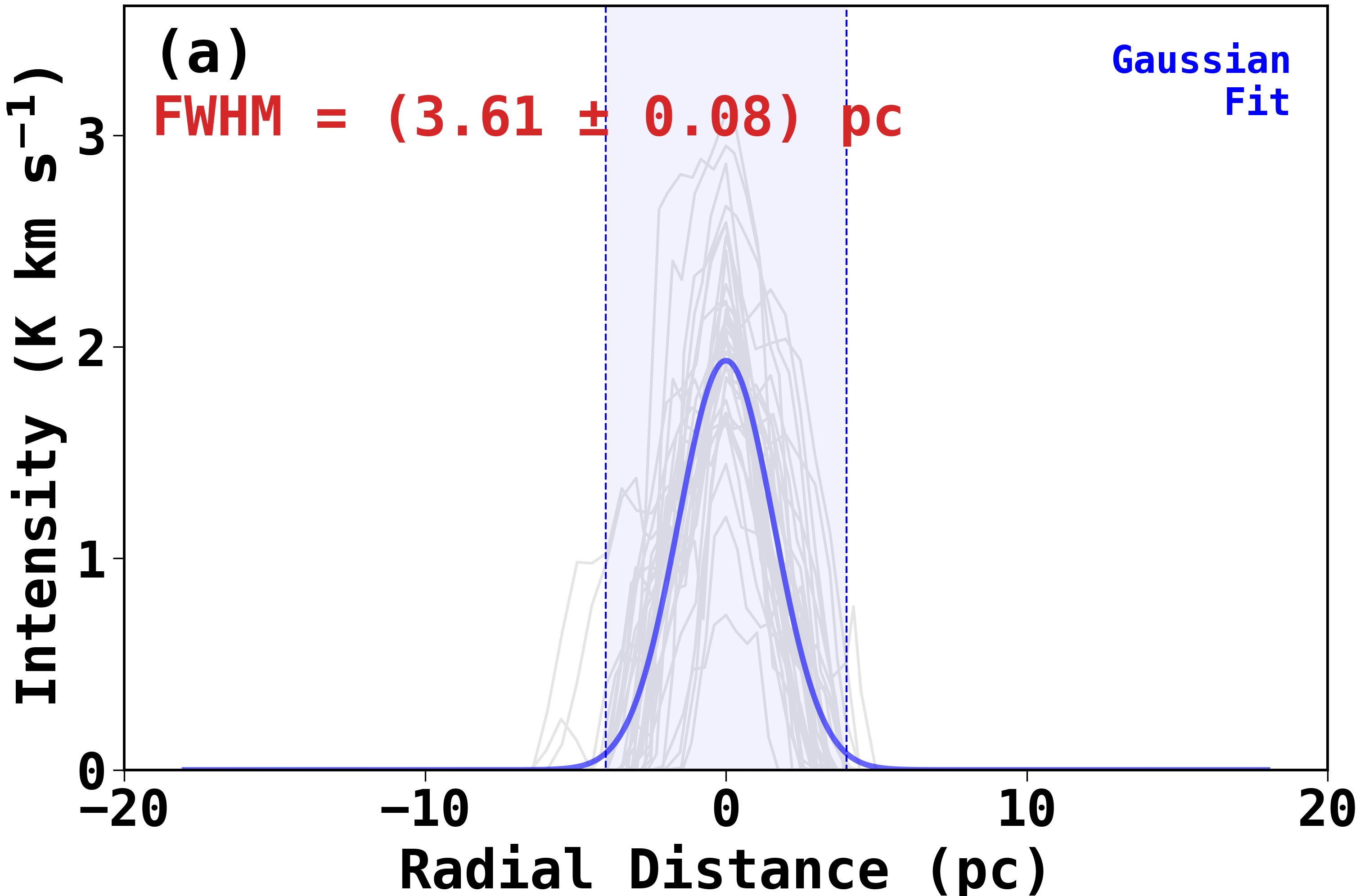}
    \includegraphics[width=6.6cm]{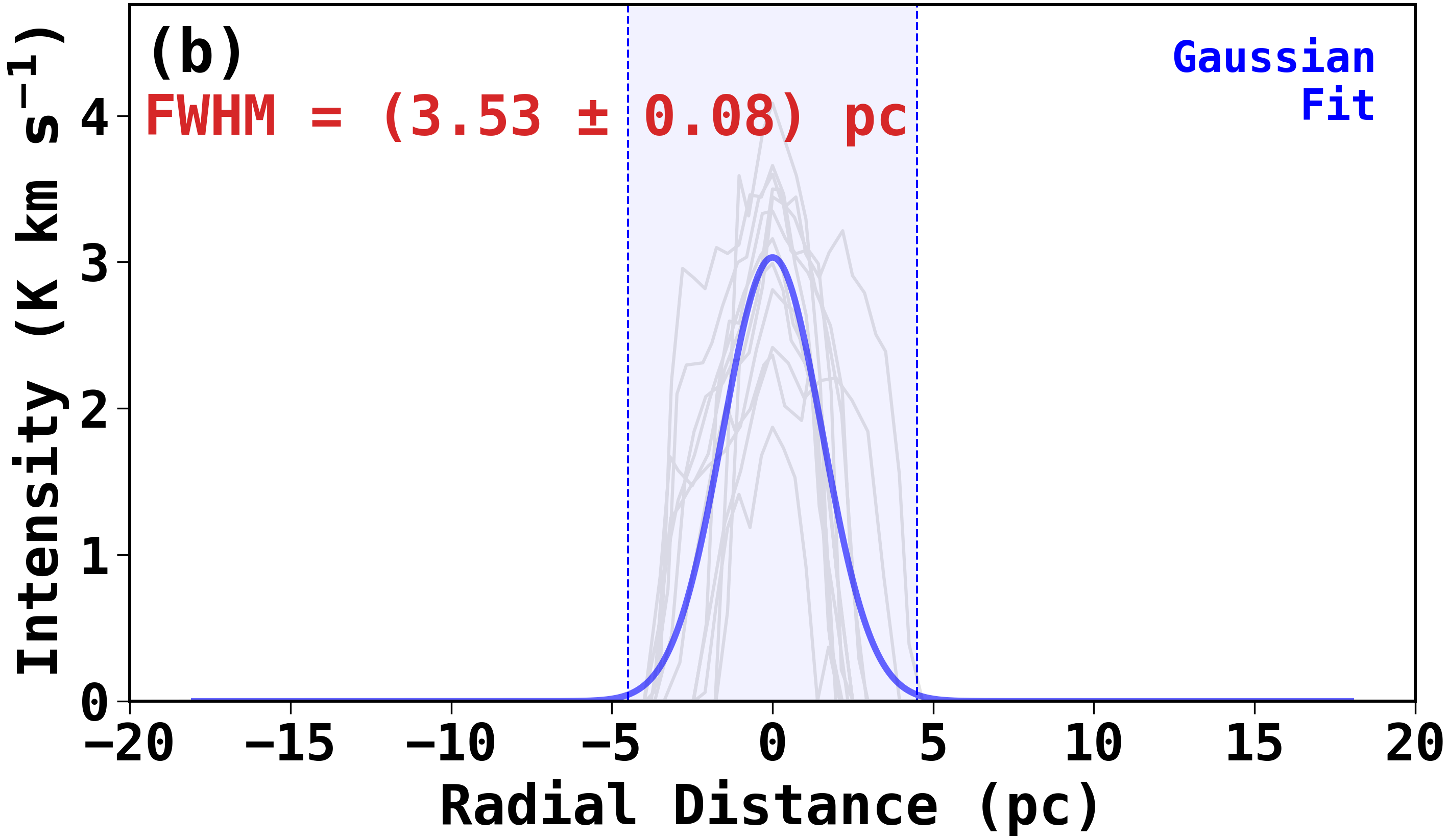}
    \includegraphics[width=6.6cm]{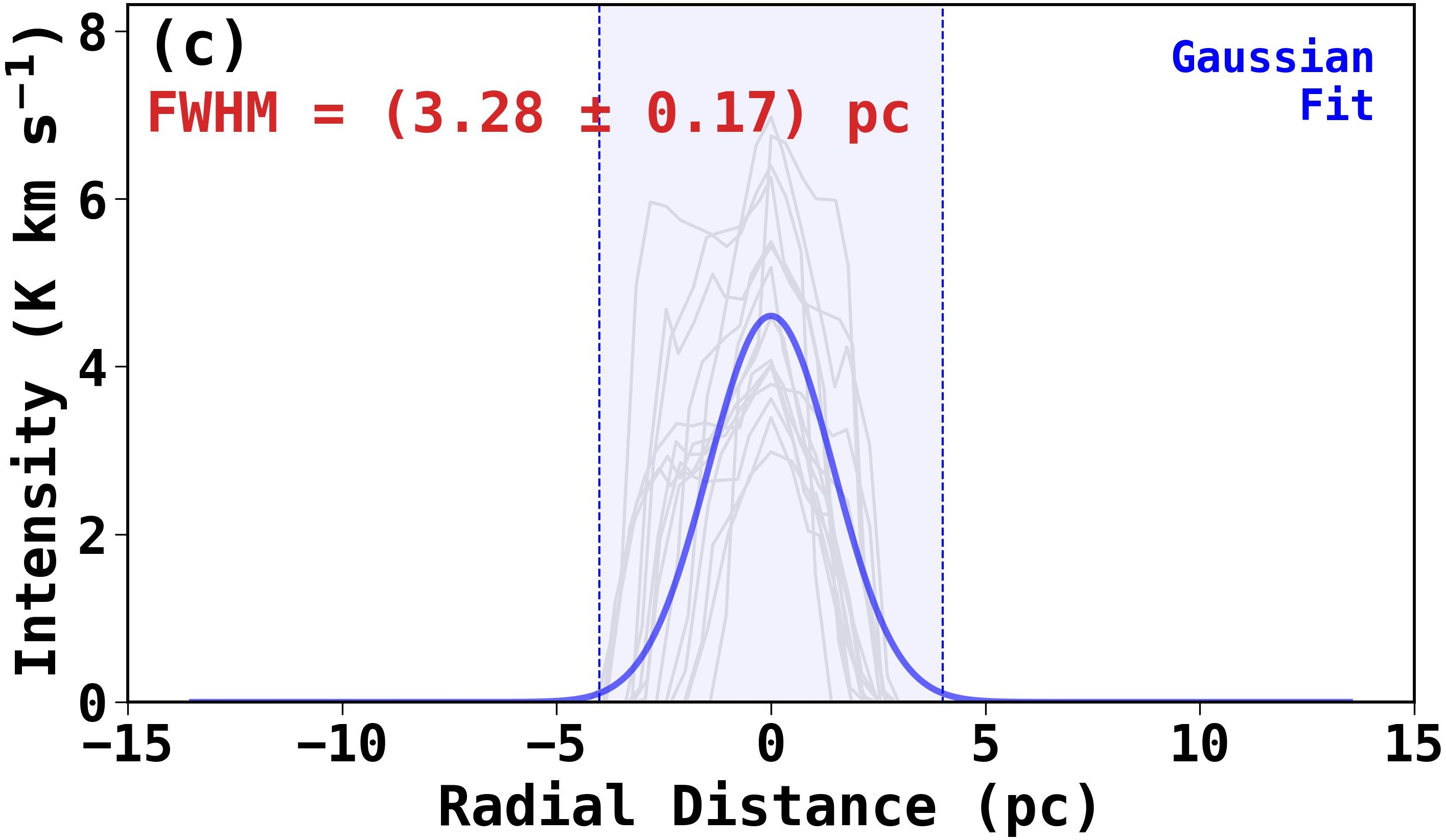}
    \includegraphics[width=6.6cm]{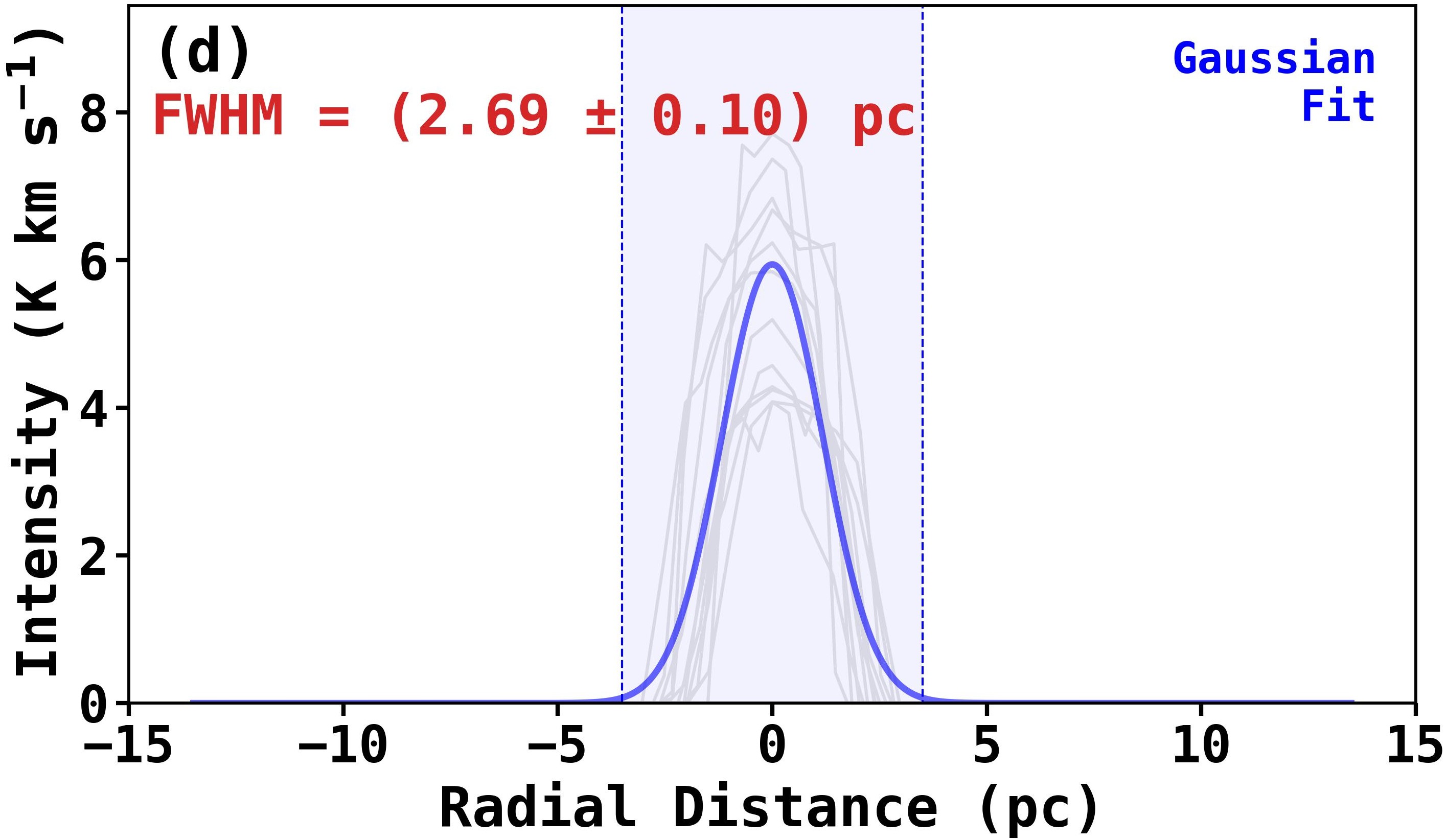}
    \includegraphics[width=6.6cm]{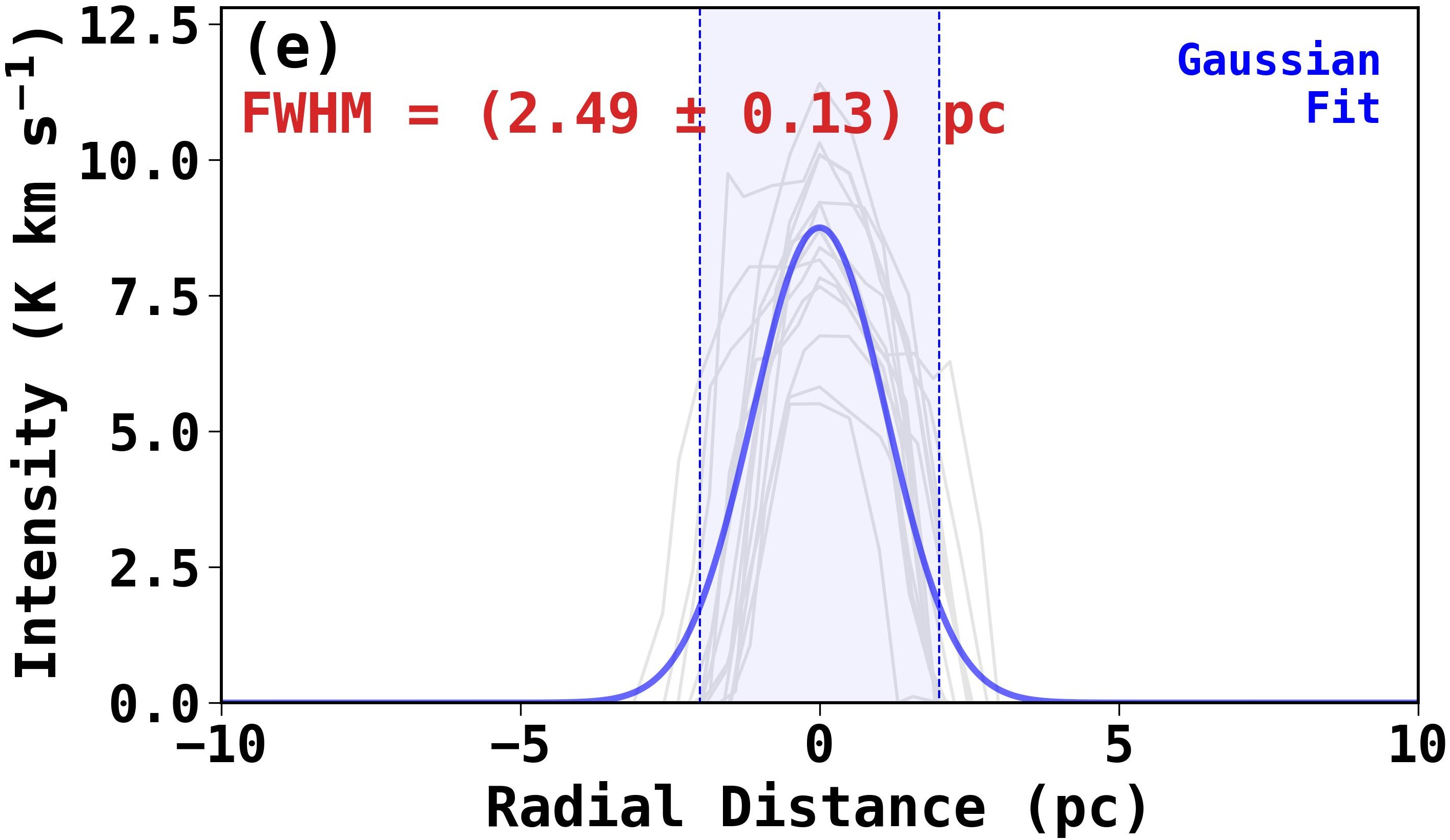}
    \caption{ The radial profiles of perpendicular cuts along the filament spines of (a) F1, (b) F3, (c) F4, (d) F5, and (e) F6, with details same as in Fig. \ref{fig_spine_cut}b. 
    }
    \label{fig_B2}
\end{figure*}

We employed $\it{RadFil}$ to find the filament radial profile, width, and other properties. %These quantities are listed in Table \ref{tab:fil_prop}. The filament lengths mentioned in the table are the sky-projected lengths of the skeletons. To obtain the width 
$\it{RadFil}$ first smooths the filament spine and then makes perpendicular cuts to the tangent lines sampled evenly across the smoothed filament (see Fig. \ref{fig_spine_cut}a and \ref{fig_B1}). %Fig. \ref{fig_spine_cut}a shows an example of the perpendicular cuts along with the smoothed spine over the filament F2. 
Each cut is shifted to the peak intensity along the cut, which is marked by the blue dots in Fig. \ref{fig_spine_cut}a and \ref{fig_B1}. Then, it computes the radial distances from the peak intensity point and the corresponding pixel intensities for each intersecting pixel along a given cut \citep[see][]{zuck-chen2018}. In this way, $RadFil$ generates the intensity profile of each cut. Following the procedure, we made equidistant cuts perpendicular to the spine using a sampling frequency of %2 pixels, which is roughly around 
1 beam size.
%($\sim$ 52\arcsec or 0.9 pc, pixel size = 30\arcsec or 0.5 pc). 
%\textcolor{red}{The centre of each cut is placed at the pixel of maximum intensity along the cut.} 
The radial profile at each cut was extracted, which gives an average profile or master profile of the filament and is shown in Fig. \ref{fig_spine_cut}b and \ref{fig_B2}. 

We identified the filament width (FWHM) by fitting a Gaussian function on the entire ensemble of the cuts. Before fitting the profile, a background was subtracted using the background subtraction estimator of $\it{RadFil}$ \citep{zuck-chen2018}. The background was estimated using the first-order polynomial for all the profiles at a given radial distance from the centre pixel (highest intensity pixel). The radial distance for background estimation is taken in the range where the observed intensity of the radial profile seems to be at a constant level for the filaments. The best-fit parameters are given in Table \ref{tab:fil_prop}. %We obtained the deconvolved FWHM by taking into account the beam size (52\arcsec) as: $\mathrm{FWHM_{decon} = \sqrt{FWHM^2 - FWHM_{bm}^2}}$ \citep{kon2015}, where FWHM$_{bm}$ is the beam size. The obtained FWHM$_{decon}$ for all the filaments are listed in Table \ref{tab:fil_prop}. An example of the Gaussian fit to the profiles of the F2 filament is shown in Fig. \ref{fig_spine_cut}b.

\end{appendix}

%%%%%%%%%%%%%%%%%%%%%%%%%%%%%%%%%%%%%%%%%%%%%%%%%%
% Don't change these lines
\bsp	% typesetting comment
\label{lastpage}

\end{document}